\documentclass{amsart}%
\usepackage{amsmath}
\usepackage{amssymb}
\usepackage{amsfonts}
\usepackage{graphicx}%
\newtheorem{theorem}{Theorem}
\theoremstyle{plain}

\newtheorem{case}{Case}

\newtheorem{conjecture}{Conjecture}

\newtheorem{definition}{Definition}
\newtheorem{example}{Example}

\newtheorem{lemma}{Lemma}

\newtheorem{problem}{Problem}

\newtheorem{remark}{Remark}

\numberwithin{equation}{section}
\begin{document}
\title[Local Zeta Functions and String Amplitudes]{Local Zeta Functions and Koba-Nielsen String Amplitudes}
\author[Bocardo-Gaspar]{M. Bocardo-Gaspar}
\address{Universidad de Guadalajara, Cucei, Departamento de Matem\'{a}ticas, Blvd.
Marcelino Garc\'{\i}a Barrag\'{a}n \#1421,~Guadalajara, Jal. 44430, M\'{e}xico}
\email{miriam.bocardo@academicos.udg.mx.}
\author[Garc\'{\i}a-Compe\'{a}n]{H. Garc\'{\i}a-Compe\'{a}n}
\address{Centro de Investigacion y de Estudios Avanzados del I.P.N., Departamento de
F\'{\i}sica, Av. Instituto Politecnico Nacional 2508, Col. San Pedro
Zacatenco, Mexico D.F., C.P. 07360, Mexico}
\email{compean@fis.cinvestav.mx}
\author[L\'{o}pez]{Edgar Y. L\'{o}pez}
\address{Centro de Investigacion y de Estudios Avanzados del I.P.N., Departamento de
F\'{\i}sica, Av. Instituto Politecnico Nacional 2508, Col. San Pedro
Zacatenco, Mexico D.F., C.P. 07360, Mexico}
\email{elopez@fis.cinvestav.mx}
\author[Z\'{u}\~{n}iga-Galindo]{W. A. Z\'{u}\~{n}iga-Galindo}
\address{University of Texas Rio Grande Valley\\
School of Mathematical \& Statistical Sciences\\
One West University Blvd\\
Brownsville, TX 78520, United States and Centro de Investigaci\'{o}n y de
Estudios Avanzados del Instituto Polit\'{e}cnico Nacional\\
Departamento de Matem\'{a}ticas, Unidad Quer\'{e}taro\\
Libramiento Norponiente \#2000, Fracc. Real de Juriquilla. Santiago de
Quer\'{e}taro, Qro. 76230\\
M\'{e}xico.}
\email{wilson.zunigagalindo@utrgv.edu, wazuniga@math.cinvestav.edu.mx}

\begin{abstract}
This article is a survey of our recent work on the connections between
Koba-Nielsen amplitudes and local zeta functions (in the sense of Gel'fand,
Weil, Igusa, Sato, Bernstein, Denef, Loeser, etc.). Our research program is
motivated by the fact that the $p$-adic strings seem to be related in some
interesting ways with ordinary strings. $p$-Adic string amplitudes share
desired characteristics with their Archimedean counterparts, such as crossing
symmetry and invariance under M\"{o}bius transformations. A direct connection
between $p$-adic amplitudes and the Archimedean ones is through the limit
$p\rightarrow1$. Gerasimov and Shatashvili studied the limit $p\rightarrow1$
of the $p$-adic effective action introduced by Brekke, Freund, Olson and
Witten. They showed that this limit gives rise to a boundary string field
theory, which was previously proposed by Witten in the context of background
independent string theory. Explicit computations in the cases of $4$ and $5$
points show that the Feynman amplitudes at the tree level of the
Gerasimov-Shatashvili Lagrangian are related with the limit $p\rightarrow1$ of
the $p$-adic Koba-Nielsen amplitudes. At a mathematical level, this phenomenon
is deeply connected with the topological zeta functions introduced by Denef
and Loeser. A Koba-Nielsen amplitude is just a new type of local zeta
function, which can be studied by using embedded resolution of singularities.
In this way, one shows the existence of a meromorphic continuations for the
Koba-Nielsen amplitudes as functions of the kinematic parameters. The
Koba-Nielsen local zeta functions are algebraic-geometric integrals that can
be defined over arbitrary local fields (for instance $\mathbb{R}$,
$\mathbb{C}$, $\mathbb{Q}_{p}$, $\mathbb{F}_{p}((T)$), and it is completely
natural to expect connections between these objects. The limit $p$ tends to
one of the Koba-Nielsen amplitudes give rise to a new amplitudes which we have
called Denef-Loeser amplitudes. Along the article, we have emphasized the
explicit calculations in the cases of $4$ and $5$ points.

\end{abstract}
\maketitle

\section{Introduction}

In the recent years, the connections between string amplitudes and arithmetic
geometry, $p$-adic analysis, combinatorics, etc. have been studied
extensively, see e.g. \cite{Schlotterer}-\cite{Zun-B-C-JHEP} and the
references therein.

The string amplitudes were introduced by Veneziano in the 60s,
\cite{Veneziano}, further generalizations were obtained by Virasoro
\cite{Virasoro}, Koba and Nielsen \cite{Koba-Nielsen}, among others. In the
80s, Freund, Witten and Volovich, among others, studied string amplitudes at
the tree level over different number fields, and suggested the existence of
connections between these amplitudes, see e.g. \cite{B-F-O-W}- \cite{Vol}. In
this framework the connection with local zeta functions appears naturally.
This article is intended as\ a survey of our recent work on the connections
between Koba-Nielsen amplitudes and local zeta functions \cite{Bocardo:2020mk}%
-\cite{Zun-B-C-JHEP} in the sense of Gel'fand, Weil, Igusa, Sato, Bernstein,
Denef, Loeser, etc.

The $p$-adic string theories have been studied over the time with some
periodic fluctuations in their interest (for some reviews, see \cite{Brekke et
al}, \cite{Hloseuk-Spect}, \cite{V-V-Z}, \cite{Dragovich:2017kge}). Recently a
considerable amount of work has been performed on this topic in the context of
the AdS/CFT correspondence
\cite{Gubser:2016guj,Heydeman:2016ldy,Gubser:2016htz,Dutta:2017bja}. String
theory with a $p$-adic world-sheet was proposed and studied for the first time
in \cite{Freund:1987kt}. Later this theory was formally known as $p$-adic
string theory. The adelic scattering amplitudes which are related with the
Archimedean ones were studied in \cite{Freund:1987ck}. The tree-level string
amplitudes were explicitly computed in the case of $p$-adic string world-sheet
in \cite{Brekke:1988dg} and \cite{Frampton:1987sp}. One can obtain these
amplitude, in a formal way, from a suitable action using general principles
\cite{Spokoiny:1988zk}. In \cite{Zabrodin:1988ep}, it was established that the
tree-level string amplitudes may be obtained starting with a discrete field
theory on a Bruhat-Tits tree. Determining the convergence of the amplitudes in
momentum space is a difficult task, both in the standard and $p$-adic case;
however, for the latter, this was precisely done for the $N$-point tree
amplitudes in \cite{Zun-B-C-LMP}. In this article we show (in a rigorous
mathematical way) that the $p$-adic open string $N$-point tree amplitudes are
bona fide integrals that admit meromorphic continuations as rational
functions, this is done by associating to them multivariate local zeta
functions (also called multivariate Igusa local zeta functions
\cite{Igusa-old}-\cite{Denef}. In \cite{Bocardo:2020mk} we establish in a
rigorous mathematical way that Koba-Nielsen amplitudes defined on any local
field of characteristic zero (for instance $\mathbb{R}$, $\mathbb{C}$,
$\mathbb{Q}_{p}$) are bona fide integrals that admit meromorphic continuations
in the kinematic parameters. In the regularization process we use techniques
of local zeta functions and embedded resolution of singularities.

The $p$-adic strings are related to ordinary strings at least in two different
ways. First, connections through the adelic relations \cite{Freund:1987ck} and
second, through the limit when $p\rightarrow1$
\cite{Gerasimov:2000zp,Spokoiny:1988zk}. In \cite{Gerasimov:2000zp} was showed
that the limit $p\rightarrow1$ of the effective action gives rise to a
boundary string field theory, this was previously proposed by Witten in the
context of background independent string theory
\cite{Witten:1992qy,Witten:1992cr}. The limit $p\rightarrow1$ in the effective
theory can be performed without any problem, since one can consider $p$ as a
real parameter and take formally the limit $p\rightarrow1$. The resulting
theory is related to a field theory describing an open string tachyon
\cite{Minahan:2000ff}. In the limit $p\rightarrow1$ also there are exact
noncommutative solitons, some of these solutions were found in
\cite{Ghoshal:2004dd}. In \cite{Ghoshal:2006te} a very interesting physical
interpretation of this limit was given in terms of a lattice discretization of
ordinary string worldsheet. In the worldsheet theory we cannot forget the
nature of $p$ as a prime number, thus the analysis of the limit is more
subtle. The correct way of taking the limit $p\rightarrow1$ involves the
introduction of finite extensions of the $p$-adic field $\mathbb{Q}_{p}$. The
totally ramified extensions gives rise to a finer discretization of the
worldsheet following the rules of the renormalization group
\cite{Ghoshal:2006te}.

In \cite{denefandloeser} Denef and Loeser showed that the limit $p$ approaches
to one of a local zeta function give rise a new object called a topological
zeta function. By using Denef-Loeser's theory of topological zeta functions,
in \ \cite{Zun-B-C-JHEP}, we show that the limit $p\rightarrow1$ of tree-level
$p$-adic string amplitudes give rise to a new amplitudes, that we have named
Denef-Loeser string amplitudes. Gerasimov and Shatashvili showed that in the
limit $p\rightarrow1$ the well-known non-local effective Lagrangian
(reproducing the tree-level $p$-adic string amplitudes) gives rise to a simple
Lagrangian with a logarithmic potential \cite{Gerasimov:2000zp}. In
\cite{Zun-B-C-JHEP}, we conjecture that the Feynman amplitudes of this last
Lagrangian are related with the Denef-Loser amplitudes.

In \cite{Compean:2020B-field} we establish rigorously the regularization of
the $p$-adic open string amplitudes, with Chan-Paton rules and a constant
$B$-field. These amplitudes were originally introduced by Ghoshal and Kawano.
In order to study these amplitudes, we introduce new multivariate local zeta
functions depending on multiplicative characters (Chan-Paton factors) and a
phase factor (the $B$-field)\ which involves an antisymmetric bilinear form.
We show that these integrals admit meromorphic continuations in the kinematic
parameters, this result allows us to regularize the Ghoshal-Kawano amplitudes,
the regularized amplitudes do not have ultraviolet divergencies.The theory is
only valid for $p$ congruent to $3$ mod $4$, this is to preserve a certain
symmetry. The limit $p\rightarrow1$ is also considered for the noncommutative
effective field theory and the Ghoshal-Kawano amplitudes.  We show that in the
case of four points, the limit $p\rightarrow1$ of the regularized
Ghoshal-Kawano amplitudes coincides with the Feynman amplitudes attached to
the limit $p\rightarrow1$ of the noncommutative Gerasimov-Shatashvili Lagrangian.

We denote by $\mathbb{K}$ a local field of characteristic zero (for instance
$\mathbb{R}$, $\mathbb{C}$, $\mathbb{Q}_{p}$), and set $\boldsymbol{f}%
:=\left(  f_{1},\ldots,f_{m}\right)  $ and $\boldsymbol{s}:=\left(
s_{1},\ldots,s_{m}\right)  \in\mathbb{C}^{m}$, where the $f_{i}(x)$ are
non-constant polynomials in the variables $x:=(x_{1},\ldots,x_{n})$ with
coefficients in $\mathbb{K}$. The multivariate local zeta function attached to
$(\boldsymbol{f},\Theta)$, where $\Theta$ is a test function, is defined as
\[
Z_{\Theta}\left(  \boldsymbol{f},\boldsymbol{s}\right)  =\int
\limits_{\mathbb{K}^{n}}\Theta\left(  x\right)  \prod\limits_{i=1}%
^{m}\left\vert f_{i}(x)\right\vert _{\mathbb{K}}^{s_{i}}%
{\displaystyle\prod\limits_{i=1}^{n}}
dx_{i},\qquad\text{ when }\operatorname{Re}(s_{i})>0\text{ for all }i\text{,}%
\]
and where $%
{\textstyle\prod\nolimits_{i=1}^{n}}
dx_{i}$ is the normalized Haar measure of $(\mathbb{K}^{n},+)$. These
integrals admit meromorphic continuations to the whole $\mathbb{C}^{m}$,
\cite{Igusa-old}, \cite{Igusa}, \cite{Loeser}, see also \cite{G-S},
\cite{Kashiwara-Takai}. The study of Archimedean and non-Archimedean local
zeta functions was started by Weil in the 60s, in connection to the
Poisson-Siegel formula. In the 70s, Igusa developed a uniform theory for local
zeta functions in characteristic zero \cite{Igusa-old}, \cite{Igusa}, see also
\cite{Loeser}, \cite{Zuniga-Veys}, \cite{Zuniga-Veys2}. In the $p$-adic
setting, the local zeta functions are connected with the number of solutions
of polynomial congruences mod $p^{l}$ and with exponential sums mod $p^{l}$
\cite{Denef}. Recently, Denef and Loeser introduced the motivic zeta functions
which constitute a vast generalization of $p$-adic local zeta functions
\cite{DL1}-\cite{DL2}.

In the 50s, Gel'fand and Shilov studied the local zeta functions attached to
real polynomials in connection with fundamental solutions, indeed, the
meromorphic continuation of Archimedean local zeta functions implies the
existence of fundamental solutions (i.e. Green functions) for differential
operators with constant coefficients. This fact was established,
independently, by Atiyah \cite{Atiyah} and Bernstein \cite{Ber}. It is
relevant to mention here that, in the $p$-adic framework, the existence of
fundamental solutions for pseudodifferential operators is also a consequence
of the fact that the Igusa local zeta functions admit a meromorphic
continuation, see \cite[Chapter 5]{Zuniga-LNM-2016}, \cite[Chapter
10]{KKZuniga}. This analogy turns out to be very important in the rigorous
construction of quantum scalar fields in the $p$-adic setting, see
\cite{M-V-Zuniga} and the references therein.

Take $N\geq4$, and complex variables $s_{1j}$ and $s_{(N-1)j}$ for $2\leq
j\leq N-2$ and $s_{ij}$ for $2\leq i<j\leq N-2$. Put $\boldsymbol{s}:=\left(
s_{ij}\right)  \in\mathbb{C}^{D}$, where $D=\frac{N(N-3)}{2}$ denotes the
total number of indices $ij$. In \cite{Bocardo:2020mk} we introduce the
multivariate local zeta function
\begin{equation}
Z_{\mathbb{K}}^{(N)}\left(  \boldsymbol{s}\right)  :=%
{\displaystyle\int\limits_{\mathbb{K}^{N-3}}}
{\displaystyle\prod\limits_{i=2}^{N-2}}
\left\vert x_{j}\right\vert _{\mathbb{K}}^{s_{1j}}\left\vert 1-x_{j}%
\right\vert _{\mathbb{K}}^{s_{(N-1)j}}\text{ }%
{\displaystyle\prod\limits_{2\leq i<j\leq N-2}}
\left\vert x_{i}-x_{j}\right\vert _{\mathbb{K}}^{s_{ij}}%
{\displaystyle\prod\limits_{i=2}^{N-2}}
dx_{i}, \label{Zeta_Function}%
\end{equation}
where ${\prod\nolimits_{i=2}^{N-2}}dx_{i}$ is the normalized Haar measure on
$\mathbb{K}^{N-3}$. We called these integrals \textit{Koba-Nielsen local zeta
functions}. In \cite[Theorems 4.1, 6.1]{Bocardo:2020mk}, we show that these
functions are bona fide integrals, which are holomorphic in an open part of
$\mathbb{C}^{D}$, containing the set given by
\begin{equation}
\frac{-2}{N-2}<\operatorname{Re}(s_{ij})<\frac{-2}{N}\text{ for \ all }ij.
\label{domain}%
\end{equation}
Furthermore, they admit meromorphic continuations to the whole $\mathbb{C}%
^{D}$.

The Koba-Nielsen open string amplitudes for $N$-points over $\mathbb{K}$ are
\textit{formally} defined as
\[
A_{\mathbb{K}}^{(N)}\left(  \boldsymbol{k}\right)  :=%
{\displaystyle\int\limits_{\mathbb{K}^{N-3}}}
{\displaystyle\prod\limits_{i=2}^{N-2}}
\left\vert x_{j}\right\vert _{\mathbb{K}}^{\boldsymbol{k}_{1}\boldsymbol{k}%
_{j}}\left\vert 1-x_{j}\right\vert _{\mathbb{K}}^{\boldsymbol{k}%
_{N-1}\boldsymbol{k}_{j}}\text{ }%
{\displaystyle\prod\limits_{2\leq i<j\leq N-2}}
\left\vert x_{i}-x_{j}\right\vert _{\mathbb{K}}^{\boldsymbol{k}_{i}%
\boldsymbol{k}_{j}}%
{\displaystyle\prod\limits_{i=2}^{N-2}}
dx_{i}\text{,}%
\]
where $\boldsymbol{k}=\left(  \boldsymbol{k}_{1},\ldots,\boldsymbol{k}%
_{N}\right)  $, $\boldsymbol{k}_{i}=\left(  k_{0,i},\ldots,k_{l,i}\right)
\in\mathbb{R}^{l+1}$, for $i=1,\ldots,N$ ($N\geq4$), is the momentum vector of
the $i$-th tachyon (with Minkowski product $\boldsymbol{k}_{i}\boldsymbol{k}%
_{j}=-k_{0,i}k_{0,j}+k_{1,i}k_{1,j}+\cdots+k_{l,i}k_{l,j}$), obeying
\[
\sum_{i=1}^{N}\boldsymbol{k}_{i}=\boldsymbol{0}\text{, \ \ \ \ \ }%
\boldsymbol{k}_{i}\boldsymbol{k}_{i}=2\text{ \ for }i=1,\ldots,N.
\]
The parameter $l$ is an arbitrary positive integer. Typically $l$ is taken to
be $25$ for bosonic strings. However, we do not require using the critical
dimension. We choose units such that the tachyon mass is $m^{2}=-2$.

In the real case, $A_{\mathbb{R}}^{(N)}\left(  \boldsymbol{k}\right)  $ is (up
to multiplication by a positive constant) the open Koba-Nielsen amplitude of
$N$-points, see \cite[ Section 8]{Brekke et al}, \cite[Section 2]{Kawai et
al}. If $N=4$, $A_{\mathbb{R}}^{(4)}\left(  \boldsymbol{k}\right)  $ is the
Veneziano amplitude \cite{Veneziano}. In the complex case, by using the
results of \cite[Section 2]{Kawai et al}, \cite{Blumenhagen et al}, the
$N$-point, closed string amplitude at the tree level is the product of
$A_{\mathbb{C}}^{(N)}\left(  \boldsymbol{k}\right)  $ times a polynomial
function in the momenta $\boldsymbol{k}$. This fact implies that the
techniques and results presented in \cite{Bocardo:2020mk} are applicable to
classical closed string amplitudes at the tree level.

We use the integrals $Z_{\mathbb{K}}^{(N)}(\boldsymbol{s})$ as regularizations
of the amplitudes $A_{\mathbb{K}}^{(N)}\left(  \boldsymbol{k}\right)  $. More
precisely, we \textit{redefine}
\begin{equation}
A_{\mathbb{K}}^{(N)}\left(  \boldsymbol{k}\right)  =Z_{\mathbb{K}}%
^{(N)}(\boldsymbol{s})\mid_{s_{ij}=\boldsymbol{k}_{i}\boldsymbol{k}_{j}},
\label{Regularization}%
\end{equation}
where $Z_{\mathbb{K}}^{(N)}(\boldsymbol{s})$ now denotes the meromorphic
continuation of (\ref{Zeta_Function}) to the whole $\mathbb{C}^{D}$, see
\cite[Theorem 6.1]{Bocardo:2020mk}. Furthermore, $A_{\mathbb{K}}^{(N)}\left(
\boldsymbol{k}\right)  $ extends to a meromorphic function to the whole
$\mathbb{C}^{(N(l+1)}$, see \cite[Theorem 7.1]{Bocardo:2020mk}.

The articles \cite{Bocardo:2020mk} and \cite{Zun-B-C-LMP} deal with the
meromorphic continuation of local zeta functions of type $Z_{\mathbb{K}}%
^{(N)}(\boldsymbol{s})$, but their results are complementary. The main result
of \cite{Zun-B-C-LMP} \ is the existence of a meromorphic continuation for
$Z_{\mathbb{K}}^{(N)}(\boldsymbol{s})$ in the case in which $\mathbb{K}$ is a
non-Archimedean local field of arbitrary characteristic. For instance
$\mathbb{Q}_{p}$ or $\mathbb{F}_{q}((T))$, the field of formal Laurent series
with coefficients in a finite field $\mathbb{F}_{q}$. The methods used are
based on the stationary phase formula, which allows the explicit calculation
of $Z_{\mathbb{K}}^{(N)}(\boldsymbol{s})$. On the other hand, the results
presented in \cite{Bocardo:2020mk} depend completely on Hironaka's resolution
of singularities theorem \cite{H}, which is available only for fields of
characteristic zero. In this framework, the results of \cite{Bocardo:2020mk}
show that for any field of characteristic zero $\mathbb{K}$, all the zeta
functions $Z_{\mathbb{K}}^{(N)}(\boldsymbol{s})$ converge on a common domain,
see (\ref{domain}), and that the possible poles can be described in a
geometric form. We propose the following conjecture:

\begin{conjecture}
\label{Conjecture A}There is a meromorphic function $\Gamma\left(
\boldsymbol{s}\right)  $ in $\mathbb{C}^{D}$ such that
\[
Z_{\mathbb{R}}^{(N)}(\boldsymbol{s})=\Gamma\left(  \boldsymbol{s}\right)
{\displaystyle\prod\limits_{p<\infty}}
Z_{\mathbb{Q}_{p}}^{(N)}(\boldsymbol{s})\text{ for any }\boldsymbol{s}%
\in\mathbb{C}^{D}\text{.}%
\]

\end{conjecture}

This conjecture and \cite[Theoreem 6.1]{Bocardo:2020mk} imply that $%
{\textstyle\prod\nolimits_{p<\infty}}
$ $Z_{\mathbb{Q}_{p}}^{(N)}(\boldsymbol{s})$ admits a meromorphic continuation
to the whole $\mathbb{C}^{D}$. By using (\ref{Regularization}), this
conjecture implies that%
\[
A_{\mathbb{R}}^{(N)}(\boldsymbol{s})=\Gamma\left(  \boldsymbol{s}\right)
{\displaystyle\prod\limits_{p<\infty}}
A_{\mathbb{Q}_{p}}^{(N)}(\boldsymbol{s})\text{ }%
\]
is meromorphic function in $\mathbb{C}^{D}$. Freund and Witten established
(formally) that the ordinary Veneziano and Virasoro-Shapiro four-particle
scattering amplitudes can be factored in terms of an infinite product of
non-Archimedean string amplitudes \cite{Freund:1987ck}. More precisely, they
established that the product of the $p$-adic string amplitudes multiplied by
the crossing-symmetric Veneziano amplitude is equal to one. Later Aref'eva,
Dragovich and Volovich noted that the infinite product used by Freund and
Witten is divergent, and established that the product of the $p$-adic string
amplitudes multiplied by the crossing-symmetric Veneziano amplitude is equal
to an expression containing the Riemann zeta function \cite{Arafeva et al}.

We already mentioned that the Denef-Loeser amplitudes $A_{top}^{(N)}%
(\boldsymbol{s})$ are the limit of the Koba-Nielsen amplitudes
$\ A_{\mathbb{Q}_{p}}^{(N)}(\boldsymbol{s})$ when $p$ tends to one. This is a
consequence of the fact that $Z_{top}^{(N)}(\boldsymbol{s})=\lim
_{p\rightarrow1}Z_{\mathbb{Q}_{p}}^{(N)}(\boldsymbol{s})$, which in turn is a
consequence of the motivic integration theory, see e.g. \cite{DL1},
\cite{Gusein-Zade}. Motivic integration is a generalization of the $p$-adic
integration and the integration with respect to the Euler characteristic. By
using this theory one can construct a function $Z_{\mathbb{C((}T\mathbb{))}%
}^{(N)}(\boldsymbol{L})$ such that
\[
Z_{\mathbb{Q}_{p}}^{(N)}(\boldsymbol{s})=Z_{\mathbb{C((}T\mathbb{))}}%
^{(N)}(\boldsymbol{L})\mid_{\boldsymbol{L}=p}\text{and }Z_{top}^{(N)}%
(\boldsymbol{s})=Z_{\mathbb{C((}T\mathbb{))}}^{(N)}(\boldsymbol{L}%
)\mid_{\boldsymbol{L}=1},
\]
which in turn implies the existence of new type of `motivic string amplitude'
at the three level $A_{\mathbb{C((}T\mathbb{))}}^{(N)}(\boldsymbol{L})$ which
is a generalization of the $p$-adic Koba-Nielsen and Denef-Loeser amplitudes.

\begin{problem}
Does the motivic string amplitudes $A_{\mathbb{C((}T\mathbb{))}}%
^{(N)}(\boldsymbol{L})$ have a physical meaning? Is there an effective action
such that the Feynman integrals at the three level are exactly the
$A_{\mathbb{C((}T\mathbb{))}}^{(N)}(\boldsymbol{L})$?
\end{problem}

In \cite{Zun-B-C-JHEP}, we verified that the four and five-point Feynman
amplitudes of the Gerasi\-mov-Shatashvili Lagrangian are related by
$\lim_{p\rightarrow1}A_{\mathbb{Q}_{p}}^{(N)}(\boldsymbol{s})$. Then we
propose the following conjecture:

\begin{conjecture}
\label{Conjecture B}The Feynman amplitudes at the three level of the
Gerasimov-Shatashvili Lagrangian are by related $\lim_{p\rightarrow
1}A_{\mathbb{Q}_{p}}^{(N)}(\boldsymbol{s})$.
\end{conjecture}

A precise formulation of this conjecture is given at the end of Section
\ref{Section_5}.

This survey is organized as follows. In Section \ref{Section_1}, we review
briefly basic aspects of string amplitudes. Section \ref{Section_2} aims to
provide an introduction to $p$-adic analysis. Section \ref{Section_3} is
dedicated to the $p$-adic Koba-Nielsen amplitudes. We compute explicitly the
amplitudes for $4$ and $5$ points. At the end of the section we provide an
outline of the proof that the $p$-adic Koba-Nielsen amplitudes admit
meromorphic continuations as rational functions in $p^{-\boldsymbol{k}%
_{i}\boldsymbol{k}_{j}}$. The Koba-Nielsen amplitudes can be defined over any
local field (i.e. $\mathbb{R}$, $\mathbb{C}$, $\mathbb{Q}_{p}$, $\mathbb{F}%
p((T))$, etc.). Section \ref{Section_4} is dedicated to amplitudes defined on
locals fields like $\mathbb{Q}_{p}$, $\mathbb{F}p((T))$. A central idea is
that the limit $p$ tends to one of the $p$-adic Koba-Nielsen amplitudes gives
a new amplitude,\ the Denef-Loeser amplitude. We compute the Denef-Loeser
amplitudes for $4$ and $5$ points. Section \ref{Section_5} deals with the
physical interpretation of the limit $p\rightarrow1$ of $p$-adic string
amplitudes. In this section we show, in the case of $4$ and $5$ particles,
that the Feynman integrals at the tree level of the Gerasimov-Shatashvili
Lagrangian are related Denef-Loeser amplitudes. We also give a precise
formulation of Conjecture \ref{Conjecture B}. Section \ref{Section_6} deals
with the $N$-point $p$-adic string amplitudes, with Chan-Paton rules and a
constant $B$-field. We emphasize the explicit calculations of these amplitudes
in the cases $N=4$, $5$. In Section \ref{Section_7}, we review briefly some
basic definitions and some central results about analytic manifolds on local
fields of characteristic zero, Hironaka's resolution of singularities theorem
and multivariate Igusa zeta functions. In Section \ref{Section_8}, we give an
outline of the proof of the fact that Koba-Nielsen amplitudes defined on local
fields of characteristic zero admit meromorphic continuations in
$\boldsymbol{k}_{i}\boldsymbol{k}_{j}\in\mathbb{C}$. In the cases $N=4$, $5$,
we show explicitly the existence of a meromorphic continuations for for
amplitudes by outlining the construction of certain embedded resolution of singularities.

\section{\label{Section_1}Generalities of String Amplitudes}

String theory is the theory of tiny one-dimensional extended objects
propagating in an underlying Riemannian or pseudo-Riemannian spacetime
manifold $M$. The theory is defined as a theory for a dynamical embedding map
$X:\Sigma_{g,b,N}\rightarrow M$, where $\Sigma_{g,b,N}$ is a compact and
oriented Riemann surface with a possible non-empty boundary ($\partial
\Sigma\not =0$) characterized by the genus $g$, their number of boundaries $b$
and $N$ marked points. $\Sigma_{g,b,N}$ has local coordinates $\sigma
=(\sigma^{0},\sigma^{1})$ ($\sigma^{a}$ with $a=0,1$). Under a Wick rotation
$\sigma^{1}\rightarrow\sigma^{1}$ and $\sigma^{2}=i\sigma^{0}$, the metric
written in terms of coordinates $(\sigma^{1},\sigma^{2})$ becomes a metric
with Euclidean signature. In addition to $X$ there is a non-dynamical filed
$h_{ab}$ on $\Sigma_{g,N}$, which is the intrinsic metric. The string may
oscillate in the target space $M$ and it has an infinite number of
\textit{quantum} modes of oscillation carrying a representation of the
Poincar\'{e} group of the target space \footnote{In the present review will be
enough to consider $M$ to be the flat Minkowskian spacetime} and characterized
by a mass and a spin. In the bosonic string the spectrum consists of a
tachyonic mode with negative square mass. There are a finite number of
massless modes described by massless fields on $M$. These modes are: the
target space metric $g_{\mu\nu}$ with Lorentzian signature $(-,+,\dots,+)$,
the anti-symmetric field $B_{\mu\nu}$, and the dilaton field $\Phi$, where
$\mu,\nu=0,\dots,D-1$. $M$ is the target space which is a Riemannian manifold
of 26 dimensions in the case of the bosonic string and 10 dimensions for the
superstring \cite{Polchinski:1998rq}, \cite{Green:2012oqa}, \cite{Tong:2009np}%
. The Euclidean worldsheet action for the bosonic open string is given by
\begin{equation}
I=S+S^{\prime}, \label{actionI}%
\end{equation}
where
\begin{equation}%
\begin{split}
S=  &  {\frac{T}{2}}\int_{\Sigma}d^{2}\sigma\sqrt{h}\left(  h^{ab}\partial
_{a}X^{\mu}\partial_{b}X^{\nu}g_{\mu\nu}(X)-2\pi i\alpha^{\prime}B_{\mu\nu
}\epsilon^{ab}\partial_{a}X^{\mu}\partial_{b}X_{\nu}\right) \\
=  &  {\frac{T}{2}}\int_{\Sigma}d^{2}\sigma\sqrt{h}h^{ab}\partial_{a}X^{\mu
}\partial_{b}X^{\nu}g_{\mu\nu}(X)-\frac{i}{2}\int_{\partial\Sigma}dlB_{\mu\nu
}X^{\mu}\partial_{t}X^{\nu},
\end{split}
\label{B-fieldAction}%
\end{equation}
where $T=\frac{1}{2\pi\alpha^{\prime}}$ is the string tension, here
$\alpha^{\prime}=\ell_{s}^{2}$ and $\ell_{s}$ stands for the string length. In
the above action the physical degrees of freedom are $X^{\mu}$ and $h_{ab}$,
$\partial_{t}$ is the tangential derivative along $\partial\Sigma$ and
$B_{\mu\nu}$ is an antisymmetric field. Moreover, the action $S^{\prime}$ is
given by
\[
S^{\prime}=\Phi_{0}\chi(\Sigma),
\]
where $\chi(\Sigma)={\frac{1}{4\pi}}\int d^{2}\sigma\sqrt{h}R+BT$ is the Euler
characteristic of $\Sigma$ and $BT$ is the Gibbons-Hawking boundary term that
we encode in $\chi(\Sigma)$ with the number of boundaries $b$. Notice that the
vacuum expectation value of the dilaton $\Phi_{0}$ can also be incorporated in
the action through this topological term. It is also worth noting that this is
the most general action that has worldsheet reparametrization invariance, in
particular Weyl invariance ($h_{ab}\rightarrow\Lambda(\sigma)h_{ab}$), and is
renormalizable by power counting (this implies that there must be two
worldsheet derivatives in each term). The second term is analogous to placing
an electromagnetic field in the background with which the string interacts; it
is a gauge field in that sense. Mathematically it is the integration of the
pullback of the $B$-field to the string worldsheet. If we define $A_{\mu
}:=B_{\mu\nu}X^{\nu}$, the second term of the second line in
(\ref{B-fieldAction}) may be written as
\begin{equation}
\frac{i}{2}\int_{\partial\Sigma}A_{\mu}\partial_{t}X^{\mu},
\label{vertexgauge}%
\end{equation}
which is exactly the term considered in \cite{Fradkin:1985qd} where the
partition function of the action was obtained explicitly at tree level.
Strings propagating in non-trivial backgrounds have been considered in other
works \cite{Abouelsaood:1986gd}.

We now consider flat space as well as a constant $B$-field. The open string
action (\ref{B-fieldAction}) determines the Neumann boundary conditions
\[
g_{\mu\nu}\partial_{\sigma}X^{\nu}+2\pi i\alpha^{\prime}B_{\mu\nu}\partial
_{t}X^{\nu}\rvert_{\partial\Sigma}=0,
\]
where $\partial_{\sigma}$ is the normal derivative to $\partial\Sigma$. We
should note that these conditions are in general complex, because we have
taken a Wick rotation to a Euclidean signature. We will be concerned to the
case when the string worldsheet $\Sigma$ is a disk, which corresponds to the
open string in the classical limit. It is known that he disk can be
transformed into the upper half plane via a conformal transformation, whose
boundary is the real line. In these variables the Neumann boundary conditions
are \cite{Seiberg:1999vs}
\[
g_{\mu\nu}(\partial-\bar{\partial})X^{\nu}+2\pi\alpha^{\prime}B_{\mu\nu
}(\partial+\bar{\partial})X^{\nu}\rvert_{z=\bar{z}}=0,
\]
where $\partial=\partial/\partial z$ and $\bar{\partial}=\partial/\partial
\bar{z}$, with $\operatorname{Im}$ $z\geq0$.

The perturbative scattering amplitudes of $N$ particles are defined as the
correlation function of $N$ vertex operators $\mathcal{V}_{\Lambda_{i}%
}(\boldsymbol{k}_{i})$. Furthermore, these amplitudes admit an expansion over
the genus $g$ of $\chi(\Sigma)=2-2g-b$ as follows:
\[
\mathcal{A}(\Lambda_{1},\boldsymbol{k}_{1},\dots,\Lambda_{N},\boldsymbol{k}%
_{N})=\sum_{g=0}^{\infty}\lambda_{s}^{2g-2+b}A_{g}(\Lambda_{1},\boldsymbol{k}%
_{1},\dots,\Lambda_{N},\boldsymbol{k}_{N}),
\]
where $\lambda_{s}:=e^{\Phi_{0}}$ is the coupling constant of string and
$A_{g}$ is the Feynman functional integral
\[
A_{g}(\Lambda_{1},\boldsymbol{k}_{1},\dots,\Lambda_{N},\boldsymbol{k}%
_{N})={\frac{1}{\mathrm{Vol}}}\int_{\mathfrak{M}_{g,b,N}}DXDh\exp
(-S)\ \prod_{i=1}^{N}\mathcal{V}_{\Lambda_{i}}(\boldsymbol{k}_{i}),
\]
here $\mathrm{Vol}$ stands for the volume of the symmetry groups leaving
invariant the action (\ref{actionI}). Here the integration is carried out over
${\mathfrak{M}_{g,b,N}}$ the moduli space of Riemann surfaces of genus $g$,
$b$ boundaries and $N$ marked points inserted on the boundary. The difficulty
of integration increases with higher values of $g$, $b$ and $N$. Fortunately
it can be carried out in a explicit form for a few simple cases with lower
values of $g$, $b$ and $N$. Thus for instance, the case for closed strings
with genus $g=0$ and $N=4$, or open strings with $g=0$, $b=1$ and $N=4$, are
explicitly performed. Vertex operators $\mathcal{V}_{\Lambda_{i}%
}(\boldsymbol{k}_{i})$ are functionals of the embedding fields $X$ and their
derivatives. They are given by
\[
\mathcal{V}_{\Lambda}(\boldsymbol{k})=\int d^{2}\sigma\sqrt{h}W_{\Lambda
}(\sigma)e^{i\boldsymbol{k}X(\sigma)},
\]
where $W_{\Lambda}(\sigma)$ represents a functional of $X$ and its derivatives
associated to the species of field in the string spectrum, $X(\sigma
)=(X^{0},\dots,X^{25})$, $\boldsymbol{k}=(k^{0},\dots,k^{25})$ are the
position and momentum vectors in target space $M$ and
\[
\boldsymbol{k}X(\sigma):=\sum_{\mu,\nu=0}^{25}k^{\mu}\eta_{\mu\nu}X^{\nu
}(\sigma)
\]
is the Minkowskian inner product. For instance for tachyons $\Lambda=t,$
$W_{t}(\sigma)=1$. For gauge fields $\Lambda=A$ and $W_{A}(\sigma
)=\varepsilon_{\mu}\partial_{t}X^{\mu}$, where $\varepsilon_{\mu}$ is the
polarization vector. The graviton is $W_{G}(\sigma)=\varepsilon_{\mu\nu}%
h^{ab}\partial_{a}X^{\mu}\partial_{b}X^{\nu}$, where $\varepsilon_{\mu\nu}$ is
the polarization tensor. For bosonic closed strings, the tachyon is the lowest
state and it has a $\boldsymbol{k}^{2}={\frac{4}{\alpha^{\prime}}}$, where
$\boldsymbol{k}^{2}:=\boldsymbol{k}\boldsymbol{k}$ is the Minkowskian norm.
The open string tachyon has $\boldsymbol{k}^{2}={\frac{1}{\alpha^{\prime}}}$.
In units where the Regge slope $\alpha^{\prime}={\frac{1}{2}}$ or where the
string tension $T={\frac{1}{\pi}}$ we have that for the closed string tachyon
$\boldsymbol{k}^{2}=8$ and for the open string $\boldsymbol{k}^{2}=2$, which
will be used in the following sections of the present survey. The massless
open string gauge field and the closed string graviton have $\boldsymbol{k}%
^{2}=0$.

For open strings one can see that in the functional integral with action
(\ref{B-fieldAction}) the expansion in Taylor's series of exponential of the
second term of (\ref{B-fieldAction}), leads to a superposition of amplitudes
with different powers of the vertex operator of the gauge field
(\ref{vertexgauge}) inserted in the boundary $\partial\Sigma$.

From now on, we restrict ourselves to the case of bosonic closed strings with
$g=0$ and bosonic open strings with $g=0$ and $b=1$, i.e. the $2$-sphere and
the disk. For closed strings, by worldsheet diffeomorphisms and Weyl
transformations of $\Sigma$, allow us to fix the worldsheet metric $h_{ab}$ to
the Euclidean flat metric $\delta_{ab}$. This is the called the
\textit{conformal gauge}. In the absence of a $B$-field, there is a remnant
$SL(2,\mathbb{C})$ symmetry which cannot be fixed by the local Weyl symmetry.
This is called the \textit{conformal Killing symmetry}. This symmetry does act
on the complex worldsheet coordinates $(z,\overline{z})$ of the 2-sphere
separately. In the holomorphic sector (for the anti-holomorphic sector a
similar argument is applied)
\[
z\mapsto{\frac{az+b}{cz+d},}%
\]
where $ad-bc=1$ and $a,b,c,d\in\mathbb{R}$. This symmetry allows to choose
three points on the sphere which could be at $0$, $1$ and $\infty$. For open
strings the remnant symmetry preserves the boundary is $SL(2,\mathbb{R})$. In
the open string case the vertex operators are inserted on the boundary of the
disk and this symmetry can be used to fix three points on the boundary that
also may be $0$, $1$ and $\infty$. Similarly to the case of closed strings,
for open strings the presence of a $B$-field spoils the $SL(2,\mathbb{R})$
symmetry avoiding the possibility to fix three points of the positions of the
vertex operators on the boundary $\partial\Sigma$.

The two-point function with these boundary conditions is given by
\begin{equation}%
\begin{split}
\langle X^{\mu}(z)X^{\nu}(z^{\prime})\rangle &  =-\alpha^{\prime}%
\Big[g^{\mu\nu}\log|z-z^{\prime}|-g^{\mu\nu}\log|z-\bar{z}^{\prime}|\\
&  +G^{\mu\nu}\log|z-\bar{z}^{\prime}|^{2}+\frac{1}{2\pi\alpha^{\prime}}%
\theta^{\mu\nu}\log\frac{z-\bar{z}^{\prime}}{\bar{z}-z^{\prime}}+D^{\mu\nu
}\Big]
\end{split}
\label{FullPropagator}%
\end{equation}
with
\begin{align}
G^{\mu\nu}  &  =\left(  \frac{1}{g+2\pi\alpha^{\prime}B}g\frac{1}{g-2\pi
\alpha^{\prime}B}\right)  ^{\mu\nu},\ \label{PropagatorDefs}\\
\ \theta^{\mu\nu}  &  =-(2\pi\alpha^{\prime})^{2}\left(  \frac{1}{g+2\pi
\alpha^{\prime}B}B\frac{1}{g-2\pi\alpha^{\prime}B}\right)  ^{\mu\nu},\nonumber
\end{align}
and where $D^{\mu\nu}$ are quantities independent on $z$. Basically what we
want are the $N$-tachyon scattering amplitudes at tree level for the action
(\ref{B-fieldAction}). This is done by inserting the following $N$ tachyon
vertex operators $\mathcal{V}(\boldsymbol{k},z)=e^{i\boldsymbol{k}X(z)}$ at
different points of the boundary of the open string worldsheet $\partial
\Sigma$ and obtaining the correlation functions of these operators.

To obtain it we must realize that the correlators in the path integral
formalism can be computed as Gaussian integrals \cite{Tong:2009np}. Consider
the following integral
\begin{equation}
\int DX\exp\left(  \int d^{2}z(X\Delta X+iJX)\right)  \sim\exp\left(  \frac
{1}{2}\int d^{2}zd^{2}z^{\prime}J(z)\boldsymbol{K}(z,z^{\prime})J(z^{\prime
})\right)  ,\label{GaussianInt}%
\end{equation}
where $\Delta$ is a differential operator. The symbol $\sim$ has been used to
indicate that there are some proportional factor not relevant to the analysis.
Here $\boldsymbol{K}$ is the inverse operator, or Green's function, that
satisfies
\[
\Delta\boldsymbol{K}(z,z^{\prime})=\delta(z-z^{\prime}).
\]
We can see that we can use this result to obtain the scattering amplitudes by
choosing the appropriate source $J(z)$.
\begin{multline}
\langle\mathcal{V}(\boldsymbol{k}_{1},z_{1})\mathcal{V}(\boldsymbol{k}%
_{2},z_{2})\dots\mathcal{V}(\boldsymbol{k}_{N},z_{N})\rangle
\label{PartitionFunction}\\
\sim\int DX\exp\left\{  -S+i\left(  \boldsymbol{k}_{1}X_{1}(\sigma
)+\cdots+\boldsymbol{k}_{N}X_{N}(\sigma)\right)  \right\}  .\nonumber
\end{multline}
Since we are interested in the vertex operators inserted at the boundary, we
should restrict the propagator to it. By taking $z=\tau$ and $z^{\prime}%
=\tau^{\prime}$ for real $\tau,\tau^{\prime}$ in (\ref{FullPropagator}) we
get
\begin{equation}
\langle X^{\mu}(\tau)X^{\nu}(\tau^{\prime})\rangle=-\alpha^{\prime}G^{\mu\nu
}\log\left\vert \tau-\tau^{\prime}\right\vert +\frac{i}{2}\theta^{\mu\nu
}\mathit{sgn}(\tau-\tau^{\prime}),\label{Propagator}%
\end{equation}
where $sgn$ is the sign function. Now we can see that it suffices to choose
the current as
\[
J(\tau)=\sum_{l}\delta(\tau-\tau_{l})\boldsymbol{k}_{l}%
\]
in (\ref{GaussianInt}). From this we can obtain the scattering amplitudes by
finally integrating the expected values of the vertex operators over the
entire boundary of the worldsheet, which now is just integrating over the real
variable $\tau$. The result is
\begin{equation}%
\begin{split}
\int d^{N}\tau\langle\mathcal{V}(\boldsymbol{k}_{1},\tau_{1})\mathcal{V}%
(\boldsymbol{k}_{2},\tau_{2})\cdots\mathcal{V}(\boldsymbol{k}_{N},\tau
_{N})\rangle\sim\int d^{N}\tau &  \exp\left(  \frac{-i}{2}\sum_{l>m}%
\boldsymbol{k}^{l}\theta\boldsymbol{k}^{m}\mathit{sgn}(\tau_{l}-\tau
_{m})\right)  \\
\times &  \exp\left(  \alpha^{\prime}\sum_{l,m}\boldsymbol{k}^{l}%
G\boldsymbol{k}^{m}\log\left\vert \tau_{l}-\tau_{m}\right\vert \right)  \\
=\int d^{N}\tau\exp\left(  \frac{-i}{2}\sum_{l>m}\boldsymbol{k}^{l}%
\theta\boldsymbol{k}^{m}\text{sgn}(\tau_{l}-\tau_{m})\right)   &  \prod
_{l,m}|\tau_{l}-\tau_{m}|^{\alpha^{\prime}\boldsymbol{k}^{l}G\boldsymbol{k}%
^{m}},
\end{split}
\label{openNpoint}%
\end{equation}
where $\boldsymbol{k}^{l}\theta\boldsymbol{k}^{m}:=\boldsymbol{k}_{\mu}%
^{l}\theta^{\mu\nu}\boldsymbol{k}_{\nu}^{m}$ and $\boldsymbol{k}%
^{l}G\boldsymbol{k}^{m}:=\boldsymbol{k}_{\mu}^{l}G^{\mu\nu}\boldsymbol{k}%
_{\nu}^{m}$. These amplitudes are of the Koba-Nielsen type. The factor
involving the sign function in the last line is actually more general, because
it only depends on the variables through the sign functions, and any
derivatives of it are $0$. This allows us to write
\[%
\begin{split}
&  \left\langle \prod_{l=1}^{N}P_{l}(\partial X(\tau_{l}),\partial^{2}%
X(\tau_{l})\dots)e^{i\boldsymbol{k}_{l}X(\tau_{l})}\right\rangle _{G,\theta}\\
&  =e^{\frac{i}{2}\sum_{l>m}k_{\mu}^{l}\theta^{\mu\nu}k_{\nu}^{m}%
\mathit{sgn}(\tau_{l}-\tau_{m})}\left\langle \prod_{l=1}^{N}P_{l}(\partial
X(\tau_{l}),\partial^{2}X(\tau_{l})\dots)e^{i\boldsymbol{k}_{l}X(\tau_{l}%
)}\right\rangle _{G,\theta=0}.
\end{split}
\]
The symbol $\langle\cdots\rangle_{G,\theta}=0$ means that the expectation
value taken on the second line is using the propagator (\ref{Propagator})
without the second term. This is achieved by writing the polynomial as an
exponential and keeping only the corresponding linear terms. For example
\cite{Kawai et al}:
\[
\mathcal{V}(\zeta,k,z)=i\zeta^{\mu}\partial X^{\mu}e^{ikX(z)}=\left[
\exp\left(  ikX+i\zeta^{\mu}\partial X^{\mu}\right)  \right]  _{\text{linear}%
},
\]
where $[\cdots]_{\text{linear}}$ indicates to only keep linear terms in
$\zeta$. So we can see two direct effects of the $B$-field on the action
(\ref{B-fieldAction}). The first is that the effective metric on spacetime
becomes $G_{\mu\nu}$ given in (\ref{PropagatorDefs}). The second is that at
tree level (Remember, this came from the fact that we are in the full upper
half plane with the simplest zero genus topology) the scattering amplitudes
acquire a phase factor that depends on the cyclic ordering of the momenta
$k^{\mu}$ and the matrix $\theta$. But the overall form of the amplitudes is
the same as without the $B$-field.

In the case of open strings for $N$ points with vanishing $B$-field and with
flat metric $G_{\mu\nu}=\eta_{\mu\nu}$ and units with $\alpha^{\prime}=1$ and
for a fixed ordering of the external momenta $\boldsymbol{k}_{i}$, from
(\ref{openNpoint}) we obtain the Koba-Nielsen amplitude, with the three fixed
points in $x_{1}=0$, $x_{N-1}=1$ and $x_{N}=\infty$,
\begin{multline*}
\widetilde{A}_{\mathbb{R}}^{(N)}\left(  \boldsymbol{k}\right)  =%
{\displaystyle\int\limits_{0<x_{2}<x_{3}<\cdots<x_{N-2}<1}}
dx_{2}dx_{3}\cdots dx_{N-2}\prod_{j=2}^{N-2}|x_{j}|^{\boldsymbol{k}%
_{1}\boldsymbol{k}_{j}}|1-x_{j}|^{\boldsymbol{k}_{N-1}\boldsymbol{k}_{j}}\\
\times\prod_{2\leq i<j\leq N-2}|x_{i}-x_{j}|^{\boldsymbol{k}_{i}%
\boldsymbol{k}_{j}},
\end{multline*}
where we have changed the variables $\tau$ to $x$'s variables in
(\ref{openNpoint}), \cite{Green:2012oqa}, \cite{Polchinski:1998rq},
\cite{Tong:2009np},\cite{Kawai et al}.

The four-point amplitude is Veneziano's amplitude
\begin{equation}
\widetilde{A}_{\mathbb{R}}^{(4)}\left(  \boldsymbol{k}\right)  =\int_{0}%
^{1}dx|x|^{\boldsymbol{k}_{1}\boldsymbol{k}_{2}}|1-x|^{\boldsymbol{k}%
_{2}\boldsymbol{k}_{3}}. \label{veneziano}%
\end{equation}
It is known that this definition of Veneziano's amplitude describes only one
of the three decay channels $s$, $t$ and $u$ corresponding to the different
inequivalent permutations of the momenta. If we consider the sum over the
three channels and performing a suitable change of variables
\cite{Fairlie:1970di} is possible to write the total amplitude as
\[
{A}_{\mathbb{R}}^{(4)}\left(  \boldsymbol{k}\right)  =\int_{\mathbb{R}%
}dx|x|^{\boldsymbol{k}_{1}\boldsymbol{k}_{2}}|1-x|^{\boldsymbol{k}%
_{2}\boldsymbol{k}_{3}}.
\]
In terms of the gamma function the above amplitude writes
\begin{multline*}
A_{\mathbb{R}}^{(4)}\left(  \boldsymbol{k}\right)  ={\frac{\Gamma
(-1-\alpha^{\prime}s)\Gamma(-1-\alpha^{\prime}t)}{\Gamma(-2-\alpha^{\prime
}(s+t))}}+{\frac{\Gamma(-1-\alpha^{\prime}s)\Gamma(-1-\alpha^{\prime}%
u)}{\Gamma(-2-\alpha^{\prime}(s+u))}}\\
+{\frac{\Gamma(-1-\alpha^{\prime}t)\Gamma(-1-\alpha^{\prime}u)}{\Gamma
(-2-\alpha^{\prime}(t+u))}}%
\end{multline*}
where $s$, $t$ and $u$ are the Mandelstam variables and they are defined as
$s:=-(p_{1}+p_{2})^{2}$, $t:=-(p_{1}+p_{3})^{2}$ and $u:=-(p_{1}+p_{4})^{2}$
and they satisfy $s+t+u=-\sum_{i}\boldsymbol{k}_{i}^{2}=\sum_{i}M_{i}^{2}$,
and
\[
\Gamma(u)=\int_{0}^{\infty}dt\ t^{u-1}e^{-t}%
\]
is the gamma function. Thus the $4$-point amplitudes whose integration is
$\mathbb{R}$ it already includes the three channels.

Moreover this procedure also works for $N$ point amplitudes \cite{Kawai et al,
B-F} and it yields
\begin{multline*}
{A}_{\mathbb{R}}^{(N)}\left(  \boldsymbol{k}\right)  =\int_{\mathbb{R}^{N-3}%
}dx_{2}dx_{3}\cdots dx_{N-2}\prod_{j=2}^{N-2}|x_{j}|^{\boldsymbol{k}%
_{1}\boldsymbol{k}_{j}}|1-x_{j}|^{\boldsymbol{k}_{N-1}\boldsymbol{k}_{j}}\\
\times\prod_{2\leq i<j\leq N-2}|x_{i}-x_{j}|^{\boldsymbol{k}_{i}%
\boldsymbol{k}_{j}}.
\end{multline*}
For completeness we write the form of closed strings in absence of $B$-field
the $N$-point amplitude is given by
\[
A_{\mathbb{C}}^{(N)}\left(  \boldsymbol{k}\right)  =\int_{\mathbb{C}^{N-2}%
}\prod d^{2}z_{i}\prod_{i<j}|z_{i}-z_{j}|^{\boldsymbol{k}_{i}\boldsymbol{k}%
_{j}/2}.
\]
The four-point amplitude is the Virasoro-Shapiro amplitude
\begin{align*}
A_{\mathbb{C}}^{(4)}\left(  \boldsymbol{k}\right)   &  =\int_{\mathbb{C}^{2}%
}d^{2}z|z|^{\boldsymbol{k}_{1}\boldsymbol{k}_{4}/2}|1-z|^{\boldsymbol{k}%
_{2}\boldsymbol{k}_{4}/2}\\
&  ={\frac{\Gamma(-1-{\frac{\alpha^{\prime}s}{4}})\Gamma(-1-{\frac
{\alpha^{\prime}t}{4}})\Gamma(-1-{\frac{\alpha^{\prime}u}{4}})}{\Gamma
(2+{\frac{\alpha^{\prime}s}{4}})\Gamma(2+{\frac{\alpha^{\prime}t}{4}}%
)\Gamma(2+{\frac{\alpha^{\prime}u}{4}})}.}%
\end{align*}
It is worth to note that the Veneziano four-point amplitude (\ref{veneziano}),
describing only one channel, can be written also as an integral over the whole
$\mathbb{R}$ if we introduce some multiplicative characters as $sign(x)$:
\[
\widetilde{A}_{\mathbb{R}}^{(4)}\left(  \boldsymbol{k}\right)  ={\frac{1}{2}%
}\int_{\mathbb{R}}dx|x|^{\boldsymbol{k}_{1}\boldsymbol{k}_{2}}%
|1-x|^{\boldsymbol{k}_{2}\boldsymbol{k}_{3}}+{\frac{1}{2}}\int_{\mathbb{R}%
}dx|x|^{\boldsymbol{k}_{1}\boldsymbol{k}_{2}}|1-x|^{\boldsymbol{k}%
_{2}\boldsymbol{k}_{3}}\mathrm{sign}(x)\ \mathrm{sign}(1-x),
\]
where the function sign$(x)$ is $+1$ of $x>0$ and is $-1$ if $x<0$,
\cite{Brekke:1988dg}.

The generalization for $N$-point amplitudes is given by
\[
\tilde{A}_{\mathbb{R}}^{(4)}\left(  \boldsymbol{k}\right)  =\int
_{\mathbb{R}^{N-3}}dx_{2}dx_{3}\cdots dx_{N-2}\prod_{j=2}^{N-2}|x_{j}%
|^{\boldsymbol{k}_{1}\boldsymbol{k}_{j}}|1-x_{j}|^{\boldsymbol{k}%
_{N-1}\boldsymbol{k}_{j}}\bigg\{\frac{1-\mathrm{sign}\big[-x_{j}%
(1-x_{j})\big]}{2}\bigg\}
\]%
\[
\times\prod_{2\leq i<j\leq N-2}|x_{i}-x_{j}|^{\boldsymbol{k}_{i}%
\boldsymbol{k}_{j}}{\frac{1}{2}}\bigg\{1-\mathrm{sign}\big[(-1)(x_{i}%
-x_{j})\big]\bigg\},
\]
\cite{Brekke:1988dg}. The open string amplitude (\ref{openNpoint}) with a
non-vanishing $B$-field can be carried out in this form by using the
multiplicative character sign$(x)$. This form is of particular interest in
connection to the Ghoshal-Kawano amplitudes \cite{Ghoshal:2004ay} studied
later in this survey.

\section{\label{Section_2}Essential Ideas of $p$\textbf{-}Adic Analysis}

In this section, we review some ideas and results on $p$-adic analysis that we
will use along this article. For an in-depth exposition, the reader may
consult \cite{Alberio et al}, \cite{Taibleson}, \cite{V-V-Z}.

\subsection{The field of $p$-adic numbers}

Throughout this article $p$ will denote a prime number. The field of $p-$adic
numbers $\mathbb{Q}_{p}$ is defined as the completion of the field of rational
numbers $\mathbb{Q}$ with respect to the $p-$adic norm $|\cdot|_{p} $, which
is defined as
\[
\left\vert x\right\vert _{p}=\left\{
\begin{array}
[c]{lll}%
0 & \text{if} & x=0\\
&  & \\
p^{-\gamma} & \text{if} & x=p^{\gamma}\frac{a}{b}\text{,}%
\end{array}
\right.
\]
where $a$ and $b$ are integers coprime with $p$. The integer $\gamma:=ord(x)
$, with $ord(0):=+\infty$, is called the\textit{\ }$p-$\textit{adic order of}
$x$. We extend the $p-$adic norm to $\mathbb{Q}_{p}^{n}$ by taking%
\[
||\boldsymbol{x}||_{p}:=\max_{1\leq i\leq n}|x_{i}|_{p},\qquad\text{for
}\boldsymbol{x}=(x_{1},\dots,x_{n})\in\mathbb{Q}_{p}^{n}.
\]
We define $ord(\boldsymbol{x})=\min_{1\leq i\leq n}\{ord(x_{i})\}$, then
$||\boldsymbol{x}||_{p}=p^{-ord(\boldsymbol{x})}$.\ The metric space $\left(
\mathbb{Q}_{p}^{n},||\cdot||_{p}\right)  $ is a complete ultrametric space. As
a topological space $\mathbb{Q}_{p}$\ is homeomorphic to a Cantor-like subset
of the real line, see e.g. \cite{Alberio et al}, \cite{V-V-Z}.

Any $p-$adic number $x\neq0$ has a unique expansion of the form
\[
x=p^{ord(x)}\sum_{i=0}^{\infty}x_{i}p^{i},
\]
where $x_{i}\in\{0,1,2,\dots,p-1\}$ and $x_{0}\neq0$.

For $r\in\mathbb{Z}$, denote by $B_{r}^{n}(\boldsymbol{a})=\{\boldsymbol{x}%
\in\mathbb{Q}_{p}^{n};||\boldsymbol{x}-\boldsymbol{a}||_{p}\leq p^{r}\}$
\textit{the ball of radius }$p^{r}$ \textit{with center at} $\boldsymbol{a}%
=(a_{1},\dots,a_{n})\in\mathbb{Q}_{p}^{n}$, and take $B_{r}^{n}(\boldsymbol{0}%
):=B_{r}^{n}$. Note that $B_{r}^{n}(\boldsymbol{a})=B_{r}(a_{1})\times
\cdots\times B_{r}(a_{n})$, where $B_{r}(a_{i}):=\{x\in\mathbb{Q}_{p}%
;|x_{i}-a_{i}|_{p}\leq p^{r}\}$ is the one-dimensional ball of radius $p^{r}$
with center at $a_{i}\in\mathbb{Q}_{p}$. The ball $B_{0}^{n}$ equals the
product of $n$ copies of $B_{0}=\mathbb{Z}_{p}$, \textit{the ring of }%
$p-$\textit{adic integers}. In addition, $B_{r}^{n}(\boldsymbol{a}%
)=\boldsymbol{a}+\left(  p^{-r}\mathbb{Z}_{p}\right)  ^{n}$. We also denote by
$S_{r}^{n}(\boldsymbol{a})=\{\boldsymbol{x}\in\mathbb{Q}_{p}^{n}%
;||\boldsymbol{x}-\boldsymbol{a}||_{p}=p^{r}\}$ \textit{the sphere of radius
}$p^{r}$ \textit{with center at} $\boldsymbol{a}\in\mathbb{Q}_{p}^{n}$, and
take $S_{r}^{n}(\boldsymbol{0}):=S_{r}^{n}$. We notice that $S_{0}%
^{1}=\mathbb{Z}_{p}^{\times}$ (the group of units of $\mathbb{Z}_{p}$), but
$\left(  \mathbb{Z}_{p}^{\times}\right)  ^{n}\subsetneq S_{0}^{n}$. The balls
and spheres are both open and closed subsets in $\mathbb{Q}_{p}^{n} $. In
addition, two balls in $\mathbb{Q}_{p}^{n}$ are either disjoint or one is
contained in the other.

As a topological space $\left(  \mathbb{Q}_{p}^{n},||\cdot||_{p}\right)  $ is
totally disconnected, i.e. the only connected \ subsets of $\mathbb{Q}_{p}%
^{n}$ are the empty set and the points. A subset of $\mathbb{Q}_{p}^{n}$ is
compact if and only if it is closed and bounded in $\mathbb{Q}_{p}^{n}$, see
e.g. \cite[Section 1.3]{V-V-Z}, or \cite[Section 1.8]{Alberio et al}. The
balls and spheres are compact subsets. Thus $\left(  \mathbb{Q}_{p}%
^{n},||\cdot||_{p}\right)  $ is a locally compact topological space.

\begin{remark}
There is a natural map, called the reduction $\operatorname{mod}p$ and denoted
as $\overline{\cdot}$, from $\mathbb{Z}_{p}$ onto $\mathbb{F}_{p}$, the finite
field with $p$ elements. More precisely, if $x=\sum_{j=0}^{\infty}x_{j}%
p^{j}\in\mathbb{Z}_{p}$, then $\overline{x}=x_{0}\in\mathbb{F}_{p}=\left\{
0,1,\ldots,p-1\right\}  $. If $\boldsymbol{a}=(a_{1},\dots,a_{n})\in
\mathbb{Z}_{p}^{n}$, then $\overline{\boldsymbol{a}}=(\overline{a}_{1}%
,\dots,\overline{a}_{n})$.
\end{remark}

\subsection{Integration on $\mathbb{Q}_{p}^{n}$}

Since $(\mathbb{Q}_{p},+)$ is a locally compact topological group, there
exists a Borel measure $dx$, called the Haar measure of $(\mathbb{Q}_{p},+)$,
unique up to multiplication by a positive constant, such that $\int_{U}dx>0 $
for every non-empty Borel open set $U\subset\mathbb{Q}_{p}$, and satisfying
$\int_{E+z}dx=\int_{E}dx$ for every Borel set $E\subset\mathbb{Q}_{p}$, see
e.g. \cite[Chapter XI]{Halmos}. If we normalize this measure by the condition
$\int_{\mathbb{Z}_{p}}dx=1$, then $dx$ is unique. From now on we denote by
$dx$ the normalized Haar measure of $(\mathbb{Q}_{p},+)$ and by $d^{n}%
\boldsymbol{x}$ the product measure on $(\mathbb{Q}_{p}^{n},+)$.

A function $\varphi:$ $\mathbb{Q}_{p}^{n}\rightarrow\mathbb{C}$ is said to be
\textit{locally constant} if for every $\boldsymbol{x}\in\mathbb{Q}_{p}^{n}$
there exists an open compact subset $U$, $\boldsymbol{x}\in U$, such that
$\varphi(\boldsymbol{x})=\varphi(\boldsymbol{u})$ for all $\boldsymbol{u}\in
U$. Any locally constant function $\varphi:$ $\mathbb{Q}_{p}^{n}%
\rightarrow\mathbb{C}$ can be expressed as a linear combination of
characteristic functions of the form $\varphi\left(  \boldsymbol{x}\right)
=\sum_{k=1}^{\infty}c_{k}{\LARGE 1}_{U_{k}}\left(  \boldsymbol{x}\right)  $,
where $c_{k}\in\mathbb{C}$ and ${\LARGE 1}_{U_{k}}\left(  \boldsymbol{x}%
\right)  $ is the characteristic function of $U_{k}$, an open compact subset
of $\mathbb{Q}_{p}^{n}$, for every $k$. If $\varphi$ has compact support, then
$\varphi\left(  \boldsymbol{x}\right)  =\sum_{k=1}^{L}c_{k}{\LARGE 1}_{U_{k}%
}\left(  \boldsymbol{x}\right)  $ and in this case
\[%
{\displaystyle\int\limits_{\mathbb{Q}_{p}^{n}}}
\varphi\left(  \boldsymbol{x}\right)  d^{n}\boldsymbol{x}=c_{1}%
{\displaystyle\int\limits_{U_{1}}}
d^{n}\boldsymbol{x}+\ldots+c_{L}%
{\displaystyle\int\limits_{U_{L}}}
d^{n}\boldsymbol{x}.
\]
A locally constant function with compact support is called a
\textit{Bruhat-Schwartz function}. These functions form a $\mathbb{C}$-vector
space denoted as $\mathcal{D}\left(  \mathbb{Q}_{p}^{n}\right)  $. By using
the fact that $\mathcal{D}\left(  \mathbb{Q}_{p}^{n}\right)  $ is a dense
subspace of $C_{c}\left(  \mathbb{Q}_{p}^{n}\right)  $, the $\mathbb{C}$-space
of continuous functions on $\mathbb{Q}_{p}^{n}$ with compact support, with the
topology of the uniform convergence, the functional \ $\varphi\rightarrow
\int_{\mathbb{Q}_{p}^{n}}\varphi\left(  \boldsymbol{x}\right)  d^{n}%
\boldsymbol{x}$, $\varphi\in\mathcal{D}\left(  \mathbb{Q}_{p}^{n}\right)  $
has a unique continuous extension to $C_{c}\left(  \mathbb{Q}_{p}^{n}\right)
$, as an unbounded linear functional. For integrating more general functions,
say locally integrable functions, \ the following notion of improper integral
will be used.

\begin{definition}
A function $\varphi\in L_{loc}^{1}$ is said to be integrable in $\mathbb{Q}%
_{p}^{n}$ if%
\[
\lim_{m\rightarrow+\infty}%
{\displaystyle\int\limits_{B_{m}^{n}}}
\varphi\left(  \boldsymbol{x}\right)  d^{n}\boldsymbol{x}=\lim_{m\rightarrow
+\infty}%
{\displaystyle\sum\limits_{j=-\infty}^{m}}
\text{ }%
{\displaystyle\int\limits_{S_{j}^{n}}}
\varphi\left(  \boldsymbol{x}\right)  d^{n}\boldsymbol{x}%
\]
exists. If the limit exists, it is denoted as $%
{\textstyle\int\nolimits_{\mathbb{Q}_{p}^{n}}}
\varphi\left(  \boldsymbol{x}\right)  d^{n}\boldsymbol{x}$, and we say that
the\textbf{\ }(improper) integral exists.
\end{definition}

\subsection{Analytic change of variables}

A function $h:U\rightarrow\mathbb{Q}_{p}$ is said to be \textit{analytic} on
an open subset $U\subset\mathbb{Q}_{p}^{n}$, if for every $\boldsymbol{b}\in
U$ there exists an open subset $\widetilde{U}\subset U$, with $\boldsymbol{b}%
\in\widetilde{U}$, and a convergent power series $\sum_{i}a_{i}\left(
\boldsymbol{x}-\boldsymbol{b}\right)  ^{i}$ for $\boldsymbol{x}\in
\widetilde{U}$, such that $h\left(  \boldsymbol{x}\right)  =\sum
_{i\in\mathbb{N}^{n}}a_{i}\left(  \boldsymbol{x}-\boldsymbol{b}\right)  ^{i}$
for $\boldsymbol{x}\in\widetilde{U}$, with $\boldsymbol{x}^{i}=x_{1}^{i_{1}%
}\cdots x_{n}^{i_{n}}$, $\boldsymbol{i}=\left(  i_{1},\ldots,i_{n}\right)  $.
In this case, $\frac{\partial}{\partial x_{l}}h\left(  \boldsymbol{x}\right)
=\sum_{i\in\mathbb{N}^{n}}a_{i}\frac{\partial}{\partial x_{l}}\left(
\boldsymbol{x}-\boldsymbol{b}\right)  ^{i}$ is a convergent power series. Let
$U$, $V$ be open subsets of $\mathbb{Q}_{p}^{n}$. A mapping $\boldsymbol{h}%
:U\rightarrow V$, $\boldsymbol{h}=\left(  h_{1},\ldots,h_{n}\right)  $ is
called \textit{analytic} if each $h_{i}$ is analytic.

Let $\varphi:V$ $\rightarrow\mathbb{C}$ be a continuous function with compact
support, and let $\boldsymbol{h}:U\rightarrow V$ $\ $be an analytic mapping.
Then
\begin{equation}%
{\textstyle\int\limits_{V}}
\varphi\left(  \boldsymbol{y}\right)  d^{n}\boldsymbol{y}=%
{\textstyle\int\limits_{U}}
\varphi\left(  \boldsymbol{h}(\boldsymbol{x})\right)  \left\vert
Jac(\boldsymbol{h}(\boldsymbol{x}))\right\vert _{p}d^{n}\boldsymbol{x}\text{,}
\label{Formula_change_Var}%
\end{equation}
where $Jac(\boldsymbol{h}(\boldsymbol{z})):=\det\left[  \frac{\partial h_{i}%
}{\partial x_{j}}\left(  \boldsymbol{z}\right)  \right]  _{\substack{1\leq
i\leq n\\1\leq j\leq n}}$, see e.g. \cite[Section 10.1.2]{Bourbaki}.

\begin{example}
In dimension one, the change of variables formula (\ref{Formula_change_Var})
implies that
\[
\int\limits_{aU}dx=|a|_{p}\int\limits_{U}dy\text{, }%
\]
for any $a\in\mathbb{Q}_{p}^{\times}$ and any Borel set $U\subseteq
\mathbb{Q}_{p}$. For example,%
\begin{equation}
\int\limits_{B_{r}}dx=\int\limits_{p^{-r}\mathbb{Z}_{p}}dx=|p^{-r}|_{p}%
\int\limits_{\mathbb{Z}_{p}}dy=p^{r}\int\limits_{\mathbb{Z}_{p}}%
dy=p^{r},\text{ }\left(  x=p^{-r}y\right)  \label{Intbola}%
\end{equation}
and
\begin{equation}
\int\limits_{S_{r}}dx=\int\limits_{B_{r}}dx-\int\limits_{B_{r-1}}%
dx=p^{r}-p^{r-1}=p^{r}(1-p^{-1}). \label{Intesfera}%
\end{equation}

\end{example}

\begin{example}
\label{Example-1} We now compute the following integral
\[
Z(s)=\int\limits_{\mathbb{Z}_{p}}|x|_{p}^{s}dx\text{, }s\in\mathbb{C}\text{,
with }\operatorname{Re}(s)>0\in\mathbb{C}.
\]
We use the partition $\mathbb{Z}_{p}=p\mathbb{Z}_{p}\sqcup\mathbb{Z}%
_{p}^{\times}$, where $\mathbb{Z}_{p}^{\times}$ is the group of units of
$\mathbb{Z}_{p}$:%
\[
\mathbb{Z}_{p}^{\times}:=\{x\in\mathbb{Z}_{p};|x|_{p}=1\}=\left\{  \sum
_{i=0}^{\infty}x_{i}p^{i}\in\mathbb{Z}_{p};x_{0}\neq0\right\}  .
\]
By using (\ref{Intesfera}),
\begin{align*}
Z(s)  &  =\int\limits_{p\mathbb{Z}_{p}}|x|_{p}^{s}dx+\int\limits_{\mathbb{Z}%
_{p}^{\times}}|x|_{p}^{s}dx=p^{-1-s}\int\limits_{\mathbb{Z}_{p}}|x|_{p}%
^{s}dx+\int\limits_{\mathbb{Z}_{p}^{\times}}dx\\
&  =p^{-1-s}\int\limits_{\mathbb{Z}_{p}}|x|_{p}^{s}dx+(1-p^{-1})=p^{-1-s}%
Z(s)+(1-p^{-1}).
\end{align*}
Therefore,
\begin{equation}
Z(s)=\frac{1-p^{-1}}{1-p^{-1-s}}\text{ for }\operatorname{Re}(s)>-1.
\label{Integral0}%
\end{equation}

\end{example}

\begin{example}
\label{Example-2} We now consider the integral
\[
Z(s):=\int\limits_{\mathbb{Z}_{p}^{\times}}|1-x|_{p}^{s}dx\text{, }%
s\in\mathbb{C}\text{, with }\operatorname{Re}(s)>0\in\mathbb{C}.
\]
By using that $\mathbb{Z}_{p}^{\times}=\sqcup_{a_{0}\in\mathbb{F}_{p}^{\times
}}\left(  a_{0}+p\mathbb{Z}_{p}\right)  $, where $\mathbb{F}_{p}^{\times
}=\left\{  1,2,...,p-1\right\}  ,$ and (\ref{Integral0}) we have%
\begin{align*}
Z(s)=%
{\displaystyle\int\limits_{1+p\mathbb{Z}_{p}}}
|1-x|_{p}^{s}dx+\sum_{a_{0}=2}^{p-1}\text{ }%
{\displaystyle\int\limits_{a_{0}+p\mathbb{Z}_{p}}}
|1-x|_{p}^{s}dx=  &  p^{-1-s}%
{\displaystyle\int\limits_{\mathbb{Z}_{p}}}
|y|_{p}^{s}dy\\
+p^{-1}\sum_{a_{0}=2}^{p-1}\text{ }%
{\displaystyle\int\limits_{\mathbb{Z}_{p}}}
|1-(a_{0}+py)|_{p}^{s}dy  &  =p^{-1-s}\frac{1-p^{-1}}{1-p^{-1-s}}+p^{-1}(p-2).
\end{align*}

\end{example}

\subsection{Some $p$-adic integrals}

In this section we compute some integrals that we will use \ later on.

\begin{example}
\label{Example-3} We consider the following integral:%
\[
Z(s):=%
{\displaystyle\int\limits_{\mathbb{Z}_{p}\times\mathbb{Z}_{p}}}
|x-y|_{p}^{s}dxdy\text{, }s\in\mathbb{C}\text{, with }\operatorname{Re}%
(s)>0\in\mathbb{C}.
\]
By using the partition $\mathbb{Z}_{p}^{2}=(p\mathbb{Z}_{p})^{2}\sqcup
S_{0}^{2}$, with $S_{0}^{2}=p\mathbb{Z}_{p}\times\mathbb{Z}_{p}^{\times}%
\sqcup\mathbb{Z}_{p}^{\times}\times p\mathbb{Z}_{p}\sqcup\mathbb{Z}%
_{p}^{\times}\times\mathbb{Z}_{p}^{\times}$, we have
\begin{multline*}
Z(s)=%
{\displaystyle\int\limits_{(p\mathbb{Z}_{p})^{2}}}
|x-y|_{p}^{s}dxdy+%
{\displaystyle\int\limits_{p\mathbb{Z}_{p}\times\mathbb{Z}_{p}^{\times}}}
|x-y|_{p}^{s}dxdy+%
{\displaystyle\int\limits_{\mathbb{Z}_{p}^{\times}\times p\mathbb{Z}_{p}}}
|x-y|_{p}^{s}dxdy\\
+%
{\displaystyle\int\limits_{\mathbb{Z}_{p}^{\times}\times\mathbb{Z}_{p}%
^{\times}}}
|x-y|_{p}^{s}dxdy=p^{-2-s}Z(s)+p^{-1}%
{\displaystyle\int\limits_{\mathbb{Z}_{p}\times\mathbb{Z}_{p}^{\times}}}
dxdy+p^{-1}%
{\displaystyle\int\limits_{\mathbb{Z}_{p}^{\times}\times\mathbb{Z}_{p}}}
dxdy\\
+%
{\displaystyle\int\limits_{\mathbb{Z}_{p}^{\times}\times\mathbb{Z}_{p}%
^{\times}}}
|x-y|_{p}^{s}dxdy.
\end{multline*}
In order to compute the last integral we use the partition
\[
(\mathbb{Z}_{p}^{\times})^{2}=%
{\textstyle\bigsqcup\limits_{a_{0},a_{1}\in\mathbb{F}_{p}^{\times}}}
\text{ }\left(  a_{0}+p\mathbb{Z}_{p}\right)  \times\left(  a_{1}%
+p\mathbb{Z}_{p}\right)
\]
as follows:%
\begin{gather*}%
{\displaystyle\int\limits_{\mathbb{Z}_{p}^{\times}\times\mathbb{Z}_{p}%
^{\times}}}
|x-y|_{p}^{s}dxdy=\sum_{\overline{a}_{0},\overline{a}_{1}\in\mathbb{F}%
_{p}^{\times}}\text{ }%
{\displaystyle\int\limits_{\left(  a_{0}+p\mathbb{Z}_{p}\right)  \times\left(
a_{1}+p\mathbb{Z}_{p}\right)  }}
|x-y|_{p}^{s}dxdy\\
=p^{-2}\sum_{\substack{a_{0},a_{1}\in\mathbb{F}_{p}^{\times}\\a_{0}\neq a_{1}%
}}\text{ }%
{\displaystyle\int\limits_{\mathbb{Z}_{p}\times\mathbb{Z}_{p}}}
|a_{0}+px-a_{1}-py|_{p}^{s}dxdy+p^{-2-s}\sum_{\substack{a_{0},a_{1}%
\in\mathbb{F}_{p}^{\times}\\a_{0}=a_{1}}}\text{ }%
{\displaystyle\int\limits_{\mathbb{Z}_{p}\times\mathbb{Z}_{p}}}
|x-y|_{p}^{s}dxdy\\
=p^{-2}(p-1)(p-2)+p^{-2-s}\left(  p-1\right)  Z(s).
\end{gather*}
Hence,
\begin{equation}
Z(s)=\frac{1-p^{-1}}{1-p^{-1-s}}. \label{Integral1}%
\end{equation}

\end{example}

\begin{example}
\label{Example-4}We now show that
\begin{equation}
Z(s):=%
{\displaystyle\int\limits_{\mathbb{Z}_{p}^{\times}\times\mathbb{Z}_{p}%
^{\times}}}
|x-y|_{p}^{s}dxdy=p^{-2}(p-1)(p-2)+p^{-2-s}\left(  p-1\right)  \frac{1-p^{-1}%
}{1-p^{-1-s}}. \label{Integral2}%
\end{equation}
Indeed, by changing variables as $x=uv,y=v$,
\[
Z(s)=%
{\displaystyle\int\limits_{\mathbb{Z}_{p}^{\times}\times\mathbb{Z}_{p}%
^{\times}}}
|u-1|_{p}^{s}dudv=\left(  1-p^{-1}\right)
{\displaystyle\int\limits_{\mathbb{Z}_{p}^{\times}}}
|u-1|_{p}^{s}du.
\]
Now the result follows from Example \ref{Example-2}.
\end{example}

\begin{example}
\label{Example-5} We now compute the following integral
\[
Z(s_{1},s_{2},s_{3}):=\int\limits_{\mathbb{Z}_{p}^{\times}\times\mathbb{Z}%
_{p}^{\times}}|1-x|_{p}^{s_{1}}|1-y|_{p}^{s_{2}}|x-y|_{p}^{s_{3}}dxdy,
\]
where $s_{1},s_{2},s_{3}\in\mathbb{C}$ satisfying $\operatorname{Re}\left(
s_{1}\right)  $, $\operatorname{Re}\left(  s_{2}\right)  $, $\operatorname{Re}%
\left(  s_{3}\right)  >0$. By using the partition
\[
(\mathbb{Z}_{p}^{\times})^{2}=%
{\textstyle\bigsqcup\limits_{a_{0},a_{1}\in\mathbb{F}_{p}^{\times}}}
\left(  a_{0}+p\mathbb{Z}_{p}\right)  \times\left(  a_{1}+p\mathbb{Z}%
_{p}\right)  ,
\]
where $\mathbb{F}_{p}^{\times}=\left\{  1,2,...,p-1\right\}  $,
\begin{align*}
Z(s_{1},s_{2},s_{3})  &  =\sum_{a_{0}=1}^{p-1}\sum_{a_{1}=1}^{p-1}\text{ }%
{\displaystyle\int\limits_{\left(  a_{0}+p\mathbb{Z}_{p}\right)  \times\left(
a_{1}+p\mathbb{Z}_{p}\right)  }}
|1-x|_{p}^{s_{1}}|1-y|_{p}^{s_{2}}|x-y|_{p}^{s_{3}}dxdy\\
&  =:\sum_{a_{0}=1}^{p-1}\sum_{a_{1}=1}^{p-1}J_{a_{0},a_{1}}(s_{1},s_{2}%
,s_{3}).
\end{align*}
By changing variables as $x=a_{0}+pu$, $y=a_{1}+pv$ we have
\[
J_{a_{0},a_{1}}(s_{1},s_{2},s_{3})=p^{-2}%
{\displaystyle\int\limits_{\mathbb{Z}_{p}^{2}}}
|1-(a_{0}+pu)|_{p}^{s_{1}}|1-(a_{1}+pv)|_{p}^{s_{2}}|(a_{0}+pu)-(a_{1}%
+pv)|_{p}^{s_{3}}dudv.
\]
Now, we compute the integrals $J_{a_{0},a_{1}}$. This calculation involves
several cases.

\textbf{ Case 1. } If $\mathbf{\ }a_{0}\mathbf{\neq}1,~a_{1}\mathbf{\neq}1$
$a_{0}\neq a_{1}$, then $|1-(a_{0}+pu)|_{p}=1$, $|1-(a_{1}+pv)|_{p}=1$, and
$|(a_{0}+pu)-(a_{1}+pv)|_{p}=1$. and
\[
J_{a_{0},a_{1}}(s_{1},s_{2},s_{3})=p^{-2}%
{\displaystyle\int\limits_{\mathbb{Z} _{p}^{2}}}
dudv=p^{-2}.
\]
There are $(p-2)(p-3)$ integrals of this type.

\textbf{Case 2.} If $a_{0}\neq1,a_{1}\neq1$ $a_{0}=a_{1},$ then,
$|1-(a_{0}+pu)|_{p}=1$, $|1-(a_{1}+pv)|_{p}=1$. By Example (\ref{Integral1}),
\[
J_{a_{0},a_{1}}(s_{1},s_{2},s_{3})=p^{-2-s_{3}}%
{\displaystyle\int\limits_{\mathbb{Z} _{p}\times\mathbb{Z} _{p}}}
|u-v|_{p}^{s_{3}}dudv=\frac{p^{-2-s_{3}}(1-p^{-1})}{1-p^{-1-s_{3}}},
\]
for $\operatorname{Re}\left(  s_{3}\right)  >-1.$ There are $p-2$ integrals of
this type.

\textbf{Case 3.} If $a_{0}=1,a_{1}\neq1,$ then $|1-(a_{1}+pv)|_{p}=1$,
$|(a_{0}+pu)-(a_{1}+pv)|_{p}=1$. By using (\ref{Integral0})
\[
J_{a_{0},a_{1}}(s_{1},s_{2},s_{3})=p^{-2-s_{1}}%
{\displaystyle\int\limits_{\mathbb{Z}_{p}^{2}}}
|u|_{p}^{s_{1}}dudv=\frac{p^{-2-s_{1}}(1-p^{-1})}{1-p^{-1-s_{1}}},
\]
for $\operatorname{Re}\left(  s_{1}\right)  >-1$. There are $p-2$ integrals of
this type.

\textbf{Case 4.} If $a_{0}\neq1,a_{1}=1$, this case is similar to Case 3,
\[
J_{a_{0},a_{1}}(s_{1},s_{2},s_{3})=p^{-2-s_{2}}%
{\displaystyle\int\limits_{\mathbb{Z}_{p}}}
|x_{3}|_{p}^{s_{2}}dx_{3}=\frac{p^{-2-s_{2}}(1-p^{-1})}{1-p^{-1-s_{2}}}%
\]
for $\operatorname{Re}(s_{2})>-1$. Therefore%
\begin{align*}
Z(s_{1},s_{2},s_{3})  &  =p^{-2}(p-2)(p-3)+(p-2)\frac{p^{-2-s_{3}}(1-p^{-1}%
)}{1-p^{-1-s_{3}}}\\
&  +(p-2)\frac{p^{-2-s_{1}}(1-p^{-1})}{1-p^{-1-s_{1}}}+(p-2)\frac{p^{-2-s_{2}%
}(1-p^{-1})}{1-p^{-1-s_{2}}}.
\end{align*}
on the region of $\mathbb{C}^{3}$ defined by $\operatorname{Re}(s_{1})>-1$,
$\operatorname{Re}(s_{2})>-1$, and $\operatorname{Re}(s_{3})>-1$.
\end{example}

\section{\label{Section_3}$p$-Adic open string amplitudes}

In \cite{B-F-O-W}, Brekke, Freund, Olson and Witten work out the $N$-point
amplitudes in explicit form and investigate how these can be obtained from an
effective Lagrangian. The $p$-adic open string $N$-point tree amplitudes are
defined as
\begin{gather}
\boldsymbol{A}^{(N)}\left(  \boldsymbol{k}\right)  =\label{Amplitude_0}\\%
{\displaystyle\int\limits_{\mathbb{Q}_{p}^{N-3}}}
{\displaystyle\prod\limits_{i=2}^{N-2}}
\left\vert x_{i}\right\vert _{p}^{\boldsymbol{k}_{1}\boldsymbol{k}_{i}%
}\left\vert 1-x_{i}\right\vert _{p}^{\boldsymbol{k}_{N-1}\boldsymbol{k}_{i}%
}\text{ }%
{\displaystyle\prod\limits_{2\leq i<j\leq N-2}}
\left\vert x_{i}-x_{j}\right\vert _{p}^{\boldsymbol{k}_{i}\boldsymbol{k}_{j}}%
{\displaystyle\prod\limits_{i=2}^{N-2}}
dx_{i}\text{,}\nonumber
\end{gather}
where $%
{\textstyle\prod\nolimits_{i=2}^{N-2}}
dx_{i}$ is the normalized Haar measure of $\mathbb{Q}_{p}^{N-3}$,
$\boldsymbol{k}=\left(  \boldsymbol{k}_{1},\ldots,\boldsymbol{k}_{N}\right)
$, $\boldsymbol{k}_{i}=\left(  k_{0,i},\ldots,k_{l,i}\right)  $,
$i=1,\ldots,N$, $N\geq4$, is the momentum vector of the $i$-th tachyon (with
Minkowski product $\boldsymbol{k}_{i}\boldsymbol{k}_{j}=-k_{0,i}%
k_{0,j}+k_{1,i}k_{1,j}+\cdots+k_{l,i}k_{l,j}$) obeying
\[
\sum_{i=1}^{N}\boldsymbol{k}_{i}=\boldsymbol{0}\text{,\ \ \ }\boldsymbol{k}%
_{i}\boldsymbol{k}_{i}=2\text{ for }i=1,\ldots,N.
\]
The parameter $l$ is an arbitrary positive integer. Typically $l$ is taken to
be $25$. However, we do not require using the critical dimension. In
\cite{Zun-B-C-LMP}, we show that the $p$-adic open string amplitudes
(\ref{Amplitude_0}) are bona fide integrals and that they have meromorphic
continuations as rational functions in the kinematic parameters. We attach to
these amplitudes a multivariate Igusa-type zeta function. Fix an integer
$N\geq4$ and set $\ T:=\left\{  2,\ldots,N-2\right\}  $, $D=\frac{\left(
N-3\right)  \left(  N-4\right)  }{2}+2\left(  N-3\right)  $ and $\mathbb{C}%
^{D}$ as
\[
\left\{
\begin{array}
[c]{lll}%
\left\{  s_{ij}\in\mathbb{C};i\in\left\{  1,N-1\right\}  ,j\in T\right\}  &
\text{if} & N=4\\
&  & \\
\left\{  s_{ij}\in\mathbb{C};i\in\left\{  1,N-1\right\}  ,j\in T\text{ or
}i,j\in T\text{ with }i<j\right\}  & \text{if} & N\geq5.
\end{array}
\right.
\]
We now put $\ \boldsymbol{s}=\left(  s_{ij}\right)  \in\mathbb{C}^{D}$ and
$\boldsymbol{k}_{i}\boldsymbol{k}_{j}=s_{ij}\in\mathbb{C}$ satisfying
$s_{ij}=s_{ji}$ for $1\leq i<j\leq N-1$. The $p$\textit{-adic open string }%
$N$\textit{-point zeta function} is defined as%
\begin{equation}
\boldsymbol{Z}^{(N)}\left(  \boldsymbol{s}\right)  =%
{\displaystyle\int\limits_{\mathbb{Q}_{p}^{N-3}\smallsetminus\Lambda}}
{\displaystyle\prod\limits_{i=2}^{N-2}}
\left\vert x_{i}\right\vert _{p}^{s_{1i}}\left\vert 1-x_{i}\right\vert
_{p}^{s_{(N-1)i}}\text{ }%
{\displaystyle\prod\limits_{2\leq i<j\leq N-2}}
\left\vert x_{i}-x_{j}\right\vert _{p}^{s_{ij}}%
{\displaystyle\prod\limits_{i=2}^{N-2}}
dx_{i}, \label{zeta_funtion_string_0}%
\end{equation}
where $\boldsymbol{s}=\left(  s_{ij}\right)  \in\mathbb{C}^{D}$, $%
{\textstyle\prod\nolimits_{i=2}^{N-2}}
dx_{i}$ is the normalized Haar measure of $\mathbb{Q}_{p}^{N-3}$, and
\[
\Lambda:=\left\{  \left(  x_{2},\ldots,x_{N-2}\right)  \in\mathbb{Q}_{p}%
^{N-3};\text{ }%
{\displaystyle\prod\limits_{i=2}^{N-2}}
x_{i}\left(  1-x_{i}\right)  \text{ }%
{\displaystyle\prod\limits_{2\leq i<j\leq N-2}}
\left(  x_{i}-x_{j}\right)  =0\right\}  .
\]
In the definition of integral (\ref{zeta_funtion_string_0}) we remove the set
$\Lambda$ from the domain of integration in order to use the formula
$a^{s}=e^{s\ln a}$ for $a>0$ and $s\in\mathbb{C}$. The reason for using the
name `zeta function' is that the integral (\ref{zeta_funtion_string_0}) is a
finite sum of multivariate local zeta functions.

Take $\phi\left(  x_{2},\ldots,x_{N-2}\right)  $ a locally constant function
with compact support, then
\begin{gather*}
Z_{\phi}^{(N)}(\boldsymbol{s})=\\%
{\displaystyle\int\limits_{\mathbb{Q}_{p}^{N-3}\smallsetminus\Lambda}}
\phi\left(  x_{2},\ldots,x_{N-2}\right)
{\displaystyle\prod\limits_{i=2}^{N-2}}
\left\vert x_{i}\right\vert _{p}^{s_{1i}}\left\vert 1-x_{i}\right\vert
_{p}^{s_{\left(  N-1\right)  i}}%
{\displaystyle\prod\limits_{2\leq i<j\leq N-2}}
\left\vert x_{i}-x_{j}\right\vert _{p}^{s_{i}{}_{j}}%
{\displaystyle\prod\limits_{i=2}^{N-2}}
dx_{i},
\end{gather*}
for $\operatorname{Re}(s_{ij})>0$ for any $ij$, is a multivariate Igusa local
zeta function. In characte\-ristic zero, a general theory for this type of
local zeta functions was elaborated by Loeser in \cite{Loeser}. In particular,
these local zeta functions admit analytic continuations as rational functions
of the variables $p^{-s_{ij}}$.

We want to highlight that the convergence of the multivariate local zeta
functions depends of the fact that the locally constant function $\phi$ has
compact support. For example, we consider the following integral:
\[
\boldsymbol{J}(s)=%
{\displaystyle\int\limits_{\mathbb{Q}_{p}}}
\left\vert x\right\vert _{p}^{s}dx\text{, }s\in\mathbb{C}.
\]
If the integral $\boldsymbol{J}(s_{0})$ exists for some $s_{0}\in\mathbb{R}$,
then the integrals
\[
\boldsymbol{J}_{0}(s_{0})=%
{\displaystyle\int\limits_{\mathbb{Z}_{p}}}
\left\vert x\right\vert _{p}^{s_{0}}dx\text{ \ and }\boldsymbol{J}_{1}(s_{0})=%
{\displaystyle\int\limits_{\mathbb{Q}_{p}\smallsetminus\mathbb{Z}_{p}}}
\left\vert x\right\vert _{p}^{s_{0}}dx\text{ }%
\]
exist. The first integral $\boldsymbol{J}_{0}(s_{0})=\frac{1-p^{-1}%
}{1-p^{-1-s_{0}}}$ for $s_{0}>-1$. For the second integral, we use that
$\left\vert x\right\vert _{p}^{s_{0}}$ is locally integrable, and thus
\[
\boldsymbol{J}_{1}(s_{0})=\sum\limits_{j=1}^{\infty}%
{\displaystyle\int\limits_{p^{-j}\mathbb{Z}_{p}^{\times}}}
\left\vert x\right\vert _{p}^{s_{0}}dx=\sum\limits_{j=1}^{\infty}p^{j+js_{0}}%
{\displaystyle\int\limits_{\mathbb{Z}_{p}^{\times}}}
dx=\left(  1-p^{-1}\right)  \sum\limits_{j=1}^{\infty}p^{j(1+s_{0})}%
<\infty\text{ }%
\]
if and only if $s_{0}<-1$. Then, integral $\boldsymbol{J}(s)$ does not exist
for any complex value $s$.

Theorem 1 in \cite{Zun-B-C-LMP} establishes that the $p$-adic open string
$N$-point zeta function is a holomorphic function in a certain domain of
$\mathbb{C}^{D}$ and that it admits an analytic continuation to $\mathbb{C}%
^{D}$ (denoted as $Z^{\left(  N\right)  }\left(  \boldsymbol{s}\right)  $) as
a rational function in the variables $p^{-s_{ij}},i,j\in\left\{
1,\ldots,N-1\right\}  $. Hence, in the theory of local zeta functions, the
convergence and the existence of a meromorphic continuation as a rational
function in the variables $p^{-s_{ij}},i,j\in\left\{  1,\ldots,N-1\right\}  $
of integrals of type (\ref{zeta_funtion_string_0}) is a new and remarkable result.

The $p$-adic open string $N$-point zeta functions $\boldsymbol{Z}%
^{(N)}(\boldsymbol{s})$ are regularizations of the amplitudes $\boldsymbol{A}%
^{(N)}\left(  \boldsymbol{k}\right)  $. More precisely, we define
\[
\boldsymbol{A}^{(N)}\left(  \boldsymbol{k}\right)  =\boldsymbol{Z}%
^{(N)}(\underline{\boldsymbol{s}})\mid_{s_{ij}=\boldsymbol{k}_{i}%
\boldsymbol{k}_{j}}\text{with }i\in\left\{  1,\ldots,N-1\right\}  \text{,
}j\in T\text{ or }i,j\in T,
\]
where $T=\left\{  2,\ldots,N-2\right\}  $. By Theorem 1 in \cite{Zun-B-C-LMP},
$\boldsymbol{A}^{(N)}\left(  \boldsymbol{k}\right)  $ are well-defined
rational functions of the variables $p^{-\boldsymbol{k}_{i}\boldsymbol{k}_{j}%
}$, $i$, $j\in\left\{  1,\ldots,N-1\right\}  $, which agree with integrals
(\ref{Amplitude}) when they converge.

In the following two sections, using the methods given in \cite{Zun-B-C-LMP},
we compute the $p$-adic open string amplitudes in the cases $N=4$, $5$.

\subsection{$p$-Adic open string 4-point amplitude}

The $4$-point string amplitude is given by
\[
\boldsymbol{A}^{(4)}\left(  \boldsymbol{k}\right)  =%
{\displaystyle\int\limits_{\mathbb{Q}_{p}}}
|x_{2}|_{p}^{\boldsymbol{k}_{1}\boldsymbol{k}_{2}}|1-x_{2}|_{p}%
^{\boldsymbol{k}_{3}\boldsymbol{k}_{2}}dx_{2}.
\]
We attach to this amplitude the local zeta function:
\[
\boldsymbol{Z}^{(4)}(\boldsymbol{s})=%
{\displaystyle\int\limits_{\mathbb{Q}_{p}}}
|x_{2}|_{p}^{s_{12}}|1-x_{2}|_{p}^{s_{32}}dx_{2}.
\]
We now divide the integration domain ($%
\mathbb{Q}
_{p}$) into sectors. Given $I\subseteqq T=\left\{  2\right\}  $, the attached
sector is defined as%
\[
Sect(I)=\left\{  x_{i}\in%
\mathbb{Q}
_{p};|x_{i}|_{p}\leq1\Longleftrightarrow i\in I\right\}  .
\]
Then, we have two sectors:%
\[%
\begin{tabular}
[c]{|c|c|c|}\hline
$I$ & $T\diagdown I$ & $Sect(I)$\\\hline
$\left\{  2\right\}  $ & $\varnothing$ & $%
\mathbb{Z}
_{p}$\\\hline
$\varnothing$ & $\left\{  2\right\}  $ & $%
\mathbb{Q}
_{p}\backslash%
\mathbb{Z}
_{p},$\\\hline
\end{tabular}
\ \ \
\]
and
\[
\boldsymbol{Z}^{(4)}(\boldsymbol{s})=\boldsymbol{Z}^{(4)}(\boldsymbol{s}%
,\varnothing)+\boldsymbol{Z}^{(4)}(\boldsymbol{s},\{2\}),
\]
where
\[
\boldsymbol{Z}^{(4)}\left(  \boldsymbol{s};\left\{  2\right\}  \right)  :=%
{\displaystyle\int\limits_{\mathbb{Z}_{p}}}
|x_{2}|_{p}^{s_{12}}|1-x_{2}|_{p}^{s_{32}}dx_{2}%
\]
and
\[
\boldsymbol{Z}^{(4)}\left(  \boldsymbol{s};\varnothing\right)  :=%
{\displaystyle\int\limits_{\mathbb{Q}_{p}\smallsetminus\mathbb{Z}_{p}}}
|x_{2}|_{p}^{s_{12}}|1-x_{2}|_{p}^{s_{32}}dx_{2}.
\]
We now compute $\boldsymbol{Z}^{(4)}\left(  \boldsymbol{s};\left\{  2\right\}
\right)  $. By using the Examples \ref{Example-1} and \ref{Example-2}:
\begin{align}
\boldsymbol{Z}^{(4)}\left(  \boldsymbol{s};\left\{  2\right\}  \right)   &  =%
{\displaystyle\int\limits_{\mathbb{Z}_{p}}}
|x_{2}|_{p}^{s_{12}}dx_{2}+%
{\displaystyle\int\limits_{\mathbb{Z}_{p}^{\times}}}
|1-x_{2}|_{p}^{s_{42}}dx_{2}\label{4point}\\
&  =1-2p^{-1}+\frac{\left(  1-p^{-1}\right)  p^{-1-s_{12}}}{1-p^{-1-s_{12}}%
}+\frac{\left(  1-p^{-1}\right)  p^{-1-s_{32}}}{1-p^{-1-s_{32}}},\nonumber
\end{align}
for $\operatorname{Re}(s_{12})>-1$ and $\operatorname{Re}(s_{32})>-1$.

We now consider integral $\boldsymbol{Z}^{(4)}\left(  \boldsymbol{s}%
;\varnothing\right)  $. By using the ultrametric inequality $|1-x_{2}%
|_{p}=|x_{2}|_{p}$ for $x_{2}\in\mathbb{Q}_{p}\smallsetminus\mathbb{Z}_{p}$,
\[
\boldsymbol{Z}^{(4)}\left(  \boldsymbol{s};\varnothing\right)  =%
{\displaystyle\int\limits_{\mathbb{Q}_{p}\smallsetminus\mathbb{Z}_{p}}}
|x_{2}|_{p}^{s_{12}+s_{32}}dx_{2}.
\]
In order to compute this last integral we proceed as follows. For
$l\in\mathbb{N\setminus}\left\{  0\right\}  $, we set%
\[
\left(  \mathbb{Q}_{p}\smallsetminus\mathbb{Z}_{p}\right)  _{-l}:=\left\{
x_{2}\in\left(  \mathbb{Q}_{p}\smallsetminus\mathbb{Z}_{p}\right)  ;-l\leq
ord(x_{2})\leq-1\right\}  ,
\]%
\[
\left(  p\mathbb{Z}_{p}\right)  _{l}:=\left\{  x_{2}\in\left(  p\mathbb{Z}%
_{p}\right)  ;1\leq ord(x_{2})\leq l\right\}  ,
\]
and
\[
\boldsymbol{I}_{-l}\left(  \boldsymbol{s}\right)  :=%
{\displaystyle\int\limits_{\left(  \mathbb{Q}_{p}\smallsetminus\mathbb{Z}%
_{p}\right)  _{-l}}}
|x_{2}|_{p}^{s_{12}+s_{32}}dx_{2}.
\]
Notice that $\left(  \mathbb{Q}_{p}\smallsetminus\mathbb{Z}_{p}\right)  _{-l}%
$, $\left(  p\mathbb{Z}_{p}\right)  _{l}$ are compact sets and that
\[%
\begin{array}
[c]{ccc}%
\left(  \mathbb{Q}_{p}\smallsetminus\mathbb{Z}_{p}\right)  _{-l} & \rightarrow
& \left(  p\mathbb{Z}_{p}\right)  _{l}\\
&  & \\
x_{2} & \rightarrow & \frac{1}{y_{2}},
\end{array}
\]
is an analytic change of variables satisfying $dx_{2}=\frac{dy_{2}}{\left\vert
y_{2}\right\vert _{p}^{2}}$, then by using this change of variables, we have
\[
\boldsymbol{I}_{-l}\left(  \boldsymbol{s}\right)  ={\int\limits_{\left(
p\mathbb{Z}_{p}\right)  _{l}}}\frac{dy_{2}}{\left\vert y_{2}\right\vert
_{p}^{s_{12}+s_{32}+2}}.
\]
Now, by using that $\int_{p\mathbb{Z}_{p}}\frac{1}{\left\vert y\right\vert
_{p}^{s}}dy$ converges for $\operatorname{Re}\left(  s\right)  <1$ and the
dominated convergence theorem, $\lim_{l\rightarrow\infty}\boldsymbol{I}%
_{-l}\left(  \boldsymbol{s}\right)  =\boldsymbol{Z}^{(4)}\left(
\boldsymbol{s};\varnothing\right)  $, i.e.
\[
\boldsymbol{Z}^{(4)}\left(  \boldsymbol{s};\left\{  \varnothing\right\}
\right)  =%
{\displaystyle\int\limits_{p\mathbb{Z}_{p}}}
\frac{dy_{2}}{\left\vert y_{2}\right\vert _{p}^{s_{12}+s_{32}+2}}%
=\frac{(1-p^{-1})p^{1+s_{12}+s_{32}}}{1-p^{1+s_{12}+s_{32}}},
\]
on $\operatorname{Re}(s_{12})+\operatorname{Re}(s_{32})<-1$. Therefore
\[
\boldsymbol{Z}^{(4)}\left(  \boldsymbol{s}\right)  =\frac{p-2}{p}+\left(
1-p^{-1}\right)  \left\{  \frac{p^{-1-s_{12}}}{1-p^{-1-s_{12}}}+\frac
{p^{-1-s_{32}}}{1-p^{\left(  -1-s_{32}\right)  }}+\frac{p^{1+s_{12}+s_{32}}%
}{1-p^{1+s_{12}+s_{32}}}\right\}  ,
\]
this function is holomorphic in%
\[
\operatorname{Re}(s_{12})>-1,\operatorname{Re}(s_{32})>-1\text{ and
}\operatorname{Re}(s_{12})+\operatorname{Re}(s_{32})<-1,
\]
which is a non-empty open subset in $\mathbb{C}^{2}$ because it contains the
open set defined by the conditions:
\[
-\frac{2}{3}<\operatorname{Re}(s_{12})<-\frac{1}{2}\quad\text{ and }-\frac
{2}{3}<\operatorname{Re}(s_{32})<-\frac{1}{2}.
\]
We denote the meromorphic continuation of integral $\boldsymbol{Z}%
^{(4)}\left(  \boldsymbol{s}\right)  $ also by $\boldsymbol{Z}^{(4)}\left(
\boldsymbol{s}\right)  $. Now, we regularize the $p$-adic open string 4-point
amplitude by using $\boldsymbol{Z}^{(4)}\left(  \boldsymbol{s}\right)  $:
\begin{align}
\boldsymbol{A}^{(4)}\left(  \boldsymbol{k}\right)   &  :=\boldsymbol{Z}%
^{(4)}(\boldsymbol{s})\mid_{s_{12}=\boldsymbol{k}_{1}\boldsymbol{k}_{2}%
,s_{32}=\boldsymbol{k}_{3}\boldsymbol{k}_{2}}\nonumber\\
&  =\frac{p-2}{p}+\left(  1-p^{-1}\right)  \left\{  \frac{p^{-1-\boldsymbol{k}%
_{1}\boldsymbol{k}_{2}}}{1-p^{-1-\boldsymbol{k}_{1}\boldsymbol{k}_{2}}}%
+\frac{p^{-1-\boldsymbol{k}_{2}\boldsymbol{k}_{3}}}{1-p^{-1-\boldsymbol{k}%
_{2}\boldsymbol{k}_{3}}}+\frac{p^{1+\boldsymbol{k}_{1}\boldsymbol{k}%
_{2}+\boldsymbol{k}_{2}\boldsymbol{k}_{3}}}{1-p^{1+\boldsymbol{k}%
_{1}\boldsymbol{k}_{2}+\boldsymbol{k}_{2}\boldsymbol{k}_{3}}}\right\}
\nonumber\\
&  =\frac{p-2}{p}+\left(  1-p^{-1}\right)  \left\{  \frac{p^{-1-\boldsymbol{k}%
_{1}\boldsymbol{k}_{2}}}{1-p^{-1-\boldsymbol{k}_{1}\boldsymbol{k}_{2}}}%
+\frac{p^{-1-\boldsymbol{k}_{2}\boldsymbol{k}_{3}}}{1-p^{-1-\boldsymbol{k}%
_{2}\boldsymbol{k}_{3}}}+\frac{p^{-1-\boldsymbol{k}_{2}\boldsymbol{k}_{4}}%
}{1-p^{-1-\boldsymbol{k}_{2}\boldsymbol{k}_{4}}}\right\}  .
\label{Amplitude_A_4}%
\end{align}
This amplitude can be rewritten as%
\[
\boldsymbol{A}^{(4)}\left(  \boldsymbol{k}\right)  =K_{4}+\frac{1}{2}%
{\textstyle\sum\limits_{i<j}}
x_{ij},
\]
with%
\[
x_{ij}:=\frac{p-1}{p}\frac{p^{-1-\boldsymbol{k}_{i}\boldsymbol{k}_{j}}%
}{1-p^{-1-\boldsymbol{k}_{i}\boldsymbol{k}_{j}}},
\]
where $\sum_{i<j}$ means \ the summation over all pairs of compatible channels
$ij$. Then
\[
\boldsymbol{A}^{(4)}\left(  \boldsymbol{k}\right)  =K_{4}+\frac{1}{2}\left\{
x_{12}+x_{13}+x_{14}+x_{23}+x_{24}+x_{34}\right\}  .
\]
Indeed, by using that $\sum_{i=1}^{4}\boldsymbol{k}_{i}=0$, $\boldsymbol{k}%
_{i}^{2}=2$, we get%
\begin{align}
\boldsymbol{k}_{1}\boldsymbol{k}_{3}+\boldsymbol{k}_{2}\boldsymbol{k}%
_{3}+\boldsymbol{k}_{3}\boldsymbol{k}_{4}+2  &  =0,\label{Eq_a}\\
\boldsymbol{k}_{1}\boldsymbol{k}_{2}+\boldsymbol{k}_{1}\boldsymbol{k}%
_{3}+\boldsymbol{k}_{1}\boldsymbol{k}_{4}+2  &  =0,\label{Eq_b}\\
\boldsymbol{k}_{1}\boldsymbol{k}_{2}+\boldsymbol{k}_{2}\boldsymbol{k}%
_{3}+\boldsymbol{k}_{2}\boldsymbol{k}_{4}+2  &  =0,\label{Eq_c}\\
\boldsymbol{k}_{1}\boldsymbol{k}_{4}+\boldsymbol{k}_{2}\boldsymbol{k}%
_{4}+\boldsymbol{k}_{3}\boldsymbol{k}_{4}+2  &  =0. \label{Eq_d}%
\end{align}
From (\ref{Eq_a})-(\ref{Eq_d}), we get%
\[
\boldsymbol{k}_{1}\boldsymbol{k}_{2}=\boldsymbol{k}_{3}\boldsymbol{k}%
_{4}\text{, }\boldsymbol{k}_{1}\boldsymbol{k}_{3}=\boldsymbol{k}%
_{2}\boldsymbol{k}_{4}\text{, }\boldsymbol{k}_{1}\boldsymbol{k}_{4}%
=\boldsymbol{k}_{2}\boldsymbol{k}_{3}.
\]
Then
\begin{equation}
\boldsymbol{A}^{(4)}\left(  \boldsymbol{k}\right)  =K_{4}+x_{12}+x_{23}%
+x_{24}, \label{Amplitude_A_4A}%
\end{equation}
which agrees with (\ref{Amplitude_A_4}).

\subsection{\label{Sect_5_points}$p$-Adic open string 5-point amplitude}

The $p$-adic open string $5$-point amplitude is given by
\[
\boldsymbol{A}^{(5)}\left(  \boldsymbol{k}\right)  =%
{\displaystyle\int\limits_{\mathbb{Q}_{p}^{2}}}
|x_{2}|_{p}^{\boldsymbol{k}_{1}\boldsymbol{k}_{2}}|x_{3}|_{p}^{\boldsymbol{k}%
_{1}\boldsymbol{k}_{3}}|1-x_{2}|_{p}^{\boldsymbol{k}_{4}\boldsymbol{k}_{2}%
}|1-x_{3}|_{p}^{\boldsymbol{k}_{4}\boldsymbol{k}_{3}}|x_{2}-x_{3}%
|_{p}^{\boldsymbol{k}_{2}\boldsymbol{k}_{3}}dx_{2}dx,
\]
and the $5$-point local zeta function is given by
\[
\boldsymbol{Z}^{(5)}\left(  \boldsymbol{s}\right)  =%
{\displaystyle\int\limits_{\mathbb{Q}_{p}^{2}}}
|x_{2}|_{p}^{s_{12}}|x_{3}|_{p}^{s_{13}}|1-x_{2}|_{p}^{s_{42}}|1-x_{3}%
|_{p}^{s_{43}}|x_{2}-x_{3}|_{p}^{s_{23}}dx_{2}dx_{3},
\]
where $s_{ij}\in\mathbb{C}$. We divide the integration domain $%
\mathbb{Q}
_{p}^{2}$ in sectors. Given $I\subseteqq T=\left\{  2,3\right\}  $ the attache
sector is defined as%
\[
Sect(I)=\left\{  (x_{1},x_{2})\in%
\mathbb{Q}
_{p}^{2};|x_{i}|_{p}\leq1\Longleftrightarrow i\in I\right\}  \text{.}%
\]
The following table shows all the sectors:
\[%
\begin{tabular}
[c]{ccc|}\hline
$I$ & $T\diagdown I$ & $Sect(I)$\\\hline
$\varnothing$ & $\left\{  2,3\right\}  $ & $\left(
\mathbb{Q}
_{p}\backslash%
\mathbb{Z}
_{p}\right)  \times\left(
\mathbb{Q}
_{p}\backslash%
\mathbb{Z}
_{p}\right)  $\\\hline
$\left\{  2\right\}  $ & $\left\{  3\right\}  $ & $%
\mathbb{Z}
_{p}\times\left(
\mathbb{Q}
_{p}\backslash%
\mathbb{Z}
_{p}\right)  $\\\hline
$\left\{  3\right\}  $ & $\left\{  2\right\}  $ & $\left(
\mathbb{Q}
_{p}\backslash%
\mathbb{Z}
_{p}\right)  \times%
\mathbb{Z}
_{p}$\\\hline
$\left\{  2,3\right\}  $ & $\varnothing$ & $%
\mathbb{Z}
_{p}\times%
\mathbb{Z}
_{p}.$\\\hline
\end{tabular}
\ \ \ \
\]
Now, by using the notation
\[
\boldsymbol{Z}^{(5)}\left(  \boldsymbol{s};I\right)  =%
{\displaystyle\int\limits_{Sect(I)}}
|x_{2}|_{p}^{s_{12}}|x_{3}|_{p}^{s_{13}}|1-x_{2}|_{p}^{s_{42}}|1-x_{3}%
|_{p}^{s_{43}}|x_{2}-x_{3}|_{p}^{s_{23}}dx_{2}dx_{3},
\]
we have%
\[
\boldsymbol{Z}^{(5)}\left(  \boldsymbol{s}\right)  =\boldsymbol{Z}%
^{(5)}\left(  \boldsymbol{s};\varnothing\right)  +\boldsymbol{Z}^{(5)}\left(
\boldsymbol{s};\{2\}\right)  +\boldsymbol{Z}^{(5)}\left(  \boldsymbol{s}%
;\{3\}\right)  +\boldsymbol{Z}^{(5)}\left(  \boldsymbol{s};\{2,3\}\right)  .
\]

\begin{case}
$Sect(\left\{  \varnothing\right\}  )=\left(
\mathbb{Q}
_{p}\backslash%
\mathbb{Z}
_{p}\right)  \times\left(
\mathbb{Q}
_{p}\backslash%
\mathbb{Z}
_{p}\right)  $.
\end{case}

By changing of variables as $x_{2}\rightarrow\frac{1}{u},$ $x_{3}%
\rightarrow\frac{1}{v}$ and by using \cite[Lemma 7]{Zun-B-C-LMP}, we have
\begin{gather}
\boldsymbol{Z}^{(5)}\left(  \boldsymbol{s};\left\{  \varnothing\right\}
\right)  =\int\limits_{(%
\mathbb{Q}
_{p}\backslash%
\mathbb{Z}
_{p})^{2}}|x_{2}|_{p}^{s_{12}+s_{42}}|x_{3}|_{p}^{s_{13}+s_{43}}|x_{2}%
-x_{3}|_{p}^{s_{23}}dx_{2}dx_{3}.\nonumber\\
=\int_{(p%
\mathbb{Z}
_{p})^{2}}|u|_{p}^{-s_{12}-s_{42}-2}|v|_{p}^{-s_{13}-s_{43}-2}|\frac{1}%
{u}-\frac{1}{v}|_{p}^{s_{23}}dudv\nonumber\\
=p^{-2+s_{12}+s_{42}+s_{23}+2+s_{13}+s_{43}+s_{23}+2-s_{23}}\int_{%
\mathbb{Z}
_{p}^{2}}\frac{|v-u|_{p}^{s_{23}}}{|u|_{p}^{s_{12}+s_{42}+s_{23}+2}%
|v|_{p}^{s_{13}+s_{43}+s_{23}+2}}dudv\nonumber\\
=\frac{p^{2+s_{12}+s_{42}+s_{23}+s_{13}+s_{43}}(1-p^{-1})}{1-p^{2+s_{12}%
+s_{42}+s_{23}+s_{13}+s_{43}}}\left[  \frac{(1-p^{-1})p^{1+s_{12}%
+s_{42}+s_{23}}}{1-p^{1+s_{12}+s_{42}+s_{23}}}\right. \nonumber\\
+\left.  \frac{(1-p^{-1})p^{1+s_{13}+s_{43}+s_{23}}}{1-p^{1+s_{13}%
+s_{43}+s_{23}}}+\frac{(1-p^{-1})p^{-1-s_{23}}}{1-p^{-1-s_{23}}}+\left(
p-2\right)  p^{-1}\right]  . \label{Sec1}%
\end{gather}
The integral $\boldsymbol{Z}^{(5)}\left(  \boldsymbol{s};\left\{
\varnothing\right\}  \right)  $ is holomorphic in the solution set of the
following conditions:%
\begin{gather}
\operatorname{Re}(s_{12})+\operatorname{Re}(s_{42})+\operatorname{Re}%
(s_{23})<-1,\label{Condition5-0}\\
\operatorname{Re}(s_{13})+\operatorname{Re}(s_{43})+\operatorname{Re}%
(s_{23})<-1,\nonumber\\
\operatorname{Re}\left(  s_{23}\right)  >-1,\nonumber\\
\operatorname{Re}(s_{12})+\operatorname{Re}(s_{42})+\operatorname{Re}%
(s_{13})+\operatorname{Re}(s_{43})+\operatorname{Re}(s_{23})<-2.\nonumber
\end{gather}

\begin{case}
$Sect(\left\{  2\right\}  )=%
\mathbb{Z}
_{p}\times\left(
\mathbb{Q}
_{p}\backslash%
\mathbb{Z}
_{p}\right)  .$
\end{case}

In this case $|x_{2}|_{p}\leq1$ and $|x_{3}|_{p}>1$, now by ultrametric
property $|1-x_{3}|_{p}=|x_{3}|_{p}$ and $|x_{2}-x_{3}|_{p}=|x_{3}|_{p}$,
then
\begin{align*}
\boldsymbol{Z}^{(5)}\left(  \boldsymbol{s};\left\{  2\right\}  \right)   &
=\int\limits_{%
\mathbb{Z}
_{p}\times\left(
\mathbb{Q}
_{p}\backslash%
\mathbb{Z}
_{p}\right)  }|x_{2}|_{p}^{s_{12}}|x_{3}|_{p}^{s_{13}}|1-x_{2}|_{p}^{s_{42}%
}|1-x_{3}|_{p}^{s_{43}}|x_{2}-x_{3}|_{p}^{s_{23}}dx_{2}dx_{3}\\
&  =\int\limits_{%
\mathbb{Z}
_{p}\times\left(
\mathbb{Q}
_{p}\backslash%
\mathbb{Z}
_{p}\right)  }|x_{2}|_{p}^{s_{12}}|x_{3}|_{p}^{s_{13}+s_{43}+s_{23}}%
|1-x_{2}|_{p}^{s_{42}}dx_{2}dx_{3}\\
&  =\int\limits_{%
\mathbb{Z}
_{p}}|x_{2}|_{p}^{s_{12}}|1-x_{2}|_{p}^{s_{42}}dx_{2}\int_{\left(
\mathbb{Q}
_{p}\backslash%
\mathbb{Z}
_{p}\right)  }|x_{3}|_{p}^{s_{13}+s_{43}+s_{23}}dx_{3}\\
&  =:\boldsymbol{Z}_{0}^{(5)}\left(  \boldsymbol{s};\left\{  2\right\}
\right)  \boldsymbol{Z}_{1}^{(5)}\left(  \boldsymbol{s};\left\{  2\right\}
\right)  .
\end{align*}
The calculation of $\boldsymbol{Z}^{(5)}\left(  \boldsymbol{s};\left\{
2\right\}  ,0\right)  $ is similar to the case of integral (\ref{4point}):%
\[
\boldsymbol{Z}^{(5)}\left(  \boldsymbol{s};\left\{  2\right\}  ,0\right)
=p^{-1}\left(  p-2\right)  +\frac{(1-p^{-1})p^{-1-s_{12}}}{1-p^{-1-s_{12}}%
}+\frac{(1-p^{-1})p^{-1-s_{42}}}{1-p^{-1-s_{42}}}.
\]
The integral $\boldsymbol{Z}^{(5)}\left(  \boldsymbol{s};\left\{  2\right\}
,0\right)  $ is holomorphic in the domain
\begin{equation}
\operatorname{Re}\left(  s_{12}\right)  >-1\text{ and }\operatorname{Re}%
\left(  s_{42}\right)  >-1. \label{Condition5-1}%
\end{equation}
The calculation of $\boldsymbol{Z}^{(5)}\left(  \boldsymbol{s};\left\{
2\right\}  ,1\right)  $ is similar to the calculation of $\boldsymbol{Z}%
^{(4)}\left(  \boldsymbol{s};\left\{  \varnothing\right\}  \right)  $ given in
the Subsection of $4$-point zeta function. By changing of variables as
$x_{3}=\frac{1}{y_{3}},dx_{3}=\frac{dy_{3}}{|y_{3}|_{p}^{2}}$, we obtain%
\begin{multline}
\boldsymbol{Z}^{(5)}\left(  \boldsymbol{s};\left\{  2\right\}  ,1\right)  =%
{\displaystyle\int\limits_{\mathbb{Q}_{p}\backslash\mathbb{Z}_{p}}}
|x_{3}|_{p}^{s_{13}+s_{43}+s_{23}}dx_{3}=%
{\displaystyle\int\limits_{p\mathbb{Z}_{p}}}
\frac{dy_{3}}{|y_{3}|_{p}^{2+s_{13}+s_{43}+s_{23}}}\label{Sec2b}\\
=\frac{(1-p^{-1})p^{1+s_{13}+s_{43}+s_{23}}}{1-p^{1+s_{13}+s_{43}+s_{23}}%
}.\nonumber
\end{multline}
The integral $\boldsymbol{Z}^{(5)}\left(  \boldsymbol{s};\left\{  2\right\}
,1\right)  $ is holomorphic in
\begin{equation}
\operatorname{Re}(s_{13})+\operatorname{Re}(s_{43})+\operatorname{Re}%
(s_{23})<-1. \label{Condition5-2}%
\end{equation}

\begin{case}
$Sect(\left\{  3\right\}  )=\left(
\mathbb{Q}
_{p}\backslash%
\mathbb{Z}
_{p}\right)  \times%
\mathbb{Z}
_{p}.$
\end{case}

In this case, $|x_{2}|_{p}>1$, by the ultrametric inequality $|1-x_{2}%
|_{p}=|x_{2}|_{p}$ and $|x_{2}-x_{3}|_{p}=|x_{2}|_{p}$, then
\begin{align*}
\boldsymbol{Z}^{(5)}\left(  \boldsymbol{s};\left\{  3\right\}  \right)   &
=\int\limits_{\left(  \mathbb{Q}_{p}\backslash\mathbb{Z}_{p}\right)
\times\mathbb{Z}_{p}}|x_{2}|_{p}^{s_{12}}|x_{3}|_{p}^{s_{13}}|1-x_{2}%
|_{p}^{s_{42}}|1-x_{3}|_{p}^{s_{43}}|x_{2}-x_{3}|_{p}^{s_{23}}dx_{2}dx_{3}\\
&  =\int\limits_{%
\mathbb{Q}
_{p}\backslash%
\mathbb{Z}
_{p}\times%
\mathbb{Z}
_{p}}|x_{2}|_{p}^{s_{12}+s_{42}+s_{23}}|x_{3}|_{p}^{s_{13}}|1-x_{3}%
|_{p}^{s_{43}}dx_{2}dx_{3}\\
&  =%
{\displaystyle\int\limits_{\mathbb{Z}_{p}}}
|x_{3}|_{p}^{s_{13}}|1-x_{3}|_{p}^{s_{43}}dx_{3}\int\limits_{%
\mathbb{Q}
_{p}\backslash%
\mathbb{Z}
_{p}}|x_{2}|_{p}^{s_{12}+s_{42}+s_{23}}dx_{2}\\
&  =:\boldsymbol{Z}_{0}^{(N)}\left(  \boldsymbol{s};\left\{  3\right\}
\right)  \boldsymbol{Z}_{1}^{(N)}\left(  \boldsymbol{s};\left\{  3\right\}
\right)  .
\end{align*}
These integrals are similar to the ones obtained in the case of $Sect(\{2\})$%
:
\begin{equation}
\boldsymbol{Z}^{(5)}\left(  \underline{\boldsymbol{s}};\left\{  3\right\}
,0\right)  =p^{-1}\left(  p-2\right)  +\frac{(1-p^{-1})p^{-1-s_{13}}%
}{1-p^{-1-s_{13}}}+\frac{(1-p^{-1})p^{-1-s_{43}}}{1-p^{-1-s_{43}}},
\label{Sec3aa}%
\end{equation}
the integral $\boldsymbol{Z}^{(5)}\left(  \boldsymbol{s};\left\{  3\right\}
,0\right)  $ is holomorphic in
\begin{equation}
\operatorname{Re}\left(  s_{13}\right)  >-1\text{ and }\operatorname{Re}%
\left(  s_{43}\right)  >-1. \label{Condition5-3}%
\end{equation}
And
\[
\boldsymbol{Z}^{(5)}\left(  \boldsymbol{s};\left\{  3\right\}  ,1\right)
=\frac{(1-p^{-1})p^{1+s_{12}+s_{42}+s_{23}}}{1-p^{1+s_{12}+s_{42}+s_{23}}},
\]
the integral $\boldsymbol{Z}^{(5)}\left(  \boldsymbol{s};\left\{  3\right\}
,1\right)  $ is holomorphic in
\begin{equation}
\operatorname{Re}(s_{12})+\operatorname{Re}(s_{42})+\operatorname{Re}%
(s_{23})<-1. \label{Condition5-4}%
\end{equation}

\begin{case}
$Sect(\left\{  2,3\right\}  )=%
\mathbb{Z}
_{p}\times%
\mathbb{Z}
_{p}$.
\end{case}

By using the partition $\mathbb{Z}_{p}^{2}=(p\mathbb{Z}_{p})^{2}\sqcup
S_{0}^{2}$, with $S_{0}^{2}=p\mathbb{Z}_{p}\times\mathbb{Z}_{p}^{\times}%
\sqcup\mathbb{Z}_{p}^{\times}\times p\mathbb{Z}_{p}\sqcup\mathbb{Z}%
_{p}^{\times}\times\mathbb{Z}_{p}^{\times}$, we have
\begin{align*}
\boldsymbol{Z}^{(5)}\left(  \boldsymbol{s};\left\{  2,3\right\}  \right)   &
=\int\limits_{%
\mathbb{Z}
_{p}\times%
\mathbb{Z}
_{p}}|x_{2}|_{p}^{s_{12}}|x_{3}|_{p}^{s_{13}}|1-x_{2}|_{p}^{s_{42}}%
|1-x_{3}|_{p}^{s_{43}}|x_{2}-x_{3}|_{p}^{s_{23}}dx_{2}dx_{3}\\
&  =:I_{31}\left(  \boldsymbol{s}\right)  +I_{32}\left(  \boldsymbol{s}%
\right)  +I_{33}\left(  \boldsymbol{s}\right)  +I_{34}\left(  \boldsymbol{s}%
\right)  ,
\end{align*}
where%
\[%
\begin{array}
[c]{l}%
I_{31}\left(  \boldsymbol{s}\right)  :=\int\limits_{\left(  p%
\mathbb{Z}
_{p}\right)  ^{2}}|x_{2}|_{p}^{s_{12}}|x_{3}|_{p}^{s_{13}}|x_{2}-x_{3}%
|_{p}^{s_{23}}dx_{2}dx_{3},\\
I_{32}\left(  \boldsymbol{s}\right)  :=\int\limits_{p%
\mathbb{Z}
_{p}\times%
\mathbb{Z}
_{p}^{\times}}|x_{2}|_{p}^{s_{12}}|1-x_{3}|_{p}^{s_{43}}dx_{2}dx_{3},\\
I_{33}\left(  \boldsymbol{s}\right)  :=\int\limits_{%
\mathbb{Z}
_{p}^{\times}\times p%
\mathbb{Z}
_{p}}|x_{3}|_{p}^{s_{13}}|1-x_{2}|_{p}^{s_{42}}dx_{2}dx_{3},\text{ }\\
I_{34}\left(  \boldsymbol{s}\right)  :=\int\limits_{%
\mathbb{Z}
_{p}^{\times}\times%
\mathbb{Z}
_{p}^{\times}}|1-x_{2}|_{p}^{s_{42}}|1-x_{3}|_{p}^{s_{43}}|x_{2}-x_{3}%
|_{p}^{s_{23}}dx_{2}dx_{3}.
\end{array}
\]
By using \cite[Lemma 4]{Zun-B-C-LMP}, we have%
\begin{align}
I_{31}\left(  \underline{\boldsymbol{s}}\right)   &  =\frac{p^{-2-s_{12}%
-s_{13}-s_{23}}}{1-p^{-2-s_{12}-s_{13}-s_{23}}}\left[  \frac{p^{-1-s_{12}%
}(1-p^{-1})^{2}}{1-p^{-1-s_{12}}}+\frac{p^{-1-s_{13}}(1-p^{-1})^{2}%
}{1-p^{-1-s_{13}}}\right. \nonumber\label{Sec4a}\\
&  \left.  +\frac{p^{-1-s_{23}}(1-p^{-1})^{2}}{1-p^{-1-s_{23}}}+\left(
p-1\right)  \left(  p-2\right)  p^{-2}\right]  .\nonumber
\end{align}
The integral $I_{31}\left(  \underline{\boldsymbol{s}}\right)  $ is
holomorphic in
\begin{equation}
\operatorname{Re}(s_{12})>-1,\ \operatorname{Re}(s_{13}%
)>-1,\ \operatorname{Re}(s_{23})>-1,\text{ and }\operatorname{Re}%
(s_{12})+\operatorname{Re}(s_{13})+\operatorname{Re}(s_{23})>-2.
\label{Condition5-5}%
\end{equation}
By using Examples \ref{Example-1} and \ref{Example-2},
\[
I_{32}\left(  \boldsymbol{s}\right)  =\frac{(1-p^{-1})p^{-1-s_{12}}%
}{1-p^{-1-s_{12}}}\left[  \frac{(1-p^{-1})p^{-1-s_{43}}}{1-p^{-1-s_{43}}%
}+p^{-1}\left(  p-2\right)  \right]  ,
\]
and
\[
I_{33}\left(  \boldsymbol{s}\right)  =\frac{(1-p^{-1})p^{-1-s_{13}}%
}{1-p^{-1-s_{13}}}\left[  \frac{(1-p^{-1})p^{-1-s_{42}}}{1-p^{-1-s_{42}}%
}+p^{-1}\left(  p-2\right)  \right]  .
\]
These integrals are holomorphic in
\begin{equation}
\operatorname{Re}\left(  s_{12}\right)  >-1,\ \operatorname{Re}\left(
s_{13}\right)  >-1,\ \operatorname{Re}\left(  s_{42}\right)  >-1,\text{ and
}\operatorname{Re}\left(  s_{43}\right)  >-1. \label{Condition5-6}%
\end{equation}
Finally, by Example \ref{Example-5},
\begin{align}
I_{34}\left(  \boldsymbol{s}\right)   &  =p^{-2}(p-2)(p-3)+(p-2)\frac
{p^{-2-s_{23}}(1-p^{-1})}{1-p^{-1-s_{23}}}\label{Sec4d}\\
&  +(p-2)\frac{p^{-2-s_{42}}(1-p^{-1})}{1-p^{-1-s_{42}}}+(p-2)\frac
{p^{-2-s_{43}}(1-p^{-1})}{1-p^{-1-s_{43}}}.\nonumber
\end{align}
This integral is holomorphic in
\begin{equation}
\operatorname{Re}(s_{42})>-1,\ \operatorname{Re}(s_{43})>-1,\ \text{ and
}\operatorname{Re}(s_{23})>-1. \label{Condition5-7}%
\end{equation}
In conclusion, the $5$-point local zeta functions is holomorphic on the region
of $\mathbb{C}^{5}$ defined by
\begin{gather*}
\operatorname{Re}\left(  s_{12}\right)  >-1,\ \operatorname{Re}\left(
s_{13}\right)  >-1,\ \operatorname{Re}\left(  s_{42}\right)  >-1,\text{ and
}\operatorname{Re}\left(  s_{43}\right)  >-1,\\
\operatorname{Re}(s_{12})+\operatorname{Re}(s_{42})+\operatorname{Re}%
(s_{23})<-1,\\
\operatorname{Re}(s_{13})+\operatorname{Re}(s_{43})+\operatorname{Re}%
(s_{23})<-1,\\
{\operatorname{Re}}(s_{12})+\operatorname{Re}(s_{13})+\operatorname{Re}%
(s_{23})>-2.\\
\operatorname{Re}(s_{12})+\operatorname{Re}(s_{42})+\operatorname{Re}%
(s_{13})+\operatorname{Re}(s_{43})+\operatorname{Re}(s_{23})<-2.
\end{gather*}
Which is a non-empty subset, since it contains the open set
\begin{gather*}
-\frac{2}{3}<\operatorname{Re}(s_{ij})<0\text{ },\\
-\frac{2}{3}<\operatorname{Re}(s_{1i})<-\frac{1}{2}\text{,}\\
-\frac{2}{3}<\operatorname{Re}(s_{\left(  N-1\right)  i})<-\frac{1}{2}.
\end{gather*}
We regularize the $p$-adic open string $5$-point amplitude by using the
meromorphic continuation of $\boldsymbol{Z}^{(5)}\left(  \boldsymbol{s}%
\right)  $ by setting
\[
\boldsymbol{A}^{(5)}\left(  \boldsymbol{k}\right)  :=\boldsymbol{Z}%
^{(5)}(\boldsymbol{s})\mid_{s_{12}=\boldsymbol{k}_{1}\boldsymbol{k}_{2}%
,s_{13}=\boldsymbol{k}_{1}\boldsymbol{k}_{3},s_{42}=\boldsymbol{k}%
_{4}\boldsymbol{k}_{2},s_{43}=\boldsymbol{k}_{4}\boldsymbol{k}_{3}%
,s_{23}=\boldsymbol{k}_{2}\boldsymbol{k}_{3}}.
\]
The amplitude $\boldsymbol{A}^{(5)}\left(  \boldsymbol{k}\right)  $ agrees
with the one computed in \cite{Brekke:1988dg} by using the Feynman rules of
the effective Lagrangian:%
\begin{equation}
\boldsymbol{A}^{(5)}\left(  \boldsymbol{k}\right)  =K_{5}+K_{4}%
{\textstyle\sum\limits_{i<j}}
x_{ij}+%
{\displaystyle\sum\limits_{_{\substack{\left\{  i,j,k,l\right\}
\subset\left\{  1,2,3,4,5\right\}  \\i<j,k<l}}}}
x_{ij}x_{kl}, \label{Amplitude_A_5A}%
\end{equation}
where $K_{5}=\frac{\left(  p-2\right)  \left(  p-3\right)  }{p^{2}}$,
$K_{4}=\frac{p-2}{p}$, and $ij$ and $kl$ are summed over all pairs of
compatible channels. By a pair of compatible channels $ij$, $kl$ we mean that
$i$, $j$, $k$ and $l$ are different, and that $i<j$, $k$ $<$ $l$.

\subsection{$p$-Adic open string $N$-point\ amplitudes}

We fix an integer $N\geq4$ and consider the general $N$-point zeta function:
\begin{equation}
\boldsymbol{Z}^{(N)}\left(  \boldsymbol{s}\right)  =%
{\displaystyle\int\limits_{\mathbb{Q}_{p}^{N-3}}}
{\displaystyle\prod\limits_{i=2}^{N-2}}
\left\vert x_{i}\right\vert _{p}^{s_{1i}}\left\vert 1-x_{i}\right\vert
_{p}^{s_{(N-1)i}}\text{ }%
{\displaystyle\prod\limits_{2\leq i<j\leq N-2}}
\left\vert x_{i}-x_{j}\right\vert _{p}^{s_{ij}}%
{\displaystyle\prod\limits_{i=2}^{N-2}}
dx_{i}. \label{NZeta}%
\end{equation}
We divide the domain of integration $\mathbb{Q}_{p}^{N-3}$ into sectors. Given
$I\subseteq T=\{2,3,\ldots,N-2\}$ the attached sector is defined as%
\[
Sect(I)=\left\{  \left(  x_{2},\ldots,x_{N-2}\right)  \in\mathbb{Q}_{p}%
^{N-3};\left\vert x_{i}\right\vert _{p}\leq1\text{ }\Leftrightarrow i\in
I\right\}  .
\]
Hence, the $N$-point zeta function (\ref{NZeta}) can be written as
\[
\boldsymbol{Z}^{(N)}\left(  \boldsymbol{s}\right)  =\sum_{I\subseteq
T}\boldsymbol{Z}^{(N)}\left(  \boldsymbol{s};I\right)  ,
\]
where
\[
\boldsymbol{Z}^{(N)}\left(  \boldsymbol{s};I\right)  :=%
{\displaystyle\int\limits_{Sect(I)}}
F\left(  \boldsymbol{s},\boldsymbol{x};N\right)
{\displaystyle\prod\limits_{i=2}^{N-2}}
dx_{i},
\]
with
\[
F\left(  \boldsymbol{s},\boldsymbol{x};N\right)  :=%
{\displaystyle\prod\limits_{i=2}^{N-2}}
\left\vert x_{i}\right\vert _{p}^{s_{1i}}\left\vert 1-x_{i}\right\vert
_{p}^{s_{(N-1)i}}\text{ }%
{\displaystyle\prod\limits_{2\leq i<j\leq N-2}}
\left\vert x_{i}-x_{j}\right\vert _{p}^{s_{ij}}%
\]
and $\boldsymbol{x}=\left(  x_{2},\ldots,x_{N-2}\right)  \in\mathbb{Q}%
_{p}^{N-3}$.

By Lemma 2 in \cite{Zun-B-C-LMP},
\begin{gather*}
Z^{(N)}\left(  \boldsymbol{s};I\right)  =p^{M(\boldsymbol{s})}\left\{
{\displaystyle\int\limits_{\mathbb{Z}_{p}^{\left\vert I\right\vert }}}
{\displaystyle\prod\limits_{i\in I}}
\left\vert x_{i}\right\vert _{p}^{s_{1i}}\left\vert 1-x_{i}\right\vert
_{p}^{s_{(N-1)i}}\text{ }%
{\displaystyle\prod\limits_{\substack{2\leq i<j\leq N-2\\i,j\in I}}}
\left\vert x_{i}-x_{j}\right\vert _{p}^{s_{i}{}_{j}}%
{\displaystyle\prod\limits_{i\in I}}
dx_{i}\right\} \\
\times\left\{
{\displaystyle\int\limits_{\mathbb{Z}_{p}^{\left\vert T\smallsetminus
I\right\vert }}}
\frac{%
{\displaystyle\prod\limits_{\substack{2\leq i<j\leq N-2\\i,j\in
T\smallsetminus I}}}
\left\vert x_{i}-x_{j}\right\vert _{p}^{s_{ij}}}{%
{\displaystyle\prod\limits_{i\in T\smallsetminus I}}
\left\vert x_{i}\right\vert _{p}^{2+s_{1i}+s_{\left(  N-1\right)  i}%
+\sum_{2\leq j\leq N-2,j\neq i}s_{ij}}}%
{\displaystyle\prod\limits_{i\in T\smallsetminus I}}
dx_{i}\right\} \\
=:p^{M(\boldsymbol{s})}Z_{0}^{(N)}\left(  \boldsymbol{s};I\right)  Z_{1}%
^{(N)}\left(  \boldsymbol{s};T\smallsetminus I\right)  ,
\end{gather*}
where%
\[
M(\boldsymbol{s}):=\left\vert T\smallsetminus I\right\vert +\sum_{i\in
T\smallsetminus I}(s_{1i}+s_{\left(  N-1\right)  i})+\sum_{\substack{2\leq
i<j\leq N-2\\i\in T\smallsetminus I,j\in T}}s_{ij}+\sum_{\substack{2\leq
i<j\leq N-2\\i\in I,j\in T\smallsetminus I}}s_{ij}.
\]
In addition $Z_{0}^{(N)}\left(  \boldsymbol{s};I\right)  $, $Z_{1}%
^{(N)}\left(  \boldsymbol{s};T\smallsetminus I\right)  $ are multivariate
local zeta functions and
\begin{equation}
Z^{(N)}\left(  \boldsymbol{s}\right)  =\sum_{I\subseteq T}p^{M(\boldsymbol{s}%
)}Z_{0}^{(N)}\left(  \boldsymbol{s};I\right)  Z_{1}^{(N)}\left(
\boldsymbol{s};T\smallsetminus I\right)  , \label{Formula_Z_N}%
\end{equation}
with the convention that $Z_{i}^{(N)}\left(  \boldsymbol{s};\varnothing
\right)  =1$ for $i=0$, $1$.

In \cite{Zun-B-C-LMP} we show that $Z^{(N)}\left(  \boldsymbol{s}\right)  $
has an analytic continuation to the whole $\mathbb{C}^{D}$ as a rational
function in the variables $p^{-s_{ij}}$ by showing that all functions that
appear on the right-hand side of (\ref{Formula_Z_N}) are holomorphic in a
region $H(\mathbb{C})$ in $\mathbb{C}^{D}$ (\cite[Definition 3 and Remarks
9-10]{Zun-B-C-LMP}) defined by
\begin{gather}
\left\vert J\right\vert +\sum_{i\in J}(\operatorname{Re}\left(  s_{1}{}%
_{i}\right)  +\operatorname{Re}\left(  s_{\left(  N-1\right)  i}\right)
)+\sum_{\substack{2\leq i<j\leq N-2\\i\in J}}\operatorname{Re}\left(
s_{ij}\right) \tag{$C1$}\\
+\sum_{\substack{2\leq i<j\leq N-2\\i\in T\smallsetminus J,j\in J}%
}\operatorname{Re}\left(  s_{ij}\right)  <0\text{ for }J\in\mathfrak{F}%
_{1};\nonumber
\end{gather}%
\begin{equation}
\left\vert K\right\vert -1+\sum_{\substack{2\leq i<j\leq N-2\\i,j\in
K}}\operatorname{Re}(s_{ij})>0\text{ for }K\in\mathfrak{F}_{2}\text{;}
\tag{$C2$}%
\end{equation}%
\begin{equation}
1+\operatorname{Re}(s_{ij})>0\text{ for }ij\in\mathcal{G\subseteq}\left\{
ij;2\leq i<j\leq N-2\right\}  , \tag{$C3$}%
\end{equation}
where $\mathfrak{F}_{1}$, $\mathfrak{F}_{2}$ are families of non-empty subsets
of $T$, and $\mathcal{G}$ is a non-empty subset\ of $\left\{  ij;2\leq i<j\leq
N-2,i,j\in T\right\}  $.%
\begin{equation}
\left\vert J\right\vert +\sum_{i\in S}\operatorname{Re}(s_{ti})+\sum
\nolimits_{2\leq i<j\leq N-2,\text{ }i,j\in J}\operatorname{Re}(s_{ij}%
)>0\text{ for }J\times S\in\mathfrak{F}_{3}\text{,} \tag{$C4$}%
\end{equation}
with $S\subseteq J$, $t\in\left\{  1,N-1\right\}  $,and $\mathfrak{F}_{3}$ a
family of non-empty subsets of $I\times I$;
\begin{equation}
\left\vert K\right\vert -1+\sum_{\substack{2\leq i<j\leq N-2\\i,j\in
K}}\operatorname{Re}(s_{ij})>0\text{ for }K\in\mathfrak{F}_{4}\text{,}
\tag{$C5$}%
\end{equation}
where $\mathfrak{F}_{4}$'s a family of non-empty subsets of $I$;%
\begin{equation}
1+\operatorname{Re}(s_{ij})>0\text{ for }ij\in G_{T}, \tag{$C6$}%
\end{equation}
where $G_{T}$ is a non-empty subset of $\left\{  2\leq i<j\leq N-2,\text{
}i,j\in J\right\}  $ with $(N-1)i$, $1i\in G_{T}$.

In \cite[Lemma 9]{Zun-B-C-LMP}, we show that region $H(\mathbb{C})$ contains
and open and connected subset of $\mathbb{C}^{D}$ defined by the conditions%
\begin{gather}
-\frac{2}{3N_{1}}<\operatorname{Re}(s_{ij})<0\text{ },\tag{$C1\prime\prime$}\\
-\frac{2}{3}<\operatorname{Re}(s_{1i})<-\frac{1}{2}\text{,}\tag{$%
C2\prime\prime$}\\
-\frac{2}{3}<\operatorname{Re}(s_{\left(  N-1\right)  i})<-\frac{1}{2},
\tag{$C3\prime\prime$}%
\end{gather}
for \textbf{\ }$N\geqslant5$, $N_{1}=\frac{(N-4)(N-3)}{2}$, $i,j\in\left\{
2,...,N-2\right\}  $. For the case $N=4$ we only consider conditions
(C2$^{\text{\textquotedblright}}$) and (C3$^{\text{\textquotedblright}}$).
Like in the cases $N=4$ and $N=5$, the key point is to reduce the integrals
$Z_{0}^{(N)}\left(  \boldsymbol{s};I\right)  $, $Z_{1}^{(N)}\left(
\boldsymbol{s};T\smallsetminus I\right)  $ to certain simple integrals, for
which admit meromorphic continuations to the whole $\mathbb{C}^{D}$ as
rational functions in the variables $p^{-s_{ij}}$.

We now state the meromorphic continuation of \ the open string $N$-point zeta function.

\begin{theorem}
[{\cite[Theorem 1]{Zun-B-C-LMP}}](1) The $p$\textit{-adic open string }%
$N$\textit{-point zeta function}, $Z^{\left(  N\right)  }\left(
\boldsymbol{s}\right)  $, gives rise to a holomorphic function on
$H(\mathbb{C})$, which contains an open and connected subset of $\mathbb{C}%
^{D}$. Furthermore, $Z^{\left(  N\right)  }\left(  \boldsymbol{s}\right)  $
admits an analytic continuation to $\mathbb{C}^{D}$, denoted also as
$Z^{\left(  N\right)  }\left(  \boldsymbol{s}\right)  $, as a rational
function in the variables $p^{-s_{ij}},i,j\in\left\{  1,\ldots,N-1\right\}  $.
The real parts of the poles of $Z^{\left(  N\right)  }\left(  \boldsymbol{s}%
\right)  $\ belong to a finite union of hyperplanes, the equations of these
hyperplanes have the form $C1$-$C6$ \ with the symbols `$<$', `$>$' replaced
by `$=$'. (2) If $\boldsymbol{s}=\left(  s_{ij}\right)  \in\mathbb{C}^{D}$,
with $\operatorname{Re}(s_{ij})\geq0$ for $i,j\in\left\{  1,\ldots
,N-1\right\}  $, then the integral $Z^{\left(  N\right)  }\left(
\boldsymbol{s}\right)  $ diverges to $+\infty$.
\end{theorem}

\section{\label{Section_4}String amplitudes over non-Archimedean local fields}

\subsection{Non-Archimedean local fields}

A non-Archimedean local field $\mathbb{K}$ is a locally compact topological
field with respect to a non-discrete topology, which comes from a norm
$\left\vert \cdot\right\vert _{\mathbb{K}}$ satisfying
\[
\left\vert x+y\right\vert _{\mathbb{K}}\leq\max\left\{  \left\vert
x\right\vert _{\mathbb{K}},\left\vert y\right\vert _{\mathbb{K}}\right\}  ,
\]
for $x,y\in\mathbb{K}$. A such norm is called an \textit{ultranorm or
non-Archimedean}. Any non-Archimedean local field $\mathbb{K}$ of
characteristic zero is isomorphic (as a topological field) to a finite
extension of $\mathbb{Q}_{p}$. The field $\mathbb{Q}_{p}$ is the basic example
of non-Archimedean local field of characteristic zero. In the case of positive
characteristic, $\mathbb{K}$ is isomorphic to a finite extension of the field
of formal Laurent series $\mathbb{F}_{q}\left(  \left(  T\right)  \right)  $
over a finite field $\mathbb{F}_{q}$, where $q$ is a power of a prime number
$p$.

The \textit{ring of integers} of $\mathbb{K}$ is defined as
\[
R_{\mathbb{K}}=\left\{  x\in\mathbb{K};\left\vert x\right\vert _{\mathbb{K}%
}\leq1\right\}  .
\]
Geometrically $R_{\mathbb{K}}$ is the unit ball of the normed space $\left(
\mathbb{K},\left\vert \cdot\right\vert _{\mathbb{K}}\right)  $. This ring is a
domain of principal ideals having a unique maximal ideal, which is given by
\[
P_{\mathbb{K}}=\left\{  x\in\mathbb{K};\left\vert x\right\vert _{\mathbb{K}%
}<1\right\}  .
\]
We fix a generator $\pi$ of $P_{\mathbb{K}}$ i.e. $P_{\mathbb{K}}=\pi
R_{\mathbb{K}}$. A such generator is also called a \textit{local uniformizing
parameter of} $\mathbb{K}$, and it plays the same role as $p$ in
$\mathbb{Q}_{p}.$

The \textit{group of units} of $R_{\mathbb{K}}$ is defined as
\[
R_{\mathbb{K}}^{\times}=\left\{  x\in R_{\mathbb{K}};\left\vert x\right\vert
_{\mathbb{K}}=1\right\}  .
\]
The natural map $R_{\mathbb{K}}\rightarrow R_{\mathbb{K}}/P_{\mathbb{K}}%
\cong\mathbb{F}_{q}$ is called the \textit{reduction mod} $P_{\mathbb{K}}$.
The quotient $R_{\mathbb{K}}/P_{\mathbb{K}}\cong\mathbb{F}_{q}$, $q=p^{f}$, is
called the \textit{residue field} of $\mathbb{K}$. Every non-zero element $x$
of $\mathbb{K}$ can be written uniquely as $x=\pi^{ord(x)}u$, $u\in
R_{\mathbb{K}}^{\times}$. We set $ord(0)=\infty$. The normalized valuation of
$\mathbb{K}$ is the mapping
\[%
\begin{array}
[c]{ccc}%
\mathbb{K} & \rightarrow & \mathbb{Z}\cup\left\{  \infty\right\} \\
x & \rightarrow & ord(x).
\end{array}
\]
Then $\left\vert x\right\vert _{\mathbb{K}}=q^{-ord(x)}$ and $\left\vert
\pi\right\vert _{\mathbb{K}}=q^{-1}$.

We fix $\mathfrak{S}\subset R_{\mathbb{K}}$ a set of representatives of
$\mathbb{F}_{q}$ in $R_{\mathbb{K}}$, i.e. $\mathfrak{S}$ is a set which is
mapped bijectively onto $\mathbb{F}_{q}$ by the reduction $\operatorname{mod}$
$P_{\mathbb{K}}$. We assume that $0\in\mathfrak{S}$. Any non-zero element $x$
of $\mathbb{K}$ can be written as
\[
x=\pi^{ord(x)}\sum\limits_{i=0}^{\infty}x_{i}\pi^{i},
\]
where $x_{i}\in\mathfrak{S}$ and $x_{0}\neq0.$ This series converges in the
norm $\left\vert \cdot\right\vert _{\mathbb{K}}$.

We extend the norm $\left\vert \cdot\right\vert _{\mathbb{K}}$ to
$\mathbb{K}^{n}$ by taking
\[
||\boldsymbol{x}||_{\mathbb{K}}:=\max_{1\leq i\leq n}|x_{i}|_{\mathbb{K}},
\]
for $\boldsymbol{x}=(x_{1},\dots,x_{n})\in\mathbb{K}^{n}.$We define
$ord(\boldsymbol{x})=\min_{1\leq i\leq n}\{ord(x_{i})\}$, then
$||\boldsymbol{x}||_{\mathbb{K}}=q^{-ord(\boldsymbol{x})}$. The metric space
$\left(  \mathbb{K}^{n},||\cdot||_{\mathbb{K}}\right)  $ is a complete
ultrametric space.

As we mentioned before, any finite extension $\mathbb{K}$ of $\mathbb{Q}_{p}$
is a non-Archimedean local field. Then
\[
pR_{\mathbb{K}}=\pi^{m}R_{\mathbb{K}},\ \ \ \ \ \ \ m\in\mathbb{N}.
\]
If $m=1$ we say that $\mathbb{K}$ is a \textit{unramified} extension of
$\mathbb{Q}_{p}.$ In other case, we say that $\mathbb{K}$ is a
\textit{ramified} extension. It is well known that for every positive integer
$e$ there exists a unique unramified extension $\mathbb{K}_{e}$ of
$\mathbb{Q}_{p}$ of degree $e$, which means that $\mathbb{K}_{e}$ is a
$\mathbb{Q}_{p}$-vector space of dimension $e$. From now on, $\pi$ denotes a
local uniformizing parameter of $\mathbb{K}_{e}$, thus $pR_{\mathbb{K}_{e}%
}=\pi R_{\mathbb{K}_{e}}$, $R_{\mathbb{K}_{e}}/P_{\mathbb{K}_{e}}%
\cong\mathbb{F}_{p^{e}}$ and $|\pi|_{\mathbb{K}_{e}}=p^{-e}$. For an in-depth
exposition of non-Archimedean local fields, the reader may consult
\cite{We,Taibleson}, see also \cite{Alberio et al,V-V-Z}.

\subsection{Open string amplitudes over non-Archimedean local fields}

The open string amplitudes can be defined over any local field. In this
section, we consider Koba-Nielsen string amplitudes on $\mathbb{K}_{e}$, the
unique unramified extension of $\mathbb{Q}_{p}$ of degree $e$ for all
$e\in\mathbb{N}\backslash\{0\}$. We recall that if $\mathbb{K}_{e}$ is the
unramified extension of degree $e$ of $\mathbb{Q}_{p}$, then $pR_{\mathbb{K}%
_{e}}=\pi R_{\mathbb{K}_{e}}$, $R_{\mathbb{K}_{e}}/P_{\mathbb{K}_{e}}%
\cong\mathbb{F}_{p^{e}}$ and $|\pi|_{\mathbb{K}_{e}}=p^{-e}$. Thus $\pi$ in
$\mathbb{K}_{e}$ plays the role of $p$ in $\mathbb{Q}_{p}$.

The Koba-Nielsen amplitudes on $\mathbb{K}_{e}$ are defined as
\begin{equation}
\boldsymbol{A}_{\mathbb{K}_{e}}^{(N)}\left(  \boldsymbol{k}\right)
=\int\limits_{\mathbb{K}_{e}^{N-3}}\prod\limits_{i=2}^{N-2}\left\vert
x_{i}\right\vert _{\mathbb{K}_{e}}^{\boldsymbol{k}_{1}\boldsymbol{k}_{i}%
}\left\vert 1-x_{i}\right\vert _{\mathbb{K}_{e}}^{\boldsymbol{k}%
_{N-1}\boldsymbol{k}_{i}}\ \prod\limits_{2\leq i<j\leq N-2}\left\vert
x_{i}-x_{j}\right\vert _{\mathbb{K}_{e}}^{\boldsymbol{k}_{i}\boldsymbol{k}%
_{j}}\ \prod\limits_{i=2}^{N-2}dx_{i}, \label{Amplitudegen}%
\end{equation}
where $\prod\nolimits_{i=2}^{N-2}dx_{i}$ is the Haar measure of $\left(
\mathbb{K}_{e}^{N-3},+\right)  $ normalized so that the measure of
$R_{\mathbb{K}_{e}}^{N-3}$ is $1$.

The procedure used to regularize the $p$-adic amplitudes extends to amplitudes
of the form (\ref{Amplitudegen}). In this case, the open string $N$-point zeta
function is defined as
\begin{equation}
\boldsymbol{Z}_{\mathbb{K}_{e}}^{(N)}\left(  \boldsymbol{s}\right)
:=\int\limits_{\mathbb{K}_{e}^{N-3}}F\left(  \boldsymbol{s},\boldsymbol{x}%
;N,\mathbb{K}_{e}\right)  {\prod\limits_{i=2}^{N-2}}dx_{i}, \label{Zeta_1}%
\end{equation}
where
\[
F\left(  \boldsymbol{s},\boldsymbol{x};N,\mathbb{K}_{e}\right)  ={\prod
\limits_{i=2}^{N-2}}\left\vert x_{i}\right\vert _{\mathbb{K}_{e}}^{s_{1i}%
}\left\vert 1-x_{i}\right\vert _{\mathbb{K}_{e}}^{s_{(N-1)i}}\text{ }%
\ {\prod\limits_{2\leq i<j\leq N-2}}\left\vert x_{i}-x_{j}\right\vert
_{\mathbb{K}_{e}}^{s_{ij}}.
\]
where $\underline{\boldsymbol{s}}=\left(  s_{ij}\right)  \in\mathbb{C}^{D}$,
with $D=\frac{\left(  N-3\right)  \left(  N-4\right)  }{2}+2\left(
N-3\right)  $. Let $T=\{2,3,\ldots,N-2\}$, then
\begin{equation}
\boldsymbol{Z}_{\mathbb{K}_{e}}^{(N)}\left(  \boldsymbol{s}\right)
=\sum_{I\subseteq T}p^{eM(\underline{\boldsymbol{s}})}\boldsymbol{Z}%
_{\mathbb{K}_{e}}^{(N)}\left(  \boldsymbol{s};I,0\right)  \ \boldsymbol{Z}%
_{\mathbb{K}_{e}}^{(N)}\left(  \boldsymbol{s};T\smallsetminus I,1\right)  ,
\label{Formula_zeta_amplitude}%
\end{equation}
where
\[
M(\boldsymbol{s}):=\left\vert T\smallsetminus I\right\vert +\sum_{i\in
T\smallsetminus I}(s_{1i}+s_{\left(  N-1\right)  i})+\sum_{\substack{2\leq
i<j\leq N-2\\i\in T\smallsetminus I,\ j\in T}}s_{ij}+\sum_{\substack{2\leq
i<j\leq N-2\\i\in I,j\in T\smallsetminus I}}s_{ij}%
\]%
\[
\boldsymbol{Z}_{\mathbb{K}_{e}}^{(N)}\left(  \boldsymbol{s};I,0\right)
={\int\limits_{R_{\mathbb{K}_{e}}^{\left\vert I\right\vert }}}{\prod
\limits_{i\in I}}\left\vert x_{i}\right\vert _{\mathbb{K}_{e}}^{s_{1i}%
}\left\vert 1-x_{i}\right\vert _{\mathbb{K}_{e}}^{s_{(N-1)i}}\text{ }%
\ {\prod\limits_{\substack{2\leq i<j\leq N-2\\i,j\in I}}}\left\vert
x_{i}-x_{j}\right\vert _{\mathbb{K}_{e}}^{s_{i}{}_{j}}\ {\prod\limits_{i\in
I}}dx_{i},
\]
and
\[
\boldsymbol{Z}_{\mathbb{K}_{e}}^{(N)}\left(  \boldsymbol{s};T\smallsetminus
I,1,\right)  ={\int\limits_{R_{\mathbb{K}_{e}}^{\left\vert T\smallsetminus
I\right\vert }}}F_{1}\left(  \boldsymbol{s},\boldsymbol{x};N,\mathbb{K}%
_{e}\right)  \ {\prod\limits_{i\in T\smallsetminus I}}dx_{i},
\]
where
\[
F_{1}\left(  \boldsymbol{s},\boldsymbol{x};N,\mathbb{K}_{e}\right)
:=\frac{{\prod\limits_{\substack{2\leq i<j\leq N-2\\i,j\in T\smallsetminus
I}}}\left\vert x_{i}-x_{j}\right\vert _{\mathbb{K}_{e}}^{s_{ij}}}%
{{\prod\limits_{i\in T\smallsetminus I}}\left\vert x_{i}\right\vert
_{\mathbb{K}_{e}}^{2+s_{1i}+s_{\left(  N-1\right)  i}+\sum_{2\leq j\leq
N-2,j\neq i}s_{ij}}}.
\]
By convention $\boldsymbol{Z}_{\mathbb{K}_{e}}^{(N)}\left(  \boldsymbol{s}%
;\varnothing,0\right)  =1$, $\boldsymbol{Z}_{\mathbb{K}_{e}}^{(N)}\left(
\boldsymbol{s};\varnothing,1\right)  =1$.

All the zeta functions appearing in the right-hand side of formula
(\ref{Formula_zeta_amplitude}) admit analytic continuations to the whole
$\mathbb{C}^{D}$ as rational functions in the variables $p^{-es_{ij}}$ and
they are holomorphic on a common domain in $\mathbb{C}^{D}$. Therefore
$\boldsymbol{Z}_{\mathbb{K}_{e}}^{(N)}\left(  \boldsymbol{s}\right)  $ is a
holomorphic function in a certain domain of $\mathbb{C}^{D}$ admitting a
meromorphic continuation to the whole $\mathbb{C}^{D}$ as a rational function
in the variables $p^{-es_{ij}}$, see \cite[Theorem 1]{Zun-B-C-LMP}.

We use $\boldsymbol{Z}_{\mathbb{K}_{e}}^{(N)}(\boldsymbol{s})$ as
regularizations of Koba-Nielsen amplitudes $\boldsymbol{A}^{(N)}\left(
\boldsymbol{k},\mathbb{K}_{e}\right)  $, more precisely, we define
\[
\boldsymbol{A}_{\mathbb{K}_{e}}^{(N)}\left(  \boldsymbol{k}\right)
=\boldsymbol{Z}_{\mathbb{K}_{e}}^{(N)}(\boldsymbol{s})\mid_{s_{ij}%
=\boldsymbol{k}_{i}\cdot\boldsymbol{k}_{j}}.
\]
Then $\boldsymbol{A}_{\mathbb{K}_{e}}^{(N)}\left(  \boldsymbol{k}\right)  $ is
a well defined rational function in the variables $p^{-e\boldsymbol{k}%
_{i}\cdot\boldsymbol{k}_{j}}$, which agree with the integral
(\ref{Amplitudegen}) when it converges.

\subsection{The limit $p$ tends to one}

The functions
\[
\boldsymbol{Z}_{\mathbb{K}_{e}}^{(N)}\left(  \boldsymbol{s};I,0\right)
\text{, \ }\boldsymbol{Z}_{\mathbb{K}_{e}}^{(N)}\left(  \boldsymbol{s}%
;T\smallsetminus I,1\right)
\]
are multivariate local zeta functions. Thus, to make mathematical sense of the
limit of $\boldsymbol{Z}_{\mathbb{Q}_{p}}^{(N)}\left(  \boldsymbol{s}\right)
$ as $p\rightarrow1$ we use the work of Denef and Loeser, see
\cite{denefandloeser} and \cite{trossmann}, and compute the limit of
$\boldsymbol{Z}_{\mathbb{K}_{e}}^{(N)}\left(  \boldsymbol{s}\right)  $ as
$e\rightarrow0$ instead of the limit of $\boldsymbol{Z}_{\mathbb{Q}_{p}}%
^{(N)}\left(  \boldsymbol{s}\right)  $ as $p\rightarrow1$. In order to compute
the limit $e\rightarrow0$ is necessary to have an explicit formula for
$\boldsymbol{Z}_{\mathbb{K}_{e}}^{(N)}\left(  \boldsymbol{s}\right)  $, so in
\cite{Zun-B-C-JHEP} we determined the explicit formula by finding explicit
formulas for integrals $\boldsymbol{Z}_{\mathbb{K}_{e}}^{(N)}\left(
\boldsymbol{s};I,0\right)  $ and $\boldsymbol{Z}_{\mathbb{K}_{e}}^{(N)}\left(
\boldsymbol{s};T\smallsetminus I,1\right)  $, see \cite[Theorem B]%
{Zun-B-C-JHEP}. After that, we define
\[
\boldsymbol{Z}_{top}^{(N)}\left(  \boldsymbol{s};I,0\right)  =\lim
_{e\rightarrow0}\boldsymbol{Z}_{\mathbb{K}_{e}}^{(N)}\left(  \boldsymbol{s}%
;I,0\right)
\]
and
\[
\boldsymbol{Z}_{top}^{(N)}\left(  \boldsymbol{s};T\smallsetminus I,1\right)
=\lim_{e\rightarrow0}\boldsymbol{Z}_{\mathbb{K}_{e}}^{(N)}\left(
\boldsymbol{s};T\smallsetminus I,1\right)  ,
\]
which are elements of $\mathbb{Q}\left(  s_{ij},i,j\in\left\{  1,\ldots
,N-1\right\}  \right)  $, the field of rational functions in the variables
$s_{ij}$, $i,j\in\{1,\dots,N-1\}$ with coefficients in $\mathbb{Q}$. Then, by
using (\ref{Formula_zeta_amplitude}), we defined the \textit{open string }%
$N$\textit{-point topological zeta function} as
\[
\boldsymbol{Z}_{top}^{(N)}\left(  \boldsymbol{s}\right)  =\sum_{I\subseteq
T}\boldsymbol{Z}_{top}^{(N)}\left(  \boldsymbol{s};I,0\right)  \boldsymbol{Z}%
_{top}^{(N)}\left(  \boldsymbol{s};T\smallsetminus I,1\right)  .
\]
The open string $N$-point topological zeta function $\boldsymbol{Z}%
_{top}^{(N)}\left(  \boldsymbol{s}\right)  $ is a rational function of
$\mathbb{Q}\left(  s_{ij},i,j\in\left\{  1,\ldots,N-1\right\}  \right)  $. We
now define the \textit{Denef-Loeser open string N-point amplitudes at the tree
level} as
\[
\boldsymbol{A}_{top}^{(N)}\left(  \boldsymbol{k}\right)  =\boldsymbol{Z}%
_{top}^{(N)}\left(  \boldsymbol{s}\right)  \mid_{s_{ij}=\boldsymbol{k}%
_{i}\cdot\boldsymbol{k}_{j}},
\]
with $i\in\left\{  1,\ldots,N-1\right\}  $, $j\in T$ or $i,j\in T$, where
$T=\left\{  2,\ldots,N-2\right\}  $. Thus the \textit{Denef-Loeser} amplitudes
are rational functions of the variables $\boldsymbol{k}_{i}\cdot
\boldsymbol{k}_{j}$, $i,j\in\left\{  1,\ldots,N\right\}  $.

\subsection{Feynman rules, explicit formulas and Denef-Loeser amplitudes}

By using the Feynman rules of the effective Lagrangian, an `explicit formula'
of the type%
\[
\boldsymbol{A}^{(N)}\left(  \boldsymbol{k}\right)  =K_{N}+%
{\textstyle\sum\limits_{l=2}^{N-3}}
K_{i_{1}...i_{l}}%
{\textstyle\sum\limits_{1\leq a<b\leq l}}
\left(  \frac{p-1}{p}\frac{p^{-k_{i_{a}}k_{i_{b}}-1}}{1-p^{-k_{i_{a}}k_{i_{b}%
}-1}}\right)
\]
was given in (\cite{Brekke:1988dg}), where the constants $K_{N}$,
$K_{i_{1}...i_{l}}\in\mathbb{Q}\left(  p\right)  $. Of course, a rigorous
demonstration of a such formula is an open problem. Taking formally the limit
$p$ tends to one, we get%
\[
\boldsymbol{A}_{top}^{(N)}\left(  \boldsymbol{k}\right)  =K_{N}^{top}+%
{\textstyle\sum\limits_{l=2}^{N-3}}
K_{i_{1}...i_{l}}^{top}%
{\textstyle\sum\limits_{1\leq a<b\leq l}}
\frac{1}{k_{i_{a}}k_{i_{b}}+1}.
\]

\subsection{\label{Section_D_L_N_4}Denef-Loeser open string $4$-point
amplitudes}

The open string $4$-point zeta function on $\mathbb{K}_{e}$ is defined as
\[
\boldsymbol{Z}_{\mathbb{K}_{e}}^{(4)}(s)=%
{\displaystyle\int\limits_{\mathbb{K}_{e}}}
|x_{2}|_{\mathbb{K}_{e}}^{s_{12}}|1-x_{2}|_{\mathbb{K}_{e}}^{s_{32}}dx_{2}.
\]
We divide the integration domain in sectors
\[
Sect(I)=\left\{  x_{i}\in%
\mathbb{Q}
:|x_{i}|_{\mathbb{K}_{e}}\leq1\Longleftrightarrow i\in I\right\}  ,\text{
}I\subseteqq T=\left\{  2,3\right\}  ,
\]%
\[%
\begin{tabular}
[c]{|c|c|c|}\hline
$I$ & $T\diagdown I$ & $Sect(I)$\\\hline
$\left\{  2\right\}  $ & $\varnothing$ & $R_{\mathbb{K}_{e}}$\\\hline
$\varnothing$ & $\left\{  2\right\}  $ & $K_{e}\backslash R_{\mathbb{K}_{e}}%
.$\\\hline
\end{tabular}
\ \ \ \ \
\]
In this way we obtain that
\begin{align*}
\boldsymbol{Z}_{\mathbb{K}_{e}}^{(4)}(s)  &  =\boldsymbol{Z}_{\mathbb{K}_{e}%
}^{(4)}(s;\left\{  2\right\}  ,0)+p^{e(1+s_{12}+s_{32})}\boldsymbol{Z}%
_{\mathbb{K}_{e}}^{(4)}(\underline{s}\left\{  2\right\}  ,1)\\
&  =\int_{R_{\mathbb{K}_{e}}}|x_{2}|_{\mathbb{K}_{e}}^{s_{12}}|1-x_{2}%
|_{\mathbb{K}_{e}}^{s_{32}}dx_{2}+p^{e(1+s_{12}+s_{32})}\int_{R_{\mathbb{K}%
_{e}}}|x_{2}|_{\mathbb{K}_{e}}^{-2-s_{12}-s_{32}}dx_{2},
\end{align*}
with%
\[
\boldsymbol{Z}_{\mathbb{K}_{e}}^{(4)}(s;\left\{  2\right\}  ,0)=1-2p^{-e}%
+\frac{\left(  1-p^{-e}\right)  p^{e\left(  -1-s_{12}\right)  }}{1-p^{e\left(
-1-s_{12}\right)  }}+\frac{\left(  1-p^{-e}\right)  p^{e\left(  -1-s_{32}%
\right)  }}{1-p^{e\left(  -1-s_{32}\right)  }},
\]
and
\[
\boldsymbol{Z}_{\mathbb{K}_{e}}^{(4)}(s;\left\{  2\right\}  ,1)=\frac{\left(
1-p^{-e}\right)  p^{e\left(  1+s_{12}+s_{32}\right)  }}{1-p^{e\left(
1+s_{12}+s_{32}\right)  }}.
\]
Taking the limit $e$ approaches to zero,
\[
\boldsymbol{Z}_{top}^{(4)}(s;\left\{  2\right\}  ,0)=-1+\frac{1}{s_{12}%
+1}+\frac{1}{s_{32}+1}%
\]
and
\[
\boldsymbol{Z}_{top}^{(4)}(s;\left\{  2\right\}  ,1)=-\frac{1}{s_{12}%
+s_{32}+1}.
\]
Consequently
\[
\boldsymbol{Z}_{top}^{\left(  4\right)  }(\boldsymbol{s})=-1+\frac{1}%
{s_{12}+1}+\frac{1}{s_{32}+1}-\frac{1}{s_{12}+s_{32}+1}.
\]
By using the kinematic relations $\boldsymbol{k}_{1}+\boldsymbol{k}%
_{2}+\boldsymbol{k}_{3}+\boldsymbol{k}_{4}=0$ and $\boldsymbol{k}_{i}^{2}=2$
we get $\boldsymbol{k}_{1}\boldsymbol{k}_{2}+\boldsymbol{k}_{3}\boldsymbol{k}%
_{2}+1=-1-\boldsymbol{k}_{2}\boldsymbol{k}_{4},$ then the Denef-Loeser string
$4$-point amplitude is given by
\begin{equation}
\boldsymbol{A}_{top}^{\left(  4\right)  }(\boldsymbol{k})=-1+\frac
{1}{\boldsymbol{k}_{1}\boldsymbol{k}_{2}+1}+\frac{1}{\boldsymbol{k}%
_{3}\boldsymbol{k}_{2}+1}+\frac{1}{\boldsymbol{k}_{2}\boldsymbol{k}_{4}+1}.
\label{Denef_Loeser_4}%
\end{equation}

\subsection{\label{Section_D_L_N_5}Denef-Loeser open string $5$-point
amplitudes}

The open string $5$-point zeta function on $\mathbb{K}_{e}$ is given by
\[
\boldsymbol{Z}_{\mathbb{K}_{e}}^{\left(  5\right)  }(\boldsymbol{s})=%
{\displaystyle\int\limits_{\mathbb{K}_{e}^{2}}}
|x_{2}|_{\mathbb{K}_{e}}^{s_{12}}|x_{3}|_{\mathbb{K}_{e}}^{s_{13}}%
|1-x_{2}|_{\mathbb{K}_{e}}^{s_{42}}|1-x_{3}|_{\mathbb{K}_{e}}^{s_{43}}%
|x_{2}-x_{3}|_{\mathbb{K}_{e}}^{s_{23}}dx_{2}dx_{3}.
\]
By dividing the integration domain in sectors as in Section
\ref{Sect_5_points}, we obtain that

\begin{table}[th]%
\begin{tabular}
[c]{|c|c|c|}\hline
$I$ & $I^{c}$ & $Sect(I)$\\\hline
$\left\{  2\right\}  $ & $\left\{  3\right\}  $ & $R_{\mathbb{K}_{e}}%
\times\mathbb{K}_{e}\backslash R_{\mathbb{K}_{e}}$\\\hline
$\left\{  3\right\}  $ & $\left\{  2\right\}  $ & $\mathbb{K}_{e}\backslash
R_{\mathbb{K}_{e}}\times R_{\mathbb{K}_{e}}$\\\hline
$\left\{  2,3\right\}  $ & $\varnothing$ & $R_{\mathbb{K}_{e}}\times
R_{\mathbb{K}_{e}}$\\\hline
$\varnothing$ & $\left\{  2,3\right\}  $ & $\mathbb{K}_{e}\backslash
R_{\mathbb{K}_{e}}\times\mathbb{K}_{e}\backslash R_{\mathbb{K}_{e}}.$\\\hline
\end{tabular}
\caption{ }%
\end{table}The open string $5$-point topological zeta function is defined as
\[
\boldsymbol{Z}_{top}^{(5)}\left(  \boldsymbol{s}\right)  =\sum_{I\subseteq
T}\boldsymbol{Z}_{top}^{(5)}\left(  \boldsymbol{s};I,0\right)  \boldsymbol{Z}%
_{top}^{(5)}\left(  \boldsymbol{s};T\smallsetminus I,1\right)  .
\]
Table 2 contains explicit formulae for all the integrals $\boldsymbol{Z}%
_{top}^{(5)}\left(  \boldsymbol{s};I,0\right)  $ and $\boldsymbol{Z}%
_{top}^{(5)}\left(  \boldsymbol{s};T\smallsetminus I,1\right)  $%
.\begin{table}[thth]
\centering%
\begin{tabular}
[c]{|c|c|c|}\hline
$I$ & $\boldsymbol{Z}_{top}^{(5)}\left(  \boldsymbol{s};I,0\right)  $ &
$\boldsymbol{Z}_{top}^{(5)}\left(  \underline{\boldsymbol{s}};T\smallsetminus
I,1\right)  $\\\hline
$\left\{  2\right\}  $ & $-1+\frac{1}{1+s_{12}}+\frac{1}{1+s_{42}}$ &
$-\frac{1}{1+s_{13}+s_{43}+s_{23}}$\\\hline
&  & \\\hline
$\left\{  3\right\}  $ & $-1+\frac{1}{1+s_{13}}+\frac{1}{1+s_{43}}$ &
$-\frac{1}{1+s_{12}+s_{42}+s_{23}}$\\\hline
&  & \\\hline
$\left\{  2,3\right\}  $ & $%
\begin{array}
[c]{c}%
\left[  \frac{1}{1+s_{12}}+\frac{1}{1+s_{13}}+\frac{1}{1+s_{23}}-1\right]
\frac{1}{2+s_{12}+s_{13}+s_{23}}\\
+\frac{1}{1+s_{12}}\left[  \frac{1}{1+s_{43}}-1\right]  +\frac{1}{1+s_{13}%
}\left[  \frac{1}{1+s_{42}}-1\right]  +\\
2-\frac{1}{1+s_{23}}-\frac{1}{1+s_{42}}-\frac{1}{1+s_{43}}+\\
\frac{1}{2+s_{42}+s_{43}+s_{23}}\left[  \frac{1}{1+s_{42}}+\frac{1}{1+s_{43}%
}+\frac{1}{1+s_{23}}-1\right]
\end{array}
$ & 1\\\hline
$\left\{  \varnothing\right\}  $ & 1 & $%
\begin{array}
[c]{c}%
-\frac{1}{2+s_{52}+s_{53}+s_{23}}\times\\
\left[
\begin{array}
[c]{c}%
\frac{1}{1+s_{12}+s_{42}+s_{23}}+\frac{1}{1+s_{13}+s_{43}+s_{23}}\\
+\frac{1}{1+s_{23}}-1
\end{array}
\right]
\end{array}
$\\\hline
\end{tabular}
\label{tab:itwo}\caption{ }%
\end{table}Therefore, the Denef-Loeser open string $5$-point amplitude is
given by%
\[
\boldsymbol{A}_{top}^{\left(  5\right)  }(\boldsymbol{k})=2-%
{\textstyle\sum\limits_{i<j}}
\frac{1}{1+\boldsymbol{k}_{i}\boldsymbol{k}_{j}}+%
{\displaystyle\sum\limits_{_{\substack{\left\{  i,j,k,l\right\}
\subset\left\{  1,2,3,4,5\right\}  \\i<j,k<l}}}}
\frac{1}{1+\boldsymbol{k}_{i}\boldsymbol{k}_{j}}\frac{1}{1+\boldsymbol{k}%
_{k}\boldsymbol{k}_{l}},
\]
where%
\begin{gather*}%
{\displaystyle\sum\limits_{_{\substack{\left\{  i,j,k,l\right\}
\subset\left\{  1,2,3,4,5\right\}  \\i<j,k<l}}}}
\frac{1}{1+\boldsymbol{k}_{i}\boldsymbol{k}_{j}}\frac{1}{1+\boldsymbol{k}%
_{k}\boldsymbol{k}_{l}}\\
=\frac{1}{1+\boldsymbol{k}_{1}\boldsymbol{k}_{2}}\frac{1}{1+\boldsymbol{k}%
_{3}\boldsymbol{k}_{4}}+\frac{1}{1+\boldsymbol{k}_{1}\boldsymbol{k}_{2}}%
\frac{1}{1+\boldsymbol{k}_{3}\boldsymbol{k}_{5}}+\frac{1}{1+\boldsymbol{k}%
_{1}\boldsymbol{k}_{2}}\frac{1}{1+\boldsymbol{k}_{4}\boldsymbol{k}_{5}}\\
+\frac{1}{1+\boldsymbol{k}_{1}\boldsymbol{k}_{3}}\frac{1}{1+\boldsymbol{k}%
_{2}\boldsymbol{k}_{4}}+\frac{1}{1+\boldsymbol{k}_{1}\boldsymbol{k}_{3}}%
\frac{1}{1+\boldsymbol{k}_{2}\boldsymbol{k}_{5}}+\frac{1}{1+\boldsymbol{k}%
_{1}\boldsymbol{k}_{3}}\frac{1}{1+\boldsymbol{k}_{4}\boldsymbol{k}_{5}}\\
+\frac{1}{1+\boldsymbol{k}_{1}\boldsymbol{k}_{4}}\frac{1}{1+\boldsymbol{k}%
_{2}\boldsymbol{k}_{3}}+\frac{1}{1+\boldsymbol{k}_{1}\boldsymbol{k}_{4}}%
\frac{1}{1+\boldsymbol{k}_{2}\boldsymbol{k}_{5}}+\frac{1}{1+\boldsymbol{k}%
_{1}\boldsymbol{k}_{4}}\frac{1}{1+\boldsymbol{k}_{3}\boldsymbol{k}_{5}}\\
+\frac{1}{1+\boldsymbol{k}_{1}\boldsymbol{k}_{5}}\frac{1}{1+\boldsymbol{k}%
_{2}\boldsymbol{k}_{3}}+\frac{1}{1+\boldsymbol{k}_{1}\boldsymbol{k}_{5}}%
\frac{1}{1+\boldsymbol{k}_{2}\boldsymbol{k}_{4}}+\frac{1}{1+\boldsymbol{k}%
_{1}\boldsymbol{k}_{5}}\frac{1}{1+\boldsymbol{k}_{3}\boldsymbol{k}_{4}}\\
+\frac{1}{1+\boldsymbol{k}_{2}\boldsymbol{k}_{3}}\frac{1}{1+\boldsymbol{k}%
_{4}\boldsymbol{k}_{5}}+\frac{1}{1+\boldsymbol{k}_{3}\boldsymbol{k}_{5}}%
\frac{1}{1+\boldsymbol{k}_{2}\boldsymbol{k}_{4}}+\frac{1}{1+\boldsymbol{k}%
_{2}\boldsymbol{k}_{5}}\frac{1}{1+\boldsymbol{k}_{3}\boldsymbol{k}_{4}}%
\end{gather*}

This amplitude agrees with the calculation%
\begin{align}
\boldsymbol{A}_{top}^{\left(  5\right)  }(\boldsymbol{k})  &  =\lim
_{p\rightarrow1}\boldsymbol{A}^{(5)}\left(  \boldsymbol{k}\right)
\label{Denef_Loeser_5}\\
=  &  K_{5}^{top}+K_{4}^{top}%
{\textstyle\sum\limits_{i<j}}
\frac{1}{\boldsymbol{k}_{i}\boldsymbol{k}_{j}+1}+%
{\textstyle\sum_{\substack{\left\{  i,j,k,l\right\}  \subset\left\{
1,2,3,4,5\right\}  \\i<j,k<l}}}
\frac{1}{\boldsymbol{k}_{i}\boldsymbol{k}_{j}+1}\frac{1}{\boldsymbol{k}%
_{k}\boldsymbol{k}_{l}+1}\nonumber
\end{align}
done using the explicit formula for given by the Feynman rules given in
\cite{Brekke:1988dg}. Here $K_{5}^{top}=-2$, $K_{4}^{top}=-1$.

\subsection{Non-Archimedean closed strings}

In Archimedean string theory, it is known that closed strings can be produced
from the scattering of open strings \cite{Polchinski:1998rq}. Therefore, for
$p$-adic string theory, it is desirable to construct a $p$-adic closed string
theory. This was first studied in \cite{F-O}, see also \cite{B-F},
\cite{Freund:Applications2005}.

The usual Archimedean closed string is described by two coordinates
$(\tau,\sigma)$ with the periodical condition $\sigma^{\prime}=\sigma+2\pi$.
That is, the string worldsheet is a cylinder, which can be conformally mapped
to the whole complex plane $\mathbb{C}$. In order to obtain tree-level
scattering amplitudes we need to insert vertex operators on $\mathbb{C}$.
Using the $\mathrm{SL}(2,\mathbb{C})$ symmetry, we are able to fix three
insertions points. As with the open string, it is conventional to fix three
points to $0,1$ and $\infty$. The simplest non trivial example is the
$4$-point tachyon scattering amplitude for closed strings, also known as the
Virasoro-Shapiro amplitude, is
\[
A_{\mathbb{C}}^{(4)}(\boldsymbol{k})=%
{\displaystyle\int\limits_{\mathbb{C}}}
|z|_{\mathbb{C}}^{\boldsymbol{k}_{1}\boldsymbol{k}_{2}/4}|1-z|_{\mathbb{C}%
}^{\boldsymbol{k}_{1}\boldsymbol{k}_{3}/4}dz=\frac{\Gamma_{\mathbb{C}}%
(-\alpha(s)/2)\Gamma_{\mathbb{C}}(-\alpha(t)/2)}{\Gamma_{\mathbb{C}}%
(-\alpha(s)/2-\alpha(t)/2)},
\]
where $\boldsymbol{k}_{i}\in\mathbb{C}^{D}$ and $\boldsymbol{k}_{i}%
\boldsymbol{k}_{j}=-k_{0,i}k_{0,j}+\cdots k_{l,i}k_{l,j}$ is the Minkowski
product; $|z|_{\mathbb{C}}:=z\bar{z}$ is the square complex norm,
$\alpha(x)=x/4+2$, and
\[
\Gamma_{\mathbb{C}}(s):=%
{\displaystyle\int\limits_{\mathbb{C}}}
\exp[2\pi i(z+\bar{z})]|z|_{\mathbb{C}}^{s-1}dz
\]
is the Gelfand-Graev gamma function over $\mathbb{C}$ \cite{B-F}. For the
closed strings case, the momenta vectors satisfy
\begin{equation}
\boldsymbol{k}_{i}^{2}=8,\qquad\sum_{i=1}^{N}\boldsymbol{k}_{i}=0.
\label{MomentaConditions}%
\end{equation}
This explains the factor of $1/4$ in the exponents relative to the open
strings case. The generalization of these amplitudes to $N$-points is given
by
\begin{equation}
A_{\mathbb{C}}^{(N)}(\boldsymbol{k})=%
{\displaystyle\int\limits_{\mathbb{C}^{N-3}}}
\prod_{i=2}^{N-2}|z_{i}|_{\mathbb{C}}^{\boldsymbol{k}_{1}\boldsymbol{k}_{i}%
/4}|1-z_{i}|_{\mathbb{C}}^{\boldsymbol{k}_{N-1}\boldsymbol{k}_{i}/4}%
\prod_{2\leq i<j\leq N-2}|z_{i}-z_{j}|_{\mathbb{C}}^{\boldsymbol{k}%
_{i}\boldsymbol{k}_{j}/4}\prod_{i=2}^{N-2}dz_{i}, \label{NptClosedArchimedean}%
\end{equation}
where $\boldsymbol{k}=(\boldsymbol{k_{1}},\dots,\boldsymbol{k}_{N})$. As we
can see, these amplitudes are very similar to the Koba-Nielsen amplitudes,
except for the fact that they are being integrated over complex variables and
the replacement $\boldsymbol{k}_{i}\rightarrow\frac{1}{2}\boldsymbol{k}_{i}$.

In order to construct $p$-adic versions of Virasoro-Shapiro amplitudes, one
can consider $\mathbb{Q}_{p}$ as the analog of $\mathbb{R}$, \ and a quadratic
extension of $\mathbb{Q}_{p}$\ as the analog of $\mathbb{C}$. However this
quadratic extension is not unique, and is not an algebraically \ closed field.
This naive approach is followed in \cite{F-O}, \cite{B-F}. This approach is
not very useful here, because we work with open string amplitudes over any
finite extension of $\mathbb{Q}_{p}$.

\begin{definition}
Given an open string $N$-point amplitude $A_{\mathbb{K}}^{(N)}(\boldsymbol{k}%
)$ defined over a non-Archimedean local field $\mathbb{K}$, we attach to it a
Virasoro-Shapiro amplitude defined as $\mathcal{A}_{\mathbb{K}_{2}}%
^{(N)}(\boldsymbol{k})=A_{\mathbb{K}_{2}}^{(N)}(\frac{1}{2}\boldsymbol{k})$,
where $\mathbb{K}_{2}$ is the unique unramified extension of $\mathbb{K}$ of
degree $2$.
\end{definition}

Then, by the results of \cite{Bocardo:2020mk}, $\mathcal{A}_{\mathbb{K}_{2}%
}^{(N)}(\boldsymbol{k})$ admits a meromorphic continuation as a rational
function in the variables $q^{-2s_{ij}}$.

\section{\label{Section_5}A physical view of the limit $p\rightarrow1$ on
$p-$adic string amplitudes}

The problem of finding a relation between the physical string amplitudes and
the $p$-adic one has been present since the initial proposal. One of the first
attempts is that of the Adelic amplitudes which can be written as an infinite
product of physical non Archimedean amplitudes \cite{Freund:1987ck},
\cite{Brekke:1988dg}.

There are other possibilities to look for a link between Arquimedian and
Non-Archimedean amplitudes. One of them is to consider the limit
$p\rightarrow1$. This limit is very intriguing and there are different
interpretations of it \cite{Spokoiny:1988zk,Gerasimov:2000zp,Ghoshal:2006te}.
For instance, in \cite{Spokoiny:1988zk} it was argued that the analytic
continuation of $p$ to the complex numbers can be reflected in the Lagrangian
and equations of motion by turning the non-local equations of motion into a
local and linear dynamical description with a logarithmic potential. This is
an ordinary theory with real physical amplitudes. In \cite{Gerasimov:2000zp}
it was argued that, similarly as found in \cite{Spokoiny:1988zk}, the
expansion of the effective action around $p\rightarrow1$ leads to a linear
theory with logarithmic potential. Moreover that this theory is deeply related
to the boundary string field theory proposed by Witten \cite{Witten:1992qy} in
the context of the developing of a background independence string theory. The
limit $p\rightarrow1$ also can be interpreted in terms of some scaling
transformations of the renormalization group for the Bruhat-Tits tree
\cite{Ghoshal:2006te}.

As we have seen in this survey, the non-Archimedean nature is encoded in the
worldsheet theory. In this context we must recall that $p$ as a prime number,
thus the analysis necessarily need to be carry out in a rigorous way. The
right way of taking the limit $p\rightarrow1$ involves the introduction of
unramified finite extensions of the $p$-adic field $\mathbb{Q}_{p}$. In
\cite{Zun-B-C-JHEP} the limit $p\rightarrow1$ was discussed for tree-level
string amplitudes, by using the topological zeta functions introduced by Denef
and Loeser \cite{denefandloeser,trossmann}. We found that the limit
$p\rightarrow1$ of $p$-adic string amplitudes leads to certain string
amplitudes (which are rational functions), that we termed the
\textit{Denef-Loeser open string amplitudes}. These Denef-Loeser amplitudes
for $N=4$, $5$ points were computed in Sections \ref{Section_D_L_N_4},
\ref{Section_D_L_N_5}, for more details see \cite{Zun-B-C-JHEP}. On the other
hand for $N=4$, $5$, in \cite{Zun-B-C-JHEP}, we computed the limit
$p\rightarrow1$ of the effective field theory of $p$-adic amplitudes, i.e. the
Gerasimov and Shatashvili Lagrangian (\ref{linearaction}) involving a
logarithmic potential. By computing the interacting generating functional at
the tree level, we verified that the corresponding amplitudes coincide exactly
with the Denef-Loeser amplitudes. Based on this fact we have formulated a
conjecture, which was already stated in the introduction of this survey.

In what follows we review the evidence of the mentioned conjecture in the
cases $N=4$, $5$ following \cite{Zun-B-C-JHEP}. First of all. we review some
basic results from \cite{Gerasimov:2000zp} starting with the effective action
which is a field theory whose perturbative analysis leads to the $p$-adic
scattering amplitudes \cite{Brekke:1988dg}, it can be written as
\begin{equation}
S(\phi)={\frac{1}{g^{2}}}{\frac{p^{2}}{p-1}}\int d^{D}x\bigg(-{\frac{1}{2}%
}\phi p^{-{\frac{1}{2}}\Delta}\phi+{\frac{1}{p+1}}\phi^{p+1}\bigg),
\label{goodaction}%
\end{equation}
where $\Delta$ is the Laplacian on $M$ and $g$ is the coupling constant. The
corresponding Euler-Lagrange equation is
\begin{equation}
p^{-{\frac{1}{2}}\Delta}\phi=\phi^{p}. \label{eomgeneral}%
\end{equation}
In the source space (and consequently in the amplitudes) $p$ is a prime
number, however, in the target space (and consequently in the Lagrangian and
in the equation of motion) $p$ is a real parameter. Thus, one can formally
proceed to approach $p$ to one and perform a Taylor expansion at $(p-1)$ of
the the expression $\exp(-{\frac{1}{2}}\Delta\log p)$ and $\exp(p\log\phi)$.
Then the resulting equation of motion is given by
\begin{equation}
\Delta\phi=-2\phi\log\phi. \label{eomlineal}%
\end{equation}
This is the motion equation of the Gerasimov-Shatashvili action
\begin{equation}
S(\phi)=\int d^{D}x\bigg((\partial\phi)^{2}-V(\phi)\bigg),
\label{linearaction}%
\end{equation}
where $(\partial\phi)^{2}=\eta^{\mu\nu}\partial_{\mu}\phi\cdot\partial_{\nu
}\phi$ and $V(\phi)$ is the potential
\[
V(\phi)=\phi^{2}\log{\frac{\phi^{2}}{e}}.
\]
We consider that action (\ref{linearaction}) is the limit $p$ tends to one of
the effective action (\ref{goodaction}).

The action corresponding to the free theory with a source is written as
\[
S_{0}(\phi)=\int d^{D}x\big[(\partial\phi)^{2}+\phi^{2}(x)+J(x)\phi(x)\big].
\]
Now, the action (\ref{linearaction}) can be conveniently rewritten as
\begin{equation}
S(\phi)=\int d^{D}x\big[(\partial\phi)^{2}+m^{2}-U(\phi)\big], \label{actionU}%
\end{equation}
where $U(\phi)=2\phi^{2}\log\phi$. Then the potential $U(\phi)$ can be
expanded in Taylor series around the origin in the form
\begin{equation}
U(\phi)=A\phi^{2}+B\phi^{3}+C\phi^{4}+D\phi^{5}+\cdots, \label{potencialU}%
\end{equation}
where $A,B,C$ and $D$ are constants and they are real numbers.

Now, we review briefly the form of obtaining the four-point and five-point
amplitudes from the Lagrangian (\ref{linearaction}) \cite{Gerasimov:2000zp}.
In quantum field theory $N$-point correlation functions of $N$ local operators
$\widehat{\phi}$ in $N$ different points $x_{1},x_{2},\dots,x_{N}$ of $M$, can
be written as
\[
\langle T(\widehat{\phi}(x_{1})\widehat{\phi}(x_{2})\cdots\widehat{\phi}%
(x_{N}))\rangle={\frac{(-i\hbar)^{N}}{\mathcal{Z}[J]}}{\frac{\delta
^{n}\mathcal{Z}[J]}{\delta J(x_{1})\delta J(x_{2})\cdots\delta J(x_{N})}%
}\bigg|_{J=0},
\]
where $\mathcal{Z}[J]$ is the generating functional constructed using
interacting Lagrangian (\ref{actionU}). The functional can be computed as
\begin{multline*}
\mathcal{Z}[J]=\exp\bigg\{-{\frac{iB}{\hbar}}\int d^{D}x\bigg(-i\hbar
{\frac{\delta}{\delta J(x)}}\bigg)^{3}\\
-{\frac{iC}{\hbar}}\int d^{D}x\bigg(-i\hbar{\frac{\delta}{\delta J(x)}%
}\bigg)^{4}-{\frac{iD}{\hbar}}\int d^{D}x\bigg(-i\hbar{\frac{\delta}{\delta
J(x)}}\bigg)^{5}+\cdots\bigg\}\mathcal{Z}_{0}[J],
\end{multline*}
where the generating functional of the correlation function for the free
theory with sources is given by
\[
\mathcal{Z}_{0}[J]=\mathcal{N}[\det(\Delta-1)]^{-1/2}\exp\bigg\{-{\frac
{i}{4\hbar}}\int d^{D}x\int d^{D}x^{\prime}J(x)G_{F}(x-x^{\prime})J(x^{\prime
})\bigg\}.
\]
Moreover $G_{F}(x-x^{\prime})$ is the Green-Feynman function of the
differential operator $(\Delta-1)$.

For the computation of the interacting $4$-point amplitudes can be obtained
through the generating functional
\[
\mathcal{Z}[J]=\cdots-iC\hbar^{3}\int d^{D}x\bigg({\frac{\delta}{\delta J(x)}%
}\bigg)^{4}\mathcal{Z}_{0}[J]+\cdots.
\]
In terms of $\mathcal{Z}[J]$, $4$-point amplitudes implies the computation of
\begin{gather}
{\frac{\delta^{4}\mathcal{Z}[J]}{\delta J(x_{1})\delta J(x_{2})\delta
J(x_{3})\delta J(x_{4})}}\bigg|_{J=0}\label{vertex}\\
=-4!iC\hbar^{3}\int d^{D}x\bigg[-{\frac{i}{2\hbar}}G_{F}(x-x_{1}%
)\bigg]\bigg[-{\frac{i}{2\hbar}}G_{F}(x-x_{2})\bigg]\times\nonumber\\
\bigg[-{\frac{i}{2\hbar}}G_{F}(x-x_{3})\bigg]\bigg[-{\frac{i}{2\hbar}}%
G_{F}(x-x_{4})\bigg]\nonumber\\
=-{\frac{3iC}{2\hbar}}\int d^{D}x\ G_{F}(x-x_{1})\ G_{F}(x-x_{2}%
)\ G_{F}(x-x_{3})\ G_{F}(x-x_{4}).\nonumber
\end{gather}
Here $G_{F}(x-y)$ is the Green-Feynman propagator. This amplitude corresponds
to the Feynman diagram with only one vertex and four external legs. following
the notation of \cite{Brekke:1988dg}, we write for it the symbol $\overline
{K}_{4}$.

The contribution to the 4-point amplitudes of $B\phi^{3}$ in (\ref{potencialU}%
) has a contribution at the second order of the expansion. The corresponding
Feynman diagrams have two vertices at points $x$ and $y$ with two external
legs, which are connected by a propagator $G_{F}(x-y)$. Thus we have
\begin{equation}
\mathcal{Z}[J]=\cdots+{\frac{B^{2}\hbar^{4}}{2}}\int d^{D}x\int d^{D}%
y\ \bigg({\frac{\delta}{\delta J(x)}}\bigg)^{3}\bigg({\frac{\delta}{\delta
J(y)}}\bigg)^{3}\mathcal{Z}_{0}[J]+\cdots. \label{cubiccubic}%
\end{equation}
Moreover the contribution to the 4-point amplitudes from the term $C\phi^{3}$
arises to second order. Thus it yields
\[
{\frac{\delta^{4}\mathcal{Z}[J]}{\delta J(x_{1})\delta J(x_{2})\delta
J(x_{3})\delta J(x_{4})}}\bigg|_{J=0}=18B^{2}\hbar^{4}\int d^{D}x\int
d^{D}y\ \bigg[-{\frac{i}{2\hbar}}G_{F}(x-y)\bigg]
\]%
\[
\times\bigg\{\bigg[-{\frac{i}{2\hbar}}G_{F}(x-x_{4})\bigg]\bigg[-{\frac
{i}{2\hbar}}G_{F}(x-x_{3})\bigg]\bigg[-{\frac{i}{2\hbar}}G_{F}(y-x_{2}%
)\bigg]\bigg[-{\frac{i}{2\hbar}}G_{F}(y-x_{1})\bigg]
\]%
\[
+\bigg[-{\frac{i}{2\hbar}}G_{F}(x-x_{4})\bigg]\bigg[-{\frac{i}{2\hbar}}%
G_{F}(y-x_{3})\bigg]\bigg[-{\frac{i}{2\hbar}}G_{F}(x-x_{2})\bigg]\bigg[-{\frac
{i}{2\hbar}}G_{F}(y-x_{1})\bigg]
\]%
\[
+\bigg[-{\frac{i}{2\hbar}}G_{F}(x-x_{4})\bigg]\bigg[-{\frac{i}{2\hbar}}%
G_{F}(y-x_{3})\bigg]\bigg[-{\frac{i}{2\hbar}}G_{F}(y-x_{2})\bigg]\bigg[-{\frac
{i}{2\hbar}}G_{F}(x-x_{1})\bigg]
\]%
\[
+\bigg[-{\frac{i}{2\hbar}}G_{F}(y-x_{4})\bigg]\bigg[-{\frac{i}{2\hbar}}%
G_{F}(y-x_{3})\bigg]\bigg[-{\frac{i}{2\hbar}}G_{F}(x-x_{2})\bigg]\bigg[-{\frac
{i}{2\hbar}}G_{F}(x-x_{1})\bigg]
\]%
\[
+\bigg[-{\frac{i}{2\hbar}}G_{F}(y-x_{4})\bigg]\bigg[-{\frac{i}{2\hbar}}%
G_{F}(x-x_{3})\bigg]\bigg[-{\frac{i}{2\hbar}}G_{F}(y-x_{2})\bigg]\bigg[-{\frac
{i}{2\hbar}}G_{F}(x-x_{1})\bigg]
\]%
\begin{equation}
+\bigg[-{\frac{i}{2\hbar}}G_{F}(y-x_{4})\bigg]\bigg[-{\frac{i}{2\hbar}}%
G_{F}(x-x_{3})\bigg]\bigg[-{\frac{i}{2\hbar}}G_{F}(x-x_{2})\bigg]\bigg[-{\frac
{i}{2\hbar}}G_{F}(y-x_{1})\bigg]\bigg\}. \label{segunda}%
\end{equation}

The sum of expressions (\ref{vertex}) and (\ref{segunda}) constitutes the
total amplitude which contains the sum of the three channels $s$, $t$ and $u$.
They are expected to be the $p$-adic 4-point amplitudes arising in the limit
$p\rightarrow1$. Thus in the Fourier space it is expected to be written
schematically as
\[
A_{GS}^{\left(  4\right)  }=K_{4}^{GS}+\sum_{i<j}x_{ij}^{GS},
\]
where $x_{ij}^{GS}$ is the propagator $x_{ij}^{GS}={\frac{1}{\boldsymbol{k}%
_{i}\boldsymbol{k}_{j}+1}}$ in momentum space. Then $A_{GS}^{\left(  4\right)
}$ agrees with $\boldsymbol{A}_{top}^{\left(  4\right)  }(\boldsymbol{k}%
)=-1+\sum_{i<j}x_{ij}^{GS}$ up to the constant $K_{4}^{GS}$, see
(\ref{Denef_Loeser_4}).

In the case of amplitudes of $5$-points, the contribution of the term
$D\phi^{5}$ in the Lagrangian is written as
\[
\mathcal{Z}[J]=\cdots-{D\hbar^{4}}\int d^{D}x\bigg({\frac{\delta}{\delta
J(x)}}\bigg)^{5}\mathcal{Z}_{0}[J]+\cdots.
\]
Thus the corresponding 5-point amplitude is
\[
{\frac{\delta^{5}\mathcal{Z}[J]}{\delta J(x_{1})\delta J(x_{2})\delta
J(x_{3})\delta J(x_{4})\delta J(x_{5})}}\bigg|_{J=0}=-5!D\hbar^{4}\int
d^{D}x\bigg[-{\frac{i}{2\hbar}}G_{F}(x-x_{1})\bigg]
\]%
\begin{equation}
\times\bigg[-{\frac{i}{2\hbar}}G_{F}(x-x_{2})\bigg]\bigg[-{\frac{i}{2\hbar}%
}G_{F}(x-x_{3})\bigg]\bigg[-{\frac{i}{2\hbar}}G_{F}(x-x_{4}%
)\bigg]\bigg[-{\frac{i}{2\hbar}}G_{F}(x-x_{5})\bigg]. \label{unouno}%
\end{equation}
The contribution will be encoded in the symbol $K_{5}^{GS}$.

Another contribution come from the term $B\phi^{3}\times C\phi^{4}$. In the
Fourier space the corresponding diagrams have precisely $2$-vertices, $5$
external legs attached to these vertices and one internal leg between the two
vertices (see, \cite[Section 3]{Brekke:1988dg}).

This latter contribution to generating functional is written as
\begin{equation}
\mathcal{Z}[J]=\cdots-iBC\hbar^{5}\int d^{D}x\int d^{D}y\ \bigg({\frac{\delta
}{\delta J(x)}}\bigg)^{3}\bigg({\frac{\delta}{\delta J(y)}}\bigg)^{4}%
\mathcal{Z}_{0}[J]+\cdots. \label{cubicquartic}%
\end{equation}
The $5$-point amplitude yields
\[
{\frac{\delta^{5}\mathcal{Z}[J]}{\delta J(x_{1})\delta J(x_{2})\delta
J(x_{3})\delta J(x_{4})\delta J(x_{5})}}\bigg|_{J=0}%
\]%
\[
={-iBC(12)^{2}\hbar^{5}}\int d^{D}x\int d^{D}y\ \bigg[-{\frac{i}{2\hbar}}%
G_{F}(x-y)\bigg]\bigg\{\bigg[-{\frac{i}{2\hbar}}G_{F}(x-x_{5}%
)\bigg]\bigg[-{\frac{i}{2\hbar}}G_{F}(x-x_{4})\bigg]
\]%
\[
\times\bigg[-{\frac{i}{2\hbar}}G_{F}(y-x_{3})\bigg]\bigg[-{\frac{i}{2\hbar}%
}G_{F}(y-x_{2})\bigg]\bigg[-{\frac{i}{2\hbar}}G_{F}(y-x_{1})\bigg]+\cdots
\]%
\[
+\bigg[-{\frac{i}{2\hbar}}G_{F}(y-x_{5})\bigg]\bigg[-{\frac{i}{2\hbar}}%
G_{F}(x-x_{4})\bigg]\bigg[-{\frac{i}{2\hbar}}G_{F}(x-x_{3})\bigg]
\]%
\begin{equation}
\times\bigg[-{\frac{i}{2\hbar}}G_{F}(y-x_{2})\bigg]\bigg[-{\frac{i}{2\hbar}%
}G_{F}(y-x_{1})\bigg]+\cdots\bigg\}. \label{fivepointstwov}%
\end{equation}

Moreover in the computation of the $5$-point amplitudes, the last contribution
comes from the term
\begin{equation}
\mathcal{Z}[J]=\cdots+{\frac{B^{3}\hbar^{6}}{3!}}\int d^{D}x\int d^{D}y\int
d^{D}z\ \bigg({\frac{\delta}{\delta J(x)}}\bigg)^{3}\cdot\bigg({\frac{\delta
}{\delta J(y)}}\bigg)^{3}\cdot\bigg({\frac{\delta}{\delta J(z)}}%
\bigg)^{3}\mathcal{Z}_{0}[J]+\cdots.\label{cubiccubiccubic}%
\end{equation}
The diagrams associated to this expansion are diagrams with three vertices.
Two of them have two external legs and the other leg is internal. The
remaining leg has one external line and two internal lines attached.

Then the total contribution is
\[
{\frac{\delta^{5}\mathcal{Z}[J]}{\delta J(x_{1})\delta J(x_{2})\delta
J(x_{3})\delta J(x_{4})\delta J(x_{5})}}\bigg|_{J=0}={\frac{B^{3}\hbar^{6}%
a}{3!}}\int d^{D}x\int d^{D}y\int d^{D}z\ \bigg[-{\frac{i}{2\hbar}}%
G_{F}(x-y)\bigg]
\]%
\[
\times\bigg[-{\frac{i}{2\hbar}}G_{F}(y-z)\bigg]\bigg\{\bigg[-{\frac{i}{2\hbar
}}G_{F}(y-x_{5})\bigg]\bigg[-{\frac{i}{2\hbar}}G_{F}(z-x_{4}%
)\bigg]\bigg[-{\frac{i}{2\hbar}}G_{F}(z-x_{3})\bigg]
\]%
\begin{equation}
\times\bigg[-{\frac{i}{2\hbar}}G_{F}(x-x_{2})\bigg]\bigg[-{\frac{i}{2\hbar}%
}G_{F}(x-x_{1})\bigg]+\cdots\bigg\}, \label{tres}%
\end{equation}
where $a$ is a constant.

Summarizing, the $5$-point amplitudes $A_{5}$ obtained from Lagrangian
(\ref{linearaction}) can be written schematically (in notation from
\cite{Brekke:1988dg}) by
\[
A_{GS}^{\left(  5\right)  }=K_{5}^{GS}+\sum_{K_{4}^{GS}i<j}x_{ij}^{GS}%
+\sum_{\substack{\left\{  i,j,k,l\right\}  \subset\left\{  1,2,3,4,5\right\}
\\i<j,\text{ }k<l}}x_{ij}^{GS}x_{kl}^{GS},
\]
where the three terms in the sum corresponds to the contributions of equations
(\ref{unouno}), (\ref{fivepointstwov}) and (\ref{tres}) respectively. Notice
that $A_{GS}^{\left(  5\right)  }$ agrees with
\[
\boldsymbol{A}_{top}^{\left(  5\right)  }(\boldsymbol{k})=2-%
{\textstyle\sum\limits_{i<j}}
x_{ij}^{GS}+%
{\displaystyle\sum\limits_{_{\substack{\left\{  i,j,k,l\right\}
\subset\left\{  1,2,3,4,5\right\}  \\i<j,\text{ }k<l}}}}
x_{ij}^{GS}x_{kl}^{GS}%
\]
up to the constants $K_{5}^{GS}$, $K_{4}^{GS}$, see (\ref{Denef_Loeser_5}).

We now give a precise formulation of Conjecture \ref{Conjecture B} announced
in the Introduction. By using the Feynman rules for the effective Lagragian
given in \cite{Brekke:1988dg},%
\[
A_{\mathbb{Q}_{p}}^{\left(  N\right)  }=K+%
{\textstyle\sum\nolimits_{l=2}^{N-3}}
K_{i_{1}...i_{l}}%
{\textstyle\sum\limits_{1\leq a<b\leq l}}
\left(  \frac{p-1}{p}\frac{p^{-k_{i_{a}}k_{i_{b}}-1}}{1-p^{-k_{i_{a}}k_{i_{b}%
}-1}}\right)  ,
\]
where the constants $K$, $K_{i_{1}...i_{l}}$ are rational functions in $p$
with rational coefficients. Furthermore, if $N\leq p+1$, the constant term $K$
is not zero. Formally, we have%
\[
A_{top}^{\left(  N\right)  }=\lim_{p\rightarrow1}A_{\mathbb{Q}_{p}}^{\left(
N\right)  }=K^{top}+%
{\textstyle\sum\nolimits_{l=2}^{N-3}}
K_{i_{1}...i_{l}}^{top}\text{ }%
{\textstyle\sum\limits_{1\leq a<b\leq l}}
\left(  \frac{1}{k_{i_{a}}k_{i_{b}}+1}\right)  .
\]
We denote by $A_{GS}^{\left(  N\right)  }$ the Feynman amplitude of Lagrangian
(\ref{linearaction}). We conjecture that%
\[
A_{GS}^{\left(  N\right)  }=K^{GS}+%
{\textstyle\sum\nolimits_{l=2}^{N-3}}
K_{i_{1}...i_{l}}^{GS}\text{ }%
{\textstyle\sum\limits_{1\leq a<b\leq l}}
x_{_{i_{a}i_{b}}}^{GS},
\]
where $x_{_{i_{a}i_{b}}}^{GS}=\frac{1}{k_{i_{a}}k_{i_{b}}+1}=\widehat{G_{F}%
}\left(  k_{i_{a}}-k_{i_{b}}\right)  $. Notice that we are not asserting the
existence of any relation between the constants $K^{top}$, $K_{i_{1}...i_{l}%
}^{top}$ and the constants $K^{GS}$, $K_{i_{1}...i_{l}}^{GS}$.

\section{\label{Section_6}$p$-Adic Open String Amplitudes coupled to a B-field
with Chan Paton Factors}

Strings propagating in a background with gauge fields was discussed many years
ago \cite{Abouelsaood:1986gd}. In particular, incorporating a Neveu-Schwarz
$B$ field in the target space leads to a noncommutative effective gauge theory
on the world-volume of D-branes \cite{Seiberg:1999vs}. In
\cite{Compean:2020B-field}, the $N$-point $p$-adic string amplitudes, with
Chan-Paton rules and a constant $B$-field (the Ghoshal-Kawano
amplitudes)\ were studied, these amplitudes were introduced in
\cite{Ghoshal:2004ay}. The study was done by attaching twisted local
multivariate zeta functions to the Ghoshal-Kawano amplitudes. In this section
we discuss the meromorphic continuation \ of the Ghoshal-Kawano amplitudes, we
also compute the $4$ and $5$ points amplitudes. Due to the need for a
particular symmetry, in this section, we take $p\equiv3\ \text{mod }4$.

\subsection{The $p$-adic sign function}

A $p$-adic sign function is a multiplicative character of $\mathbb{Q}%
_{p}^{\times}$ that takes values in $\left\{  \pm1\right\}  $. We set $\left[
\mathbb{Q}_{p}^{\times}\right]  ^{2}$ as the multiplicative subgroup of
squares in $\mathbb{Q}_{p}^{\times}$, i.e.
\[
\left[  \mathbb{Q}_{p}^{\times}\right]  ^{2}=\{a\in\mathbb{Q}_{p}%
;a=b^{2}\text{ for some }b\in\mathbb{Q}_{p}^{\times}\}.
\]
Let $\varepsilon\in\{1,\ldots,p-1\}$ such that $(\frac{\varepsilon}{p})=-1$,
where $(\frac{\cdot}{\cdot})$ is the \textit{Legendre symbol} (see for
instance the Appendix of \cite{Compean:2020B-field}). For $p\neq2$ we have
\[
\mathbb{Q}_{p}^{\times}/\left[  \mathbb{Q}_{p}^{\times}\right]  ^{2}%
=\{1,\varepsilon,p,\varepsilon p\},
\]
this means that any nonzero $p$-adic number can be written uniquely as
\[
x=\tau a^{2},\text{ with }a\in\mathbb{Q}_{p}^{\times}\text{ and }\tau
\in\mathbb{Q}_{p}^{\times}/\left[  \mathbb{Q}_{p}^{\times}\right]  ^{2}.
\]
For a fixed $\tau\in\{\varepsilon,p,\varepsilon p\}$, and $x\in\mathbb{Q}%
_{p}^{\times}$, we define the $p$-adic sign function
\begin{equation}
\mathrm{sgn}_{\tau}(x):=%
\begin{cases}
1 & \text{if}\ x=a^{2}-\tau b^{2}\ \text{for}\ a,b\in\mathbb{Q}_{p}\\
-1 & \text{otherwise.}%
\end{cases}
\label{signdef}%
\end{equation}
The following is the list of all the possible $p$-adic sign functions:%
\begin{equation}%
\begin{tabular}
[c]{|l|l|}\hline
$p\equiv1$ $\operatorname{mod}$ $4$ & $p\equiv3$ $\operatorname{mod}$
$4$\\\hline
$\mathrm{sgn}_{\varepsilon}(x)=\left(  -1\right)  ^{ord(x)}$ & $\mathrm{sgn}%
_{\varepsilon}(x)=\left(  -1\right)  ^{ord(x)}$\\\hline
$\mathrm{sgn}_{p}(x)=\left(  \frac{x_{0}}{p}\right)  $ & $\mathrm{sgn}%
_{p}(x)=\left(  -1\right)  ^{ord(x)}\left(  \frac{x_{0}}{p}\right)  $\\\hline
$\mathrm{sgn}_{\varepsilon p}(x)=\left(  -1\right)  ^{ord(x)}\left(
\frac{x_{0}}{p}\right)  $ & $\mathrm{sgn}_{\varepsilon p}(x)=\left(
\frac{x_{0}}{p}\right)  ,$\\\hline
\end{tabular}
\ \ \ \ \label{Table}%
\end{equation}
see \cite{Gubser:2018cha}. The function\ $\mathrm{sgn}_{\tau}$ is a
multiplicative character, which means that
\[
\mathrm{sgn}_{\tau}(xy)=\mathrm{sgn}_{\tau}(x)\mathrm{sgn}_{\tau}(y).
\]
Also $\mathrm{sgn}_{\tau}$ is a locally constant function in $\mathbb{Q}%
_{p}^{\times}$, which means that, $\mathrm{sgn}_{\tau}(x-y)=\mathrm{sgn}%
_{\tau}(x)$ if $|y|_{p}<|x|_{p}$.

We need the symmetry $\mathrm{sgn}_{\tau}(-y)=-\mathrm{sgn}_{\tau}(y)$ (i.e.
$\mathrm{sgn}_{\tau}(-1)=-1$), this requires $p\equiv3$ $\operatorname{mod}$
$4$ and $\tau\in\left\{  p,\varepsilon p\right\}  $. Finally, for any
$x\in\mathbb{Q}_{p}^{\times}$, we define the $p$-adic Heaviside step function
as
\[
H_{\tau}(x)=\frac{1}{2}(1+\mathrm{sgn}_{\tau}(x))=\left\{
\begin{array}
[c]{ll}%
1 & \text{if \textrm{sgn}$_{\tau}(x)=1$}\\
& \\
0 & \text{if \textrm{sgn}$_{\tau}(x)=-1.$}%
\end{array}
\right.
\]

\subsection{A class of twisted multivariate local zeta functions}

Let
\[
\boldsymbol{f}:=\left(  f_{1},\ldots,f_{m}\right)  \text{, \allowbreak
\ }\boldsymbol{\chi}:=\left(  \chi_{1},\ldots,\chi_{m}\right)
\]
be vectors of non-constant polynomials and multiplicative characters,
respectively. And $\boldsymbol{s}:=\left(  s_{1},\ldots,s_{m}\right)
\in\mathbb{C}^{m}\text{.}$ The twisted multivariate local zeta functions have
the form:
\begin{equation}
Z_{\Theta}\left(  \boldsymbol{s},\boldsymbol{\chi},\boldsymbol{f}\right)
=\int\limits_{\mathbb{Q}_{p}^{n}\smallsetminus\cup_{i=1}^{m}f^{-1}\left(
0\right)  }\Theta\left(  x\right)  \prod\limits_{i=1}^{m}\chi_{i}\left(
ac\left(  f_{i}(x)\right)  \right)  \left\vert f_{i}(x)\right\vert _{p}%
^{s_{i}}{\prod\limits_{i=1}^{n}}dx_{i}\text{, } \label{Zeta Function General}%
\end{equation}
with $\operatorname{Re}(s_{i})>0\ $\ for all$\ i$, $ac$ stands for the angular
component and $\Theta(x)$ is a test function. Integrals of type
(\ref{Zeta Function General}) are holomorphic functions in $\boldsymbol{s}$,
which admit meromorphic continuations as rational functions in the variables
$p^{-s_{1}},\ldots,p^{-s_{m}}$ to the whole $\mathbb{C}^{m}$,
\cite[Th\'{e}or\`{e}me 1.1.4.]{Loeser}, see also \cite{Igusa}. The case when
$\chi_{i}$ is the trivial character has been studied previously, see e.g.
\cite[Lemma 8.2.1]{Igusa}. The case when $\chi_{i}=\mathrm{sgn}_{\tau}$ is new.

By using Hironaka's resolution of singularities theorem \cite{H}, in
\cite{Compean:2020B-field} we show that $Z_{\Theta}\left(  \boldsymbol{s}%
,\boldsymbol{\chi},\boldsymbol{f}\right)  $ \ admits a meromorphic
continuation as a rational function in the variables $p^{-s_{1}}%
,\ldots,p^{-s_{m}}$. More precisely,%
\begin{equation}
Z_{\Theta}\left(  \boldsymbol{s},\boldsymbol{\chi},\boldsymbol{f}\right)
=\frac{L_{\Theta,\boldsymbol{\chi}}\left(  \boldsymbol{s}\right)  }%
{\prod\limits_{j\in\mathcal{T}}\left(  1-p^{-\sum_{i=1}^{m}N_{i,j}s_{j}-v_{j}%
}\right)  }, \label{Multizeta-special-case}%
\end{equation}
where $L_{\Theta,\boldsymbol{\chi}}\left(  \boldsymbol{s}\right)  $ is a
polynomial in the variables $p^{-s_{1}},\ldots,p^{-s_{m}}$, and the real parts
of its poles belong to the finite union of hyperplanes
\[
{\sum\limits_{i=1}^{m}}N_{i,j}\operatorname{Re}\left(  s_{i}\right)
+v_{j}=0,\quad\text{ for }j\in\mathcal{T}.
\]
This result is a variation of \cite[Th\'{e}or\`{e}me 1.1.4.]{Loeser}.

\subsection{The Ghoshal-Kawano local zeta function}

In \cite{Ghoshal:2004ay} Ghoshal and Kawano proposed the following amplitude
(for the $N$-point tree-level, $p$-adic open string amplitude, with Chan-Paton
rules in a constant $B$-field):
\begin{align}
A^{(N)}  &  \left(  \boldsymbol{k},\theta,\tau;x_{1},x_{N-1}\right)
:={\int\limits_{\mathbb{Q}_{p}^{N-3}\smallsetminus\Lambda}}\text{{ }}%
{\prod\limits_{i=2}^{N-2}}\left\vert x_{i}\right\vert _{p}^{\boldsymbol{k}%
_{1}\boldsymbol{k}_{i}}\left\vert 1-x_{i}\right\vert _{p}^{\boldsymbol{k}%
_{N-1}\boldsymbol{k}_{i}}H_{\tau}\left(  x_{i}\right)  H_{\tau}\left(
1-x_{i}\right) \nonumber\\
&  \times\text{ }{\prod\limits_{2\leq i<j\leq N-2}}\left\vert x_{i}%
-x_{j}\right\vert _{p}^{\boldsymbol{k}_{i}\boldsymbol{k}_{j}}H_{\tau}\left(
x_{i}-x_{j}\right)  \text{ }\nonumber\\
&  \times\exp\left\{  -\frac{\sqrt{-1}}{2}\left(  {\sum\limits_{1\leq i<j\leq
N-1}}(\boldsymbol{k}_{i}\theta\boldsymbol{k}_{j})\mathrm{sgn}_{\tau}%
(x_{i}-x_{j})\right)  \right\}  {\prod\limits_{i=2}^{N-2}}dx_{i}\text{,}
\label{G-K-Amplitudes}%
\end{align}
where $N\geq4$, $\boldsymbol{k}=\left(  \boldsymbol{k}_{1},\ldots
,\boldsymbol{k}_{N}\right)  $, $\boldsymbol{k}_{i}=\left(  k_{0,i}%
,\ldots,k_{l,i}\right)  $, $i=1,\ldots,N$, is the momentum vector of the
$i$-th tachyon (with Minkowski product $\boldsymbol{k}_{i}\boldsymbol{k}%
_{j}=-k_{0,i}k_{0,j}+k_{1,i}k_{1,j}+\cdots+k_{l,i}k_{l,j}$), $\theta$ is a
fixed antisymmetric bilinear form that is built with the inverse of the
$B$-field; ${\prod\nolimits_{i=2}^{N-2}}dx_{i}$ is the normalized Haar measure
of $\mathbb{Q}_{p}^{N-3}$,
\[
\Lambda:=\left\{  \left(  x_{2},\ldots,x_{N-2}\right)  \in\mathbb{Q}_{p}%
^{N-3};\text{{ }}{\prod\limits_{i=2}^{N-2}}x_{i}\left(  1-x_{i}\right)  \text{
}{\prod\limits_{2\leq i<j\leq N-2}}\left(  x_{i}-x_{j}\right)  \text{
=}0\right\}  ,
\]
and the momentum vectors obey
\begin{equation}
\sum_{i=1}^{N}\boldsymbol{k}_{i}=\boldsymbol{0}\text{, \ \ \ \ }%
\boldsymbol{k}_{i}\boldsymbol{k}_{i}=2\text{ \ for }i=1,\ldots,N.
\label{momentumconservation}%
\end{equation}
In order to preserve the symmetry under the exchange of external momentum
vectors, we require that $\mathrm{sgn}_{\tau}(x_{i}-x_{j})=-\mathrm{sgn}%
_{\tau}(x_{j}-x_{i})$, or equivalently $\mathrm{sgn}_{\tau}(-1)=-1$. Then for
the rest of this section we assume that $\tau\in\left\{  p,\varepsilon
p\right\}  $.

To simplify the notation, we introduce the variables $s_{ij}\in\mathbb{C}$,
and $\tilde{s}_{ij}\in\mathbb{R}$ for $1\leq i<j\leq N-1$. We also set
\begin{multline*}
F(\boldsymbol{x},\boldsymbol{s},\tau):=\prod_{i=2}^{N-2}|x_{i}|_{p}^{s_{1i}%
}|1-x_{i}|_{p}^{s_{(N-1)i}}H_{\tau}(x_{i})H_{\tau}(1-x_{i})\\
\times\prod_{2\leq i<j\leq N-2}|x_{i}-x_{j}|_{p}^{s_{ij}}H_{\tau}(x_{i}%
-x_{j}),
\end{multline*}
and
\begin{equation}
E(\boldsymbol{x},\widetilde{\boldsymbol{s}},\tau;x_{1},x_{N-1}):=\exp\left\{
\frac{-\sqrt{-1}}{2}\left(  {\sum\limits_{1\leq i<j\leq N-1}}\widetilde
{s}_{ij}\mathrm{sgn}_{\tau}(x_{1}-x_{j})\right)  \right\}  . \label{Formula_E}%
\end{equation}
Now, we define the Ghoshal-Kawano local zeta function as
\begin{equation}
Z^{\left(  N\right)  }(\boldsymbol{s},\widetilde{\boldsymbol{s}},\tau
;x_{1},x_{N-1})={\int\limits_{\mathbb{Q}_{p}^{N-3}\smallsetminus\Lambda}%
}F(\boldsymbol{x},\boldsymbol{s},\tau)E(\boldsymbol{x},\widetilde
{\boldsymbol{s}},\tau;x_{1},x_{N-1}){\prod\limits_{i=2}^{N-2}}dx_{i}.
\label{Zeta_G_K}%
\end{equation}
For the sake of simplicity, we use $\mathbb{Q}_{p}^{N-3}$ as domain of
integration in (\ref{Zeta_G_K}) from now on. We now consider the convergence
of the amplitudes (\ref{G-K-Amplitudes}). By using that $\left\vert
E(\boldsymbol{x},\widetilde{\boldsymbol{s}},\tau;x_{1},x_{N-1})\right\vert
=1$, $\left\vert H_{\tau}(x_{i})\right\vert \leq1$, $\left\vert H_{\tau
}(1-x_{i})\right\vert \leq1$, for any $i$, and that $\left\vert H_{\tau}%
(x_{i}-x_{j})\right\vert \allowbreak\leq1$, for any $i$, $j$, we have%
\begin{multline*}
\left\vert Z^{\left(  N\right)  }(\boldsymbol{s},\widetilde{\boldsymbol{s}%
},\tau;x_{1},x_{N-1})\right\vert \\
\leq{\int\limits_{\mathbb{Q}_{p}^{N-3}}}\text{{ }}{\prod\limits_{i=2}^{N-2}%
}\left\vert x_{i}\right\vert _{p}^{\operatorname{Re}(s_{1i})}\left\vert
1-x_{i}\right\vert _{p}^{\operatorname{Re}\left(  s_{(N-1)i}\right)  }\text{
}{\prod\limits_{2\leq i<j\leq N-2}}\left\vert x_{i}-x_{j}\right\vert
_{p}^{\operatorname{Re}\left(  s_{ij}\right)  }{\prod\limits_{i=2}^{N-2}%
}dx_{i}\\
=Z^{(N)}\left(  \operatorname{Re}(\boldsymbol{s})\right)  ,
\end{multline*}
where $Z^{(N)}\left(  \boldsymbol{s}\right)  $ is the Koba-Nielsen string
amplitude studied in \cite{Zun-B-C-LMP}, \cite{Bocardo:2020mk}. By applying
the results of \cite{Zun-B-C-LMP}, \cite{Bocardo:2020mk}, integral
$Z^{(N)}\left(  \operatorname{Re}(\boldsymbol{s})\right)  $ is holomorphic in
an open set $\mathcal{K}\subset$ $\mathbb{C}^{D}$, therefore
\[
Z^{\left(  N\right)  }(\boldsymbol{s},\widetilde{\boldsymbol{s}},\tau)\text{
is holomorphic in }\boldsymbol{s\in}\mathcal{K}\text{ for any }\widetilde
{\boldsymbol{s}}\text{, }\tau\text{, }x_{1}\text{, }x_{N-1}\text{.}%
\]
It is important to notice that if any of the integration variables is in
$\mathbb{Q}_{p}\smallsetminus\mathbb{Z}_{p}$, then due to the local constancy
of $\mathrm{sgn}_{\tau}$, the factor $H_{\tau}(x)H_{\tau}(-x)$ appears in
$F(\boldsymbol{x},\boldsymbol{s},\tau)$, and $H_{\tau}(x)H_{\tau}(-x)=0$. For
this reason, we redefine the Ghoshal-Kawano local zeta function as
\begin{equation}
Z^{\left(  N\right)  }(\boldsymbol{s},\widetilde{\boldsymbol{s}},\tau
;x_{1},x_{N-1}):={\int\limits_{\mathbb{Z}_{p}^{N-3}}}F(\boldsymbol{x}%
,\boldsymbol{s},\tau)E(\boldsymbol{x},\widetilde{\boldsymbol{s}},\tau
;x_{1},x_{N-1}){\prod\limits_{i=2}^{N-2}}dx_{i}.
\label{Ghoshal-Kawano-zeta-funct}%
\end{equation}

\subsection{Ghoshal-Kawano Amplitude as a local zeta function}

For $\tilde{s}\in\mathbb{R}$ we have
\begin{equation}
\exp\left\{  \frac{-\sqrt{-1}\widetilde{s}}{2}\mathrm{sgn}_{\tau}(x)\right\}
=\cos\left(  \frac{\widetilde{s}}{2}\right)  -\sqrt{-1}\text{ }\mathrm{sgn}%
_{\tau}(x)\sin\left(  \frac{\widetilde{s}}{2}\right)  . \label{Formula_1}%
\end{equation}
By using this identity and the convention $x_{1}=0$, $x_{N-1}=1$, we have
\begin{equation}
E(\boldsymbol{x},\widetilde{\boldsymbol{s}},\tau)={\sum\limits_{I,J,K}%
}E_{I,J,K}(\widetilde{\boldsymbol{s}})\text{ }{\prod\limits_{j\in I}%
}\mathrm{sgn}_{\tau}(x_{j})\text{\ }{\prod\limits_{j\in J}}\mathrm{sgn}_{\tau
}(1-x_{j}){\prod\limits_{i,j\in K}}\mathrm{sgn}_{\tau}(x_{i}-x_{j}).
\label{Formula_2}%
\end{equation}
In a similar way, we obtain that%
\begin{align}
&  \prod_{i=2}^{N-2}H_{\tau}(x_{i})H_{\tau}(1-x_{i})\prod_{2\leq i<j\leq
N-2}H_{\tau}(x_{i}-x_{j})\nonumber\\
&  ={\sum\limits_{I,J,K}}e_{I,J,K}{\prod\limits_{j\in I}}\mathrm{sgn}_{\tau
}(x_{j})\ {\prod\limits_{j\in J}}\mathrm{sgn}_{\tau}(1-x_{j})\text{{ }}%
{\prod\limits_{i,j\in K}}\mathrm{sgn}_{\tau}(x_{i}-x_{j}), \label{Formula_3}%
\end{align}
By using formulas (\ref{Formula_2})-(\ref{Formula_3}), and assuming $x_{1}=0$,
$x_{N-1}=1$, $x_{N}=\infty$, we can write the Ghoshal-Kawano zeta function as
a finite sum of integrals of type%
\begin{multline*}
C(\widetilde{\boldsymbol{s}})%
{\textstyle\int\limits_{\mathbb{Z}_{p}^{N-3}}}
\prod_{i=2}^{N-2}|x_{i}|_{p}^{s_{1i}}|1-x_{i}|_{p}^{s_{(N-1)i}}\prod_{2\leq
i<j\leq N-2}|x_{i}-x_{j}|_{p}^{s_{ij}}%
{\textstyle\prod\limits_{j\in I}}
\chi_{\tau}(x_{j})%
{\textstyle\prod\limits_{j\in J}}
\chi_{\tau}(1-x_{j})\\
\times%
{\textstyle\prod\limits_{i,j\in K}}
\chi_{\tau}(x_{i}-x_{j})%
{\displaystyle\prod\limits_{i=2}^{N-2}}
dx_{i},
\end{multline*}
where $C(\widetilde{\boldsymbol{s}})$ is an $\mathbb{R}$-analytic function and
$\chi_{\tau}$ is a trivial character or $\mathrm{sgn}_{\tau}$. This expression
is a local zeta functions of the type (\ref{Zeta Function General}), which
implies that amplitudes (\ref{G-K-Amplitudes}) are equal to a finite sum of
local zeta functions and consequently they admit a meromorphic continuation in
the kinematic parameters. Finally, we point out that the meromorphic
continuation of the Ghoshal-Kawano local zeta function
(\ref{Ghoshal-Kawano-zeta-funct}) is also valid without taking the
normalization $x_{1}=0$, $x_{N-1}=1$, $x_{N}=\infty$, see section V-C of
\cite{Compean:2020B-field}.

\subsection{Explicit computation of $Z^{\left(  4\right)  }(\boldsymbol{s}%
,\widetilde{\boldsymbol{s}},\tau,)$}

We first notice that the exponential factor $E^{(4)}(\tilde{s},\tau)$ can be
taken outside the integral, i.e. the four point Ghoshal-Kawano zeta function
is
\begin{multline*}
Z^{\left(  4\right)  }(\boldsymbol{s},\widetilde{\boldsymbol{s}},\tau
)=\exp\left\{  \frac{i}{2}\left(  \widetilde{s}_{13}+\widetilde{s}%
_{12}+\widetilde{s}_{23}\right)  \right\} \\
\times%
{\textstyle\int\limits_{\mathbb{Z}_{p}}}
\left\vert x_{2}\right\vert _{p}^{s_{12}}\left\vert 1-x_{2}\right\vert
_{p}^{s_{32}}H_{\tau}\left(  x_{2}\right)  H_{\tau}\left(  1-x_{2}\right)
dx_{2}.
\end{multline*}
The computation of this integral is based on the calculation of the following
integrals: assume that $S\subset\mathbb{Z}_{p}\smallsetminus\left\{
0\right\}  $ satisfies $-S=S$, then%
\[%
{\textstyle\int\limits_{S}}
\left\vert x_{2}\right\vert _{p}^{s_{12}}\mathrm{sgn}_{\tau}(x_{2}%
)dx_{2}=0\text{, for }\tau\in\left\{  p,\varepsilon p\right\}  \text{,}%
\]
and\textbf{ }
\begin{align*}
I(\boldsymbol{s},\tau)  &  =%
{\textstyle\int\limits_{\mathbb{Z}_{p}}}
\left\vert x_{2}\right\vert _{p}^{s_{12}}\left\vert 1-x_{2}\right\vert
_{p}^{s_{32}}H_{\tau}\left(  x_{2}\right)  H_{\tau}\left(  1-x_{2}\right)
dx_{2}\\
&  =\frac{p-3}{4p}+\frac{p^{-1-s_{12}}\left(  1-p^{-1}\right)  }{2\left(
1-p^{-1-s_{12}}\right)  }+\frac{p^{-1-s_{32}}\left(  1-p^{-1}\right)
}{2\left(  1-p^{-1-s_{32}}\right)  }.
\end{align*}
For further details the reader may consult \cite{Compean:2020B-field}. The
explicit expression for the $4$-point amplitude, also reported in
\cite{Ghoshal:2004ay}, is
\begin{multline*}
Z^{\left(  4\right)  }(\boldsymbol{s},\widetilde{\boldsymbol{s}},\tau
)=\exp\left\{  \frac{i}{2}\left(  \widetilde{s}_{13}+\widetilde{s}%
_{12}+\widetilde{s}_{23}\right)  \right\} \\
\times\left(  \frac{p-3}{4p}+\frac{p^{-1-s_{12}}\left(  1-p^{-1}\right)
}{2\left(  1-p^{-1-s_{12}}\right)  }+\frac{p^{-1-s_{32}}\left(  1-p^{-1}%
\right)  }{2\left(  1-p^{-1-s_{32}}\right)  }\right)
\end{multline*}
which is holomorphic in
\[
\operatorname{Re}(s_{12})>-1\text{ and }\operatorname{Re}(s_{32})>-1.
\]

\subsection{Explicit computation of $Z^{(5)}(\boldsymbol{s},\widetilde
{\boldsymbol{s}},\tau)$}

After some simple considerations involving sign functions,
\[
Z^{(5)}(\boldsymbol{s},\widetilde{\boldsymbol{s}},\tau)=E^{(5)}(\widetilde
{\boldsymbol{s}})L(\boldsymbol{s},\tau):=E^{(5)}(\widetilde{\boldsymbol{s}%
})\int_{\mathbb{Z}_{p}^{2}}F^{(5)}(x_{2},x_{3},\boldsymbol{s},\tau
)dx_{2}dx_{3}.
\]
where
\begin{align*}
F^{(5)}(x_{2},x_{3},\boldsymbol{s},\tau)  &  =|x_{2}|_{p}^{s_{12}}|x_{3}%
|_{p}^{s_{13}}|1-x_{2}|_{p}^{s_{42}}|1-x_{3}|_{p}^{s_{43}}|x_{2}-x_{3}%
|_{p}^{s_{23}}\\
&  \times H_{\tau}(x_{2})H_{\tau}(x_{3})H_{\tau}(1-x_{2})H_{\tau}%
(1-x_{3})H_{\tau}(x_{2}-x_{3}).
\end{align*}
and
\[
E^{(5)}(\widetilde{\boldsymbol{s}})=\exp\left\{  \frac{\sqrt{-1}}{2}\left(
\widetilde{s}_{14}+\widetilde{s}_{12}+\widetilde{s}_{13}+\widetilde{s}%
_{24}+\tilde{s}_{34}+\tilde{s}_{32}\right)  \right\}  ,
\]
see \cite{Compean:2020B-field} for further details. To compute the integral
$L(\boldsymbol{s},\tau)$, we subdivide the domain of integration as
\[
\mathbb{Z}_{p}^{2}=\bigsqcup_{i,j=0}^{p-1}\left(  i+p\mathbb{Z}_{p}\right)
\times\left(  j+p\mathbb{Z}_{p}\right)  ,
\]
then $L(\boldsymbol{s},\tau)$ becomes
\[
L(\boldsymbol{s},\tau)=\sum_{i,j=0}^{p-1}L_{ij}(\boldsymbol{s},\tau
)=:\sum_{i,j=0}^{p-1}\int_{i+p\mathbb{Z}_{p}\times j+p\mathbb{Z}_{p}}%
F^{(5)}(x_{2},x_{3},\boldsymbol{s},\tau)dx_{2}dx_{3}.
\]
The computation of the integrals $L_{ij}(\boldsymbol{s},\tau)$ is a simple but
technical calculation, see \cite{Compean:2020B-field} for further details. The
explicit expression for the $5$-point Ghoshal-Kawano zeta function is
\begin{gather*}
Z^{(5)}(\boldsymbol{s},\widetilde{\boldsymbol{s}})=E^{(5)}(\widetilde
{\boldsymbol{s}})\left\{  \frac{(p-3)(p-7)}{32p^{2}}+\left(  \frac{p-3}%
{8p}\right)  \left[  \frac{p^{-1-s_{23}}(1-p^{-1})}{(1-p^{-1-s_{23}})}\right.
\right. \\
\left.  +\frac{p^{-1-s_{13}}(1-p^{-1})}{(1-p^{-1-s_{13}})}+\frac{p^{-1-s_{42}%
}(1-p^{-1})}{(1-p^{-1-s_{42}})}\right]  +\frac{\left(  1-p^{-1}\right)  ^{2}%
}{4}\frac{p^{-2-s_{13}-s_{42}}}{\left(  1-p^{-1-s_{13}}\right)  \left(
1-p^{-1-s_{42}}\right)  }\\
+\frac{1}{4}\frac{p^{-s_{12}-s_{13}-s_{23}-2}\left(  1-p^{-1}\right)
}{1-p^{-s_{12}-s_{13}-s_{23}-2}}\left[  \frac{1}{2}-\frac{3}{2p}%
+\frac{p^{-1-s_{23}}\left(  1-p^{-1}\right)  }{1-p^{-1-s_{23}}}+\frac
{p^{-1-s_{13}}\left(  1-p^{-1}\right)  }{1-p^{-1-s_{13}}}\right] \\
\left.  +\frac{1}{4}\frac{p^{-s_{42}-s_{43}-s_{23}-2}\left(  1-p^{-1}\right)
}{1-p^{-s_{42}-s_{43}-s_{23}-2}}\left[  \frac{1}{2}-\frac{3}{2p}%
+\frac{p^{-1-s_{23}}\left(  1-p^{-1}\right)  }{1-p^{-1-s_{23}}}+\frac
{p^{-1-s_{42}}\left(  1-p^{-1}\right)  }{1-p^{-1-s_{42}}}\right]  \right\}  .
\end{gather*}
This function is holomorphic in {\normalsize
\begin{align*}
&  \mathrm{Re}(s_{13})>-1;\quad\mathrm{Re}(s_{23})>-1;\quad\mathrm{Re}%
(s_{42})>-1;\\
&  \mathrm{Re}(s_{12}+s_{13}+s_{23})>-2;\quad\mathrm{Re}(s_{42}+s_{43}%
+s_{23})>-2.
\end{align*}
}

\section{\label{Section_7}Resolution of singularities and Multivariate Igusa
zeta functions}

\subsection{Local fields}

We take $\mathbb{K}$ to be a non-discrete$\ $locally compact field of
characteristic zero. Then $\mathbb{K}$ is $\mathbb{R}$, $\mathbb{C}$, or a
finite extension of $\mathbb{Q}_{p}$, the field of $p$-adic numbers. If
$\mathbb{K}$ is $\mathbb{R}$ or $\mathbb{C}$, we say that $\mathbb{K}$ is an
$\mathbb{R}$\textit{-field}, otherwise we say that $\mathbb{K}$ is a
$p$\textit{-field}.

For $a\in\mathbb{K}$, we define the \textit{modulus} $\left\vert a\right\vert
_{\mathbb{K}}$ of $a$ by%
\[
\left\vert a\right\vert _{\mathbb{K}}=\left\{
\begin{array}
[c]{l}%
\text{the rate of change of the Haar measure in }(\mathbb{K},+)\text{ under
}x\rightarrow ax\text{ }\\
\text{for }a\neq0,\\
\\
0\ \text{ for }a=0\text{.}%
\end{array}
\right.
\]

\noindent\ is well-known that, if $\mathbb{K}$ is an $\mathbb{R}$-field, then
$\left\vert a\right\vert _{\mathbb{R}}=\left\vert a\right\vert $ and
$\left\vert a\right\vert _{\mathbb{C}}=\left\vert a\right\vert ^{2}$, where
$\left\vert \cdot\right\vert $ denotes the usual absolute value in
$\mathbb{R}$ or $\mathbb{C}$, and, if $\mathbb{K}$ is a $p$-field, then
$\left\vert \cdot\right\vert _{\mathbb{K}}$ is the normalized absolute value
in $\mathbb{K}$.

\subsection{$\mathbb{K}$-analytic manifolds an resolution of singularities}

We review the basic definitions of $\mathbb{K}$-analytic manifolds following
Igusa's book \cite{Igusa}.

Let $\mathbb{K}$ be a local field of characteristic zero, and let $V$ be a
non-empty open subset of $\mathbb{K}^{n}$ and let $f:V\rightarrow\mathbb{K}$
be a function. We say that $f$ is a $\mathbb{K}$-analytic function on $V$, if
for every point $a=(a_{1},\ldots,a_{n})\in V$, there exists an element
$f_{a}(x)$ of $\mathbb{K}\langle\langle x-a\rangle\rangle=\mathbb{K}%
\langle\langle x_{1}-a_{1},\ldots,x_{n}-a_{n}\rangle\rangle$ (the ring of
convergent power series around $a$) such that $f(x)=f_{a}(x)$ for any point
$x$ near to $a$. A map $g=(g_{1},\ldots,g_{m}):V\rightarrow\mathbb{K}^{m}$ is
called $\mathbb{K}$- analytic mapping on $V$ if each $g_{i}$ is an analytic
function on $V$. Let $X$ Hausdorff space and let $n$ be a fixed non-negative.
Let $U$ be a non-empty open subset of $X$, if $\phi_{U}:U\rightarrow\phi
_{U}(U)$ is an homeomorphism, where $\phi_{U}(U)$ is a non-empty open subset
of $\mathbb{K}^{n}$, then we say that the pair $(U,\phi_{U}(U))$ is a
\textit{chart}. For a variable point $x\in U$, $\phi(x)=(x_{1},\ldots,x_{n})$
are called the local coordinates of $x$. A collection of charts $\{U,\phi
_{U}\}$ is called \textit{an atlas} of $X$ if the union of all open sets $V$
is $X$ and for every $U$, $U^{\prime}$ such that $U\cap U^{\prime}%
\neq\varnothing$ the map
\[
\phi_{U^{\prime}}\circ\phi_{U}^{-1}:\phi_{U}(U\cap U^{\prime})\rightarrow
\phi_{U^{\prime}}(U\cap U^{\prime})
\]
is $\mathbb{K}$- analytic. There is an equivalence relation over the set of
atlases on $X$. Two atlases are equivalent if their union is also an atlas.
Thus any equivalence class is called \textit{an }$n$\textit{-dimensional }%
$K$\textit{-analytic structure} on $X$. Hence we say that $X$ is a
$\mathbb{K}$\textit{-analytic manifold} and we write $dim(X)=n$.

Suppose that $X$ and $Y$ are two $\mathbb{K}$- analytical manifolds defined by
the atlases $\{(U,\phi_{U}(U))\}$ and $\{(V,\psi_{V}(V))\}$, respectively, and
$f:X\rightarrow Y$ a map. If for every $U$, $V$ such that $U\cap f^{-1}%
(V)\neq\varnothing$, the map
\[
\psi_{V}\circ f\circ\phi_{U}^{-1}:\phi_{U}(U\cap f^{-1}(V))\rightarrow
\mathbb{K}^{dim(Y)}%
\]
is $\mathbb{K}$- analytic, then $f$ is called a $K$-analytic map. If
$Y=\mathbb{K}$ we say that $f$ is a $\mathbb{K}$-analytic function on $X$.
This definition does not depend on the choice of atlases into the equivalence class.

Let $X$ be a $\mathbb{K}$-analytic manifold defined by an atlas $\{(U,\phi
_{U})\}$ and let $Y$ be a non-empty open set of $X$. If $U^{\prime}=U\cap
Y\neq\varnothing$ we put $\phi_{U^{\prime}}=\phi_{U^{\prime}}=\phi
_{U}|_{U^{\prime}}$. Then $\{(U^{\prime},\phi_{U^{\prime}})\}$ is an atlas for
$Y$. Hence, $Y$ is an open $\mathbb{K}$- analytic submanifold of $X$.
Furthermore, $dim(Y)=dim(X)$.

Now let $Y$ be a non-empty closed subset of a $\mathbb{K}$-analytic manifold
$X$ of dimension $n$. Suppose that $X$ is defined by an atlas $\{(U,\phi
_{U})\}$ with the following property: If $\phi_{U}(x)=(x_{1},\ldots,x_{n})$
and $U^{\prime}=Y\cap U\neq\varnothing$, there exist $F_{1},\ldots,F_{m}$
$\mathbb{K}$-analytic functions on $U$ with $0<m\leq n$ such that $U^{\prime}$
becomes the set of all $x$ in $U$ such that $F_{1}(x)=\ldots=F_{m}(x)=0$ and
\[
\det\left[  \frac{\partial F_{i}}{\partial x_{j}}\right]  _{\substack{1\leq
i\leq m\\1\leq j\leq m}}\left(  a\right)  \neq0,
\]
for every $a\in U^{\prime}$. By the implicit function theorem, the mapping
\[
x\mapsto(F_{1}(x),\ldots,F_{m}(x),x_{m+1},\ldots,x_{n})
\]
gives a $\mathbb{K}$-bianalytic map from a neighborhood of $a$ in $U$ to its
image in $K^{n}$. If we denote by $V$ the intersection of such neighborhood
and $Y$, and put $\psi_{V}(x)=(x_{m+1},\ldots,x_{n})$ for every $x\in V$. Then
$(V,\psi_{V})$ for all $V$ and for each $U$ as above, gives an atlas on $Y$.
Thus $Y$ becomes a $\mathbb{K}$\textit{-analytic closed submanifold} of $X$ of
dimension $n-m$.

Assume that $U$ and $V$ are non-empty open sets that contain a point $a$ of
$X$, and let $f$, $g$ be two $\mathbb{K}$- analytic functions respectively on
$U$, $V$ such that $f|_{W}=g|_{W}$ for some non-empty open set $W$ such that
$a\in W\subseteq U\cap V$. Then we say that $f$ and $g$ are equivalent. An
equivalence class is called a \textit{germ of analytic function} at $a$. The
set of such equivalence classes becomes a local ring $\mathcal{O}_{X,a}$ with
maximal ideal $m_{X,a}=\{f\in\mathcal{O}_{X,a}:f(a)=0\}$. In addition,
$\mathbb{K}\subseteq\mathcal{O}_{X,a}$.

\section{Multivariate Igusa zeta functions}

Let $\mathbb{K}$ be a local field. If $\mathbb{K}$ is a $p$-field, resp. an
$\mathbb{R}$-field, we denote by $\mathcal{D}(\mathbb{K}^{n})$ the
$\mathbb{C}$-vector space consisting of all $\mathbb{C}$-valued locally
constant functions, resp. all smooth functions, on $\mathbb{K}^{n}$, with
compact support. An element of $\mathcal{D}(\mathbb{K}^{n})$ is called a
\textit{test function}.

Let $f_{i}(\boldsymbol{x})\in\mathbb{K}\left[  x_{1},\ldots,x_{n}\right]  $ be
a non-constant polynomial for $i=1,\ldots,m$. We set $\boldsymbol{f}=\left(
f_{1},\ldots,f_{m}\right)  $ and $\boldsymbol{s}=\left(  s_{1},\ldots
,s_{m}\right)  \in\mathbb{C}^{m}$. Let $\Phi$ be a test function. The
multivariate local zeta function attached to $\left(  \boldsymbol{f}%
,\Phi\right)  $ is defined as
\begin{equation}
Z_{\Phi}\left(  \boldsymbol{f},\boldsymbol{s}\right)  =\int\limits_{\mathbb{K}%
^{n}\smallsetminus D_{\mathbb{K}}}\Phi\left(  x\right)  \prod\limits_{i=1}%
^{m}\left\vert f_{i}(x)\right\vert _{\mathbb{K}}^{s_{i}}%
{\displaystyle\prod\limits_{i=1}^{n}}
dx_{i}\text{, \qquad when }\operatorname{Re}(s_{i})>0\text{ for all }i\text{.}
\label{Zeta Function}%
\end{equation}
where $D_{\mathbb{K}}:=\cup_{i=1}^{m}f_{i}^{-1}(0)$ is the divisor attached to
$f_{i}(\boldsymbol{x})$, $i=1,\ldots,m$. In the multivariate case i.e. for
$m\geq1$, the local zeta functions over local fields of zero characteristic
were studied by Loeser \cite{Loeser}. In the case of zero characteristic, the
main tool to show the existence of a meromorphic continuation of the
multivariate local function is the Hironaka's resolution of singularities
theorem. By applying this theorem to the divisor $D_{\mathbb{K}}$, the
mutivariate local zeta function is reduced to the case of monomial integrals
\cite{Igusa-old}, \cite{Igusa}, and \cite{Loeser}. Currently, the methods used
by Igusa are not available in positive characteristic, so the problem of the
meromorphic continuations in this setting it is still an open problem.

\begin{theorem}
[Hironaka, \cite{H}]\label{thresolsing} Let $\mathbb{K}$ be a local field of
characteristic zero. There exists an embedded resolution $\sigma
:X\rightarrow\mathbb{K}^{n}$ of $D_{\mathbb{K}}$, that is,

\noindent(i) $X$ is an $n$-dimensional $\mathbb{K}$-analytic manifold,
$\sigma$ is a proper $\mathbb{K}$-analytic map which is a composition of a
finite number of blow-ups at closed submanifolds, and which is an isomorphism
outside of $\sigma^{-1}(D_{\mathbb{K}})$;

\noindent(ii) $\sigma^{-1}\left(  D_{\mathbb{K}}\right)  $ is a normal
crossings divisor, meaning that $\sigma^{-1}\left(  D_{\mathbb{K}}\right)
=\cup_{i\in T}E_{i}$, where the $E_{i}$\ are closed submanifolds of $X$ of
codimension one, each equipped with an $m$-tuple of nonnegative integers
$\left(  N_{f_{1},i},\ldots,N_{f_{m},i}\right)  $ and a positive integer
$v_{i}$, satisfying the following. At every point $b$ of $X$ there exist local
coordinates $\left(  y_{1},\ldots,y_{n}\right)  $ on $X$ around $b$ such that,
if $E_{1},\ldots,E_{r}$ are the $E_{i}$ containing $b$, we have on some open
neighborhood $V$ of $b$ that $E_{i}$ is given by $y_{i}=0$ for $i\in
\{1,\ldots,r\}$,
\begin{equation}
\sigma^{\ast}\left(  dx_{1}\wedge\ldots\wedge dx_{n}\right)  =\eta\left(
y\right)  \left(  \prod_{i=1}^{r}y_{i}^{v_{i}-1}\right)  dy_{1}\wedge
\ldots\wedge dy_{n}, \label{for2}%
\end{equation}
and
\begin{equation}
f_{j}^{\ast}(y):=\left(  f_{j}\circ\sigma\right)  \left(  y\right)
=\varepsilon_{f_{j}}\left(  y\right)  \prod_{i=1}^{r}y_{i}^{N_{f_{j},i}%
},\qquad\text{ for }j=1,\ldots,m, \label{for4}%
\end{equation}
where $\eta\left(  y\right)  $ and the $\varepsilon_{f_{j}}\left(  y\right)  $
belong to $\mathcal{O}_{X,b}^{\times}$, the group of units of the local ring
of $X$\ at $b$.
\end{theorem}

The Hironaka resolution theorem allows expressing a multivariate local zeta
function as a linear combination of monomial integrals, through a finite
sequence of changes of variables. We have the following theorem on
multivariate local zeta functions over local fields of zero characteristic.

\begin{theorem}
\cite[Lemma 6.4, Remark 2]{Bocardo:2020mk}\label{thm: num data and poles} Let
$f_{1}(x),\ldots,f_{m}(x)\in\mathbb{K}\left[  x_{1},\ldots,x_{n}\right]  $ be
non-constant polynomials and $\Phi:\mathbb{K}^{n}\rightarrow\mathbb{C}$ a
smooth function with compact support, to which we associate the multivariate
local zeta function $Z_{\Phi}\left(  \boldsymbol{f},\boldsymbol{s}\right)  $.
Fix an embedded resolution $\sigma:X\rightarrow\mathbb{K}$ of $D_{\mathbb{K}%
}=\cup_{i=1}^{m}f_{i}^{-1}(0)$ as in Theorem \ref{thresolsing} . Then

\noindent(i) $Z_{\Phi}\left(  \boldsymbol{f},\boldsymbol{s}\right)  $ is
convergent and defines a holomorphic function in the region
\[
\sum_{j=1}^{m}N_{f_{j},i}\operatorname{Re}(s_{j})+v_{i}>0,\qquad\text{ for
}i\in T ;
\]

\noindent(ii) $Z_{\Phi}\left(  \boldsymbol{f},\boldsymbol{s}\right)  $ admits
a meromorphic continuation to the whole $\mathbb{C}^{m}$, with poles belonging
to
\[
\bigcup_{i\in T}\text{ }\bigcup_{t\in\mathbb{N}}\left\{  \sum_{j=1}%
^{m}N_{f_{j},i}s_{j}+v_{i}+t=0\right\}
\]
with $t\in\mathbb{N}$ if $\mathbb{K}=\mathbb{R}$ and $t=\frac{1}{2}\mathbb{N}$
if $\mathbb{K}=\mathbb{C}$, and with poles belonging to
\[
\bigcup_{1\leq i\leq r}\text{ }\left\{  \sum_{j=1}^{m}a_{j,i}s_{j}%
+b_{i}=0\right\}  ,
\]
in the $p$-field case. In addition, in the $p$-adic case the multivariate
local zeta function has a meromorphic continuation as a rational function
\[
Z_{\Phi}\left(  \boldsymbol{f},\boldsymbol{s}\right)  =\frac{P_{\Phi}\left(
\boldsymbol{s}\right)  }{\prod\limits_{i\in T}\left(  1-q^{-\left(
{\textstyle\sum\limits_{j=1}^{m}}
N_{f_{j},i}s_{j}+v_{i}\right)  }\right)  }%
\]
in $q^{-s_{1}},\ldots,q^{-s_{m}}$, where $P_{\Phi}\left(  \boldsymbol{s}%
\right)  $ is a polynomial in the variables $q^{-s_{i}}$.
\end{theorem}

\section{\label{Section_8}Meromorphic Continuation of Koba-Nielsen Amplitudes
Defined on Local Fields of Characteristic Zero}

The Koba-Nielsen open string amplitudes for $N$-points over a local field
$\mathbb{K}$ of characteristic zero are defined as
\begin{equation}
A_{\mathbb{K}}^{(N)}\left(  \boldsymbol{k}\right)  :={\int\limits_{\mathbb{K}%
^{N-3}}}{\prod\limits_{i=2}^{N-2}}\left\vert x_{j}\right\vert _{\mathbb{K}%
}^{\boldsymbol{k}_{1}\boldsymbol{k}_{j}}\left\vert 1-x_{j}\right\vert
_{\mathbb{K}}^{\boldsymbol{k}_{N-1}\boldsymbol{k}_{j}}\text{ }{\prod
\limits_{2\leq i<j\leq N-2}}\left\vert x_{i}-x_{j}\right\vert _{\mathbb{K}%
}^{\boldsymbol{k}_{i}\boldsymbol{k}_{j}}{\prod\limits_{i=2}^{N-2}}%
dx_{i}\text{,} \label{Amplitude}%
\end{equation}
where $\boldsymbol{k}=\left(  \boldsymbol{k}_{1},\ldots,\boldsymbol{k}%
_{N}\right)  $, $\boldsymbol{k}_{i}=\left(  k_{0,i},\ldots,k_{l,i}\right)
\in\mathbb{C}^{l+1}$, for $i=1,\ldots,N$, is the momentum vector of the $i$-th
tachyon (with Minkowski product $\boldsymbol{k}_{i}\boldsymbol{k}_{j}%
=-k_{0,i}k_{0,j}+k_{1,i}k_{1,j}+\cdots+k_{l,i}k_{l,j}$), obeying
\begin{equation}
\sum_{i=1}^{N}\boldsymbol{k}_{i}=\boldsymbol{0}\text{, \ \ \ \ \ }%
\boldsymbol{k}_{i}\boldsymbol{k}_{i}=2\text{ \ for }i=1,\ldots,N.
\label{momenta-conservation}%
\end{equation}
The parameter $l$ is typically taken to be $25$. However, we do not require
using the critical dimension, thus $l$ can be any positive number. These
amplitudes were introduced by Brekke, Freund, Olson and Witten, among others
in their works about string amplitudes, see e.g. \cite[ Section 8]{B-F}. In
the case $N=4$ and $\mathbb{K}=\mathbb{R}$, the amplitude (\ref{Amplitude}) is
the Veneziano amplitude, see \cite{Veneziano}. In \cite[Section 2]{Kawai et
al} and \cite{Blumenhagen et al}, it was established that the $N$-point closed
string amplitude at the tree level is the product of $A_{\mathbb{C}}%
^{(N)}\left(  \boldsymbol{k}\right)  $ times a polynomial function in the
momenta $\boldsymbol{k}$. Hence, the results established in
\cite{Bocardo:2020mk} are still valid for closed string amplitudes at tree level.

If we take the Minkowski products of the kinematic parameters as follows
$\boldsymbol{k}_{i}\boldsymbol{k}_{j}=s_{ij}\in\mathbb{C}$, the string
amplitude (\ref{Amplitude}) becomes to the following integral which is a type
of multivariate local zeta function
\begin{equation}
Z_{\mathbb{K}}^{(N)}\left(  \boldsymbol{s}\right)  :=%
{\displaystyle\int\limits_{\mathbb{K}^{N-3}\smallsetminus D_{N}}}
{\displaystyle\prod\limits_{i=2}^{N-2}}
\left\vert x_{j}\right\vert _{\mathbb{K}}^{s_{1j}}\left\vert 1-x_{j}%
\right\vert _{\mathbb{K}}^{s_{(N-1)j}}\text{ }%
{\displaystyle\prod\limits_{2\leq i<j\leq N-2}}
\left\vert x_{i}-x_{j}\right\vert _{\mathbb{K}}^{s_{ij}}%
{\displaystyle\prod\limits_{i=2}^{N-2}}
dx_{i}, \label{zeta_funtion_string}%
\end{equation}
where ${\prod\nolimits_{i=2}^{N-2}}dx_{i}$ is the normalized Haar measure on
$\mathbb{K}^{N-3}$, $\boldsymbol{s}:=\left(  s_{ij}\right)  \in\mathbb{C}^{D}%
$, with $D=\frac{N(N-3)}{2}$ denotes the total number of indices $ij$, and
\[
D_{N}:=\left\{  x\in\mathbb{K}^{N-3};%
{\displaystyle\prod\limits_{i=2}^{N-2}}
x_{i}\text{ }%
{\displaystyle\prod\limits_{i=2}^{N-2}}
\left(  1-x_{i}\right)  \text{ }%
{\displaystyle\prod\limits_{2\leq i<j\leq N-2}}
\left(  x_{i}-x_{j}\right)  =0\right\}  .
\]
We have called integrals of type (\ref{zeta_funtion_string})
\textit{Koba-Nielsen local zeta functions}. For simplicity of notation, we put
$\mathbb{K}^{N-3}$ instead of $\mathbb{K}^{N-3}\smallsetminus D_{N}$ in
(\ref{zeta_funtion_string}).

In \cite{Bocardo:2020mk}, we establish, in a rigorous mathematical way, that
the Koba-Nielsen string amplitudes defined on any local field of
characteristic zero are bona fide integrals and that they can be extended to
meromorphic functions in the kinematic parameters. In order to prove the
meromorphic continuation of (\ref{zeta_funtion_string}), we express it as
linear combinations of local zeta functions. These computations were first
made in the case of $\mathbb{K}=\mathbb{R}$, but they can be easily extended
to other local fields of characteristic zero. Thus we only review the real
case. We consider $\mathbb{R}^{N-3}$ as an $\mathbb{R}$-analytic manifold,
with $N\geq4$, and use $\left\{  x_{2},\ldots,x_{N-2}\right\}  $ as a
coordinate system. In order to regularize the integral
(\ref{zeta_funtion_string}), we use a partition of $\mathbb{R}^{N-3}$
constructed using a smooth function $\chi:\mathbb{R}\rightarrow\mathbb{R}$
satisfying%
\[
\chi\left(  x\right)  =\left\{
\begin{array}
[c]{ccc}%
1 & \text{if} & x\in\left[  -2,2\right] \\
&  & \\
0 & \text{if} & x\in\left(  -\infty,-2-\epsilon\right]  \cup\left[
2+\epsilon,+\infty\right)  ,
\end{array}
\right.
\]
for some fixed positive $\epsilon$ sufficiently small. This function is
well-known, see e.g. \cite[Section 1.4]{Hormander}, \cite[Section 5.2]{Igusa}.
The number $2$ was chosen in an arbitrary form, the key point is that the
interval $\left[  0,1\right]  $ is included in the locus where $\chi\equiv1$.

Now, we can write
\begin{equation}
Z^{(N)}(\boldsymbol{s})=\sum_{I}Z_{I}^{(N)}(\boldsymbol{s}) \label{Eq_2A}%
\end{equation}
with
\begin{equation}
Z_{I}^{(N)}(\boldsymbol{s}):=%
{\displaystyle\int\limits_{\mathbb{R}^{N-3}}}
\varphi_{I}\left(  x\right)
{\displaystyle\prod\limits_{j=2}^{N-2}}
\left\vert x_{j}\right\vert ^{s_{1j}}%
{\displaystyle\prod\limits_{j=2}^{N-2}}
\left\vert 1-x_{j}\right\vert ^{s_{\left(  N-1\right)  j}}%
{\displaystyle\prod\limits_{2\leq i<j\leq N-2}}
\left\vert x_{i}-x_{j}\right\vert ^{s_{ij}}\text{ }%
{\displaystyle\prod\limits_{i=2}^{N-2}}
dx_{i}, \label{Eq_3}%
\end{equation}
where the functions $\{\varphi_{I}\}$ are defined as
\begin{equation}
\varphi_{I}:\mathbb{R}^{N-3}\rightarrow\mathbb{R};\text{ }x\mapsto%
{\displaystyle\prod\limits_{i\in I}}
\chi\left(  x_{i}\right)
{\displaystyle\prod\limits_{i\notin I}}
\left(  1-\chi\left(  x_{i}\right)  \right)  , \label{Function_phi}%
\end{equation}
for $I\subseteq\left\{  2,\ldots,N-2\right\}  $, including the empty set, with
the convention that $%
{\textstyle\prod\nolimits_{i\in\emptyset}}
\cdot\equiv1$. Notice that $\varphi_{I}\in C^{\infty}\left(  \mathbb{R}%
^{N-3}\right)  $ and $\sum_{I}\varphi_{I}\left(  x\right)  \equiv1$, for
$x\in\mathbb{R}^{N-3}$, that is, the functions $\{\varphi_{I}\}$ form a
partition of the unity.

In the case $I=\left\{  2,\ldots,N-2\right\}  $, $Z_{I}^{(N)}(\boldsymbol{s})$
is a classical multivariate Igusa local zeta function (since $\varphi
_{I}\left(  x\right)  $ has compact support). It is well known that these
integrals are holomorphic functions in a region including $\operatorname{Re}%
\left(  s_{ij}\right)  >0$ for all $ij$, and they admit meromorphic
continuations to the whole $\mathbb{C}^{D}$, see \cite[Theorem 3.2]%
{Bocardo:2020mk}.

In the case $I\neq\left\{  2,\ldots,N-2\right\}  $, by changing variables in
(\ref{Eq_3}) as $x_{i}\rightarrow\frac{1}{x_{i}}$ for $i\not \in I$, and
$x_{i}\rightarrow x_{i}$ for $i\in I$, we have $%
{\textstyle\prod\nolimits_{i=2}^{N-2}}
dx_{i}\rightarrow%
{\textstyle\prod\nolimits_{i\not \in I}}
\frac{1}{\left\vert x_{i}\right\vert ^{2}}$ $%
{\textstyle\prod\nolimits_{i=2}^{N-2}}
dx_{i}$, and by setting $\widetilde{\chi}\left(  x_{i}\right)  :=1-\chi\left(
\frac{1}{x_{i}}\right)  $ for $i\not \in I$, i.e.,%
\[
\widetilde{\chi}\left(  x_{i}\right)  =\left\{
\begin{array}
[c]{ccc}%
1 & \text{if} & \left\vert x_{i}\right\vert \leq\frac{1}{2+\epsilon}\\
&  & \\
0 & \text{if} & \left\vert x_{i}\right\vert \geq\frac{1}{2},
\end{array}
\right.
\]
we have that supp $\widetilde{\chi}\subseteq\left[  -\frac{1}{2},\frac{1}%
{2}\right]  $ and $\widetilde{\chi}\in C^{\infty}\left(  \mathbb{R}\right)  $.
Now setting $\widetilde{\varphi}_{I}\left(  x\right)  :=\prod
\nolimits_{i\not \in I}\widetilde{\chi}\left(  x_{i}\right)  $ $\prod
\nolimits_{i\in I}\chi\left(  x_{i}\right)  $, and
\begin{align*}
F_{I}\left(  x,\boldsymbol{s}\right)   &  :=\prod\limits_{j\in I}\left\vert
x_{j}\right\vert ^{s_{1j}}\text{ }\prod\limits_{j=2}^{N-2}\left\vert
1-x_{j}\right\vert ^{s_{\left(  N-1\right)  j}}\prod\limits_{\substack{2\leq
i<j\leq N-2\\i,\text{ }j\in I}}\left\vert x_{i}-x_{j}\right\vert ^{s_{ij}%
}\text{ }\\
\times &  \prod\limits_{\substack{2\leq i<j\leq N-2\\i,\text{ }j\not \in
I}}\left\vert x_{i}-x_{j}\right\vert ^{s_{ij}}\text{ }\prod
\limits_{\substack{2\leq i<j\leq N-2\\i\not \in I,\text{ }j\in I}}\left\vert
1-x_{i}x_{j}\right\vert ^{s_{ij}}\prod\limits_{\substack{2\leq i<j\leq
N-2\\i\in I,\text{ }j\not \in I}}\left\vert 1-x_{i}x_{j}\right\vert ^{s_{ij}},
\end{align*}
we have
\begin{equation}
Z_{I}^{(N)}(\boldsymbol{s})=%
{\displaystyle\int\limits_{\mathbb{R}^{N-3}\smallsetminus D_{I}}}
\text{ }\frac{\widetilde{\varphi}_{I}\left(  x\right)  F_{I}\left(
x,\boldsymbol{s}\right)  }{\prod\limits_{i\not \in I}\left\vert x_{i}%
\right\vert ^{s_{1i}+s_{\left(  N-1\right)  i}+\sum_{\substack{2\leq j\leq
N-2\\j\neq i}}s_{ij}+2}\text{ }}%
{\displaystyle\prod\limits_{i=2}^{N-2}}
dx_{i}, \label{Eq 4}%
\end{equation}
where $D_{I}$ is the divisor defined by the polynomial
\begin{multline*}
\prod\limits_{i=2}^{N-2}x_{i}\text{ }\prod\limits_{i=2}^{N-2}\left(
1-x_{i}\right)  \text{ }\prod\limits_{\substack{2\leq i<j\leq N-2\\i,\text{
}j\in I}}\left(  x_{i}-x_{j}\right)  \prod\limits_{\substack{2\leq i<j\leq
N-2\\i,\text{ }j\notin I}}\left(  x_{i}-x_{j}\right) \\
\times\text{ }\prod\limits_{\substack{2\leq i<j\leq N-2\\i\notin I,\text{
}j\in I}}\left(  1-x_{i}x_{j}\right)  \text{ }\prod\limits_{\substack{2\leq
i<j\leq N-2\\\text{ }i\in I,\text{ }j\notin I}}\left(  1-x_{i}x_{j}\right)  .
\end{multline*}
The integrals $Z_{I}^{(N)}(\boldsymbol{s})$, with $I\neq\left\{
2,\ldots,N-2\right\}  $, are not classical multivariate local zeta functions
thus we don't apply the theory of local zeta functions. Thus, in
\cite{Bocardo:2020mk} we show that they define holomorphic functions on some
nonempty open in $\mathbb{C}^{D}$, and admit meromorphic continuations to the
whole $\mathbb{C}^{D}$.

\begin{lemma}
{\cite[Lemma 4.1]{Bocardo:2020mk}} \label{Lemma1} For any $I\subseteq\left\{
2,\ldots,N-2\right\}  $, the function $Z_{I}^{(N)}(\boldsymbol{s})$ is
holomorphic in $\boldsymbol{s}$ on the solution set $\mathcal{H}(I)$ of a
system of inequalities of the form
\begin{equation}
\mathcal{H}(I):=\left\{  s_{ij}\in\mathbb{C}^{D};\text{ }\sum_{ij\in M\left(
I\right)  }N_{ij,k}\left(  I\right)  \operatorname{Re}\left(  s_{ij}\right)
+\gamma_{k}\left(  I\right)  >0,\text{ for }k\in T(I)\right\}  \neq\emptyset,
\label{Eq_7}%
\end{equation}
where $N_{ij,k}\left(  I\right)  ,\gamma_{k}\left(  I\right)  \in\mathbb{Z}$,
and $M(I)$, $T(I)$ are finite sets. More precisely, for each $k$, either all
numbers $N_{ij,k}\left(  I\right)  $ are equal to $0$ or $1$ and $\gamma
_{k}\left(  I\right)  >0$, or all numbers $N_{ij,k}\left(  I\right)  $ are
equal to $0$ or $-1$ and $\gamma_{k}\left(  I\right)  <0$.

In addition, $Z_{I}^{(N)}(\boldsymbol{s})$ admits an analytic continuation to
the whole $\mathbb{C}^{D}$, as a meromorphic function with poles belonging to
\begin{equation}
\mathcal{P}(I):=\bigcup\limits_{t\in\mathbb{N}}\bigcup\limits_{k\in
T(I)}\left\{  s_{ij}\in\mathbb{C}^{D};\sum_{ij\in M\left(  I\right)  }%
N_{ij,k}(I)s_{ij}+\gamma_{k}(I)+t=0\right\}  . \label{Eq_7A}%
\end{equation}

\end{lemma}

In order to show the existence of the meromorphic continuation of
$Z^{(N)}(\boldsymbol{s})$ it is necessary that all the integrals $Z_{I}%
^{(N)}(\boldsymbol{s})$ be holomorphic in a common domain, and then formula
(\ref{Eq_2A}) allows us to construct a meromorphic continuation of
$Z^{(N)}(\boldsymbol{s})$. We show that $\cap_{I}\mathcal{H}(I)$ contains a
non-empty open subset of $\mathbb{C}^{D}$ by studying the possible poles of
integrals
\[
Z_{I}^{(N)}(s):=Z_{I}^{(N)}(\boldsymbol{s})\mid_{s_{ij}=s}%
\]
for any $I$, and proving that $Z^{(N)}(s):=Z^{(N)}(\boldsymbol{s})\mid
_{s_{ij}=s}$ is a holomorphic function in the region
\[
-\frac{2}{N-2}<\operatorname{Re}(s)<-\frac{2}{N},
\]
This fact was proved in \cite[Theorem 4.1]{Bocardo:2020mk}.

Furthermore, the Koba-Nielsen local zeta function $Z^{(N)}(\boldsymbol{s})$ is
convergent and holomorphic in the region determined by the following
inequalities:
\[
\operatorname{Re}(s_{ij})>-1
\]
for all $\frac{N(N-3)}{2}$ variables $s_{ij}$,
\[
\sum_{j\in J}\operatorname{Re}(s_{1j})+\sum_{\substack{i,j\in J\\i<j}%
}\operatorname{Re}(s_{ij})>-\#J
\]
for all subsets $J\subset\{2,\dots,N-2\}$ with $\#J\geq2$,
\[
\sum_{j\in J}\operatorname{Re}(s_{(N-1)j})+\sum_{\substack{i,j\in
J\\i<j}}\operatorname{Re}(s_{ij})>-\#J
\]
for all subsets $J\subset\{2,\dots,N-2\}$ with $\#J\geq2$,
\[
\sum_{\substack{i,j\in J\\i<j}}\operatorname{Re}(s_{ij})>-\#J+1
\]
for all subsets $J\subset\{2,\dots,N-2\}$ with $\#J\geq3$,
\[
\sum_{j\in J}\operatorname{Re}(s_{1j})+\sum_{j\in J}\operatorname{Re}%
(s_{\left(  N-1\right)  j})+\sum_{\substack{i\in\{2,\dots,N-2\}\setminus
J\\j\in J}}\operatorname{Re}(s_{ij})+\sum_{\substack{i,j\in J\\i<j}%
}\operatorname{Re}(s_{ij})<-\#J
\]
for all subsets $J\subset\{2,\dots,N-2\}$ with $\sharp J\geq1$. This region
contains the open subset defined by
\[
-\frac{2}{N-2}<\operatorname{Re}(s_{ij})<-\frac{2}{N},
\]
for all indices $ij$.

As in the $p$-adic case, we use the meromorphic continuation of
(\ref{zeta_funtion_string}) to the whole $\mathbb{C}^{D}$, which is denoted by
$Z_{\mathbb{K}}^{(N)}(\boldsymbol{s})$, as regularizations of the amplitudes
$A_{\mathbb{K}}^{(N)}\left(  \boldsymbol{k}\right)  $ by redefining
\[
A_{\mathbb{K}}^{(N)}\left(  \boldsymbol{k}\right)  =Z_{\mathbb{K}}%
^{(N)}(\boldsymbol{s})\mid_{s_{ij}=\boldsymbol{k}_{i}\boldsymbol{k}_{j}},
\]
see \cite[Theorem 6.1]{Bocardo:2020mk}. It is important to mention here that,
in the regularization of $A_{\mathbb{K}}^{(N)}\left(  \boldsymbol{k}\right)
$, we do not use the kinematic restrictions (\ref{momenta-conservation}).

Furthermore, in \cite{Bocardo:2020mk} we show that $A_{\mathbb{K}}%
^{(N)}\left(  \boldsymbol{k}\right)  $ converges on some open subset of
$\mathbb{C}^{(N-1)(l+1)}$ by showing that this open is mapped into the domain
of convergency of $Z_{\mathbb{K}}^{(N)}\left(  \boldsymbol{s}\right)  $ by
$\boldsymbol{k}\rightarrow s_{ij}=\boldsymbol{k}_{i}\boldsymbol{k}_{j}$. In
addition, we prove that $A_{\mathbb{K}}^{(N)}\left(  \boldsymbol{k}\right)  $
extends to a meromorphic function to the whole $\mathbb{C}^{(N-1)(l+1)}$, and
that its polar set is contained in the inverse image of the polar set of
$Z_{\mathbb{K}}^{(N)}\left(  \boldsymbol{s}\right)  $ under that mapping,
where the possible poles are described in terms of numerical data of suitable
resolutions of singularities:

\begin{theorem}
{\cite[Theorem 7.1]{Bocardo:2020mk}} Let $\mathbb{K}$ be a local field of
characteristic zero. The integral $A_{\mathbb{K}}^{(N)}\left(  \boldsymbol{k}%
\right)  $ converges and is holomorphic in the open set $\mathcal{U}%
\subset\mathbb{C}^{\left(  N-1\right)  (l+1)}$. It extends to a meromorphic
function in $\boldsymbol{k}$ on the whole $\mathbb{C}^{\left(  N-1\right)
(l+1)}$.

If $\mathbb{K}$ is an $\mathbb{R}$-field, then the possible poles of
$A_{\mathbb{K}}^{(N)}\left(  \boldsymbol{k}\right)  $ belong to
\begin{equation}%
{\displaystyle\bigcup\limits_{I\subseteq\left\{  2,\ldots,N-2\right\}  }}
\bigcup\limits_{t\in\mathbb{N}}\bigcup\limits_{r\in T(I)}\left\{
\boldsymbol{k}\in\mathbb{C}^{\left(  N-1\right)  (l+1)};\sum_{ij\in M\left(
I\right)  }N_{ij,r}(I)\boldsymbol{k}_{i}\boldsymbol{k}_{j}+\gamma_{r}%
(I)+\frac{t}{\left[  \mathbb{K}:\mathbb{R}\right]  }=0\right\}  ,
\label{polarlocusR}%
\end{equation}
where $N_{ij,r}\left(  I\right)  ,\gamma_{r}\left(  I\right)  \in\mathbb{Z}$,
and $M(I)$, $T(I)$ are finite sets, and $\left[  \mathbb{K}:\mathbb{R}\right]
=1$ if $\mathbb{K=R}$, and $\left[  \mathbb{K}:\mathbb{R}\right]  =2$ if
$\mathbb{K=C}$. If $\mathbb{K}$ is a $p$-adic field, then $A_{\mathbb{K}%
}^{(N)}\left(  \boldsymbol{k}\right)  $ is a rational function in the
variables $q^{-\boldsymbol{k}_{i}\boldsymbol{k}_{j}}$, and its possible poles
belong to
\[%
{\displaystyle\bigcup\limits_{I\subseteq\left\{  2,\ldots,N-2\right\}  }}
\bigcup\limits_{r\in T(I)}\left\{  \boldsymbol{k}\in\mathbb{C}^{\left(
N-1\right)  (l+1)};\sum_{ij\in M\left(  I\right)  }N_{ij,r}%
(I)\operatorname{Re}(\boldsymbol{k}_{i}\boldsymbol{k}_{j})+\gamma
_{r}(I)=0\right\}  .
\]
More precisely, for each $r$, either all numbers $N_{ij,r}\left(  I\right)  $
are equal to $0$ or $1$ and $\gamma_{r}\left(  I\right)  >0$, or all numbers
$N_{ij,r}\left(  I\right)  $ are equal to $0$ or $-1$ and $\gamma_{r}\left(
I\right)  <0$.
\end{theorem}

We now show explicitly the existence of a meromorphic continuation for
$Z_{\mathbb{R}}^{(N)}\left(  \boldsymbol{s}\right)  $\ in the cases $N=4$,
$5$, by using Hironaka's resolution of singularities theorem.

\subsection{Example: $4$-point Koba-Nielsen string
amplitude\label{Example_N_4}}

The $4$-point Koba-Nielsen open string amplitude is defined as
\[
Z^{(4)}(s)=%
{\displaystyle\int\limits_{\mathbb{R}}}
|x_{2}|^{s_{12}}|1-x_{2}|^{s_{32}}dx_{2}.
\]
By using the function
\begin{equation}
\chi\left(  x_{2}\right)  =\left\{
\begin{array}
[c]{ccc}%
1 & \text{if} & x_{2}\in\left[  -2,2\right] \\
&  & \\
0 & \text{if} & x_{2}\in\left(  -\infty,-2-\epsilon\right]  \cup\left[
2+\epsilon,+\infty\right)  ,
\end{array}
\right.  \label{Basic function}%
\end{equation}
where $\epsilon>0$ is sufficiently small, we construct the following partition
of the unity:
\[%
\begin{array}
[c]{ccccl}%
\varphi_{\{2\}}: & \mathbb{R} & \rightarrow & \mathbb{R}; & x_{2}\mapsto
\chi\left(  x_{2}\right) \\
\varphi_{\varnothing}: & \mathbb{R} & \rightarrow & \mathbb{R}; & x_{2}%
\mapsto1-\chi\left(  x_{2}\right)
\end{array}
.
\]
Notice that $\varphi_{\varnothing},\varphi_{\{2\}}\in C^{\infty}\left(
\mathbb{R}\right)  $ and $\varphi_{\{2\}}\left(  x\right)  +\varphi
_{\varnothing}\left(  x\right)  \equiv1$, for $x\in\mathbb{R}$. Hence,
\[
Z^{(4)}(\boldsymbol{s})=Z_{\{2\}}^{(4)}(\boldsymbol{s})+Z_{\varnothing}%
^{(4)}(\boldsymbol{s}),
\]
with
\begin{equation}
Z_{\{2\}}^{(4)}(\boldsymbol{s})=%
{\displaystyle\int\limits_{\mathbb{R}}}
\chi(x_{2})|x_{2}|^{s_{12}}|1-x_{2}|^{s_{32}}dx_{2} \label{4-1ex}%
\end{equation}
and
\begin{equation}
Z_{\varnothing}^{(4)}(\boldsymbol{s})=%
{\displaystyle\int\limits_{\mathbb{R}}}
(1-\chi(x_{2}))|x_{2}|^{s_{12}}|1-x_{2}|^{s_{32}}dx_{2}. \label{4-2Ex}%
\end{equation}
The integral (\ref{4-1ex}) is a multivariate local zeta function since
$\chi(x_{2})$ has compact support. By a classical result of local zeta
functions $Z_{\{2\}}^{(4)}(\boldsymbol{s})$ converges when $\operatorname{Re}%
(s_{12})>-1$ and $\operatorname{Re}(s_{32})>-1.$ Furthermore, it admits a
meromorphic continuation to the whole $\mathbb{C}^{2}$, see \cite[Theorem
3.2]{Bocardo:2020mk}.

The second integral (\ref{4-2Ex}) is not a multivariate local zeta function
but it can be transformed in one by changing of variables as $x_{2}%
\rightarrow\frac{1}{x_{2}}$, then $dx_{2}\rightarrow\frac{1}{\left\vert
x_{2}\right\vert ^{2}}$ $dx_{2}$, and by setting $\widetilde{\chi}\left(
x_{2}\right)  :=1-\chi\left(  \frac{1}{x_{2}}\right)  $, we have that
\[
\widetilde{\chi}\left(  x_{2}\right)  =\left\{
\begin{array}
[c]{ccc}%
1 & \text{if} & \left\vert x_{2}\right\vert \leq\frac{1}{2+\epsilon}\\
&  & \\
0 & \text{if} & \left\vert x_{2}\right\vert \geq\frac{1}{2},
\end{array}
\right.  \text{.}%
\]
Notice that the support of $\widetilde{\chi}$ is contained in $\left[
-\frac{1}{2},\frac{1}{2}\right]  $ and $\widetilde{\chi}\in C^{\infty}\left(
\mathbb{R}\right)  $. Thus integral (\ref{4-2Ex}) becomes
\[
Z_{\emptyset}^{(4)}(\boldsymbol{s})=\int_{\mathbb{R}}\widetilde{\chi}%
(x_{2})|x_{2}|^{-s_{12}-s_{32}-2}|1-x_{2}|^{s_{32}}dx_{2},
\]
which is analytic when $-\operatorname{Re}(s_{12})-\operatorname{Re}%
(s_{32})-1>0$ and $\operatorname{Re}(s_{32})+1>0$. We concluded that
$Z^{(4)}(s)$ is analytic in the region
\[
\operatorname{Re}(s_{12})>-1,\quad\operatorname{Re}(s_{32})>-1,\quad
\operatorname{Re}(s_{12})+\operatorname{Re}(s_{32})<-1.
\]
Which contains the open set $-1<\operatorname{Re}(s_{12})<-\frac{1}{2}$ and
$-1<\operatorname{Re}(s_{32})<-\frac{1}{2}$.

\subsection{Example: $5$-point Koba-Nielsen string
amplitude\label{Example_N_5}}

Fix $N=5$. By using the function $\chi\left(  x\right)  $, see
(\ref{Basic function}), we define the following partition of the unity:
\[%
\begin{array}
[c]{ccccl}%
\varphi_{\{2,3\}}: & \mathbb{R}^{2} & \rightarrow & \mathbb{R}; & (x_{2}%
,x_{3})\mapsto\chi\left(  x_{2}\right)  \chi\left(  x_{3}\right)  ,\\
\varphi_{\{2\}}: & \mathbb{R}^{2} & \rightarrow & \mathbb{R}; & (x_{2}%
,x_{3})\mapsto\chi\left(  x_{2}\right)  (1-\chi\left(  x_{3}\right)  ),\\
\varphi_{\{3\}}: & \mathbb{R}^{2} & \rightarrow & \mathbb{R}; & (x_{2}%
,x_{3})\mapsto\chi\left(  x_{3}\right)  (1-\chi\left(  x_{2}\right)  ),\\
\varphi_{\varnothing}: & \mathbb{R}^{2} & \rightarrow & \mathbb{R}; &
(x_{2},x{3})\mapsto(1-\chi\left(  x_{2}\right)  )(1-\chi\left(  x_{3}\right)
).
\end{array}
\]
Then $Z^{(5)}(\boldsymbol{s})=Z_{\{2,3\}}^{(5)}(\boldsymbol{s})+Z_{\{3\}}%
^{(5)}(\boldsymbol{s})+Z_{\{2\}}^{(5)}(\boldsymbol{s})+Z_{\emptyset}%
^{(5)}(\boldsymbol{s})$. We consider the first integral
\begin{equation}
Z_{\{2,3\}}^{(5)}(\boldsymbol{s})=%
{\displaystyle\int\limits_{\mathbb{R}^{2}}}
\chi(x_{2})\chi(x_{3})|x_{2}|^{s_{12}}|x_{3}|^{s_{13}}|1-x_{2}|^{s_{42}%
}|1-x_{3}|^{s_{43}}|x_{2}-x_{3}|^{s_{23}}dx_{2}dx_{3}. \label{Z23}%
\end{equation}
Since $\varphi_{\{2,3\}}$ has compact support, then integral (\ref{Z23}) is a
local zeta function. Thus, we use resolution of singularities of the divisor
$D_{5}$ defined by $x_{2}x_{3}(1-x_{2})(1-x_{3})(x_{2}-x_{3})=0$. This
arrangement is not locally monomial only at the points $(0,0)$ and $(1,1)$.
Hence, we pick a partition of the unity, $\sum_{i=0}^{2}$ $\Omega_{i}%
(x_{2},x_{3})=1$, and we write
\[
Z_{\{2,3\}}^{(5)}(s)=\sum_{j=0}^{2}Z_{\Omega_{j}}^{(5)}(s),
\]
where
\[
Z_{\Omega_{j}}^{(5)}(s)=\int_{\mathbb{R}^{2}}\Omega_{j}(x_{2},x_{3}%
)|x_{2}|^{s_{12}}|x_{3}|^{s_{13}}|1-x_{2}|^{s_{42}}|1-x_{3}|^{s_{43}}%
|x_{2}-x_{3}|^{s_{23}}dx_{2}dx_{3}.
\]
We assume that $\Omega_{0}$ and $\Omega_{1}$ are smooth functions with support
in a small neighborhood of $(0,0)$ and $(1,1)$, respectively. So we only
compute the integrals $Z_{\Omega_{0}}^{(5)}(s)$ and $Z_{\Omega_{1}}^{(5)}(s)$.

In terms of convergence and holomorphy, around $(0,0)$, the factor
$|1-x_{2}|^{s_{42}}|1-x_{3}|^{s_{43}}$ can be neglected, hence we only need an
embedded resolution of $x_{2}x_{3}(x_{2}-x_{3})=0$, which is obtained by a
blow-up at the origin. It means to make the changes of variables
\begin{align*}
\sigma_{0}:\mathbb{R}^{2}\rightarrow\mathbb{R}^{2};  &  \ u_{2}\mapsto
x_{2}=u_{2}\\
&  \ u_{3}\mapsto x_{3}=u_{2}u_{3}.
\end{align*}
Hence integral $Z_{\Omega_{0}}^{(5)}(s)$ becomes
\[
{\int\limits_{\mathbb{R}^{2}}}\left(  \Omega_{0}\circ\sigma_{0}\right)
\left(  u_{2},u_{3}\right)  \left\vert u_{2}\right\vert ^{s_{12}+s_{13}%
+s_{23}+1}|u_{3}|^{s_{13}}|1-u_{3}|^{s_{23}}g(u,\boldsymbol{s})du_{2}du_{3},
\]
where $g(u,s)$ is invertible on the support of $\Omega_{0}\circ\sigma_{0}$ and
can thus be neglected from the point of view of convergence and holomorphy.
Hence, by \cite[Lemma 3.1]{Bocardo:2020mk}, we obtain the convergence
conditions:
\begin{equation}
\operatorname{Re}(s_{12})+\operatorname{Re}(s_{13})+\operatorname{Re}%
(s_{23})+2>0,\quad\operatorname{Re}(s_{23})+1>0,\quad\operatorname{Re}%
(s_{13})+1>0. \label{ex1}%
\end{equation}
We can take other chart of the blow-up, i.e. the change of variables
\begin{align*}
\sigma_{0}^{\prime}:\mathbb{R}^{2}\rightarrow\mathbb{R}^{2};  &
\ u_{2}\mapsto x_{2}=u_{2}u_{3}\\
&  \ u_{3}\mapsto x_{3}=u_{3},
\end{align*}
with this change yields the same first and second condition and also
\begin{equation}
\operatorname{Re}(s_{12})+1>0. \label{ex2}%
\end{equation}
By symmetry we can consider any of these change of variables. Completely
similarly, for the convergence of $Z_{\Omega_{1}}^{(5)}(s)$, we need also the
new conditions
\begin{equation}
\operatorname{Re}(s_{42})+\operatorname{Re}(s_{43})+\operatorname{Re}%
(s_{23})+2>0,\quad\operatorname{Re}(s_{42})+1>0,\quad\operatorname{Re}%
(s_{43})+1>0. \label{ex3}%
\end{equation}
The conditions coming from the locally monomial integral $Z_{\Omega_{2}}%
^{(5)}(s)$ are already included.

The integral $Z_{\{3\}}^{(5)}(\boldsymbol{s})$ is not a multivariate local
zeta function, so we take the following change of variables $x_{2}%
\rightarrow\frac{1}{x_{2}}$, then we have $dx_{2}\rightarrow\frac
{1}{\left\vert x_{2}\right\vert ^{2}}$ $dx_{2}$, and by setting $\widetilde
{\chi}\left(  x_{2}\right)  :=1-\chi\left(  \frac{1}{x_{2}}\right)  $, we have
that
\[
\widetilde{\chi}\left(  x_{2}\right)  =\left\{
\begin{array}
[c]{ccc}%
1 & \text{if} & \left\vert x_{2}\right\vert \leq\frac{1}{2+\epsilon}\\
&  & \\
0 & \text{if} & \left\vert x_{2}\right\vert \geq\frac{1}{2},
\end{array}
\right.  \text{.}%
\]
Then
\begin{multline*}
Z_{\{3\}}^{(5)}(\boldsymbol{s})=\int_{\mathbb{R}^{2}}\widetilde{\chi}%
(x_{2})\chi(x_{3})|x_{2}|^{-s_{12}-s_{42}-s_{23}-2}|x_{3}|^{s_{13}}%
|1-x_{2}|^{s_{42}}|1-x_{3}|^{s_{43}}\times\\
|1-x_{2}x_{3}|^{s_{23}}dx_{2}dx_{3}.
\end{multline*}
Since $x_{2}x_{3}(1-x_{2})(1-x_{3})(1-x_{2}x_{3})$ is locally monomial in the
support of $\widetilde{\chi}(x_{2})\chi(x_{3})$, the only new condition is
\begin{equation}
-\operatorname{Re}(s_{12})-\operatorname{Re}(s_{42})-\operatorname{Re}%
(s_{23})-1>0. \label{ex4}%
\end{equation}
Completely analogously, $Z_{\{2\}}^{(5)}(\boldsymbol{s})$ induces the extra
condition
\begin{equation}
-\operatorname{Re}(s_{13})-\operatorname{Re}(s_{43})-\operatorname{Re}%
(s_{23})-1>0. \label{ex5}%
\end{equation}
For the last integral, we setting $\widetilde{\chi}\left(  x_{2}\right)
:=1-\chi\left(  \frac{1}{x_{2}}\right)  $ and $\widetilde{\chi}\left(
x_{3}\right)  :=1-\chi\left(  \frac{1}{x_{3}}\right)  $. Thus,
\begin{multline*}
Z_{\emptyset}^{(5)}(\boldsymbol{s})=\int_{\mathbb{R}^{2}}\widetilde{\chi
}(x_{2})\widetilde{\chi}(x_{3})|x_{2}|^{-s_{12}-s_{42}-s_{23}-2}%
|x_{3}|^{-s_{13}-s_{43}-s_{23}-2}|1-x_{2}|^{s_{42}}|1-x_{3}|^{s_{43}}\times\\
|x_{2}-x_{3}|^{s_{23}}dx_{2}dx_{3}.
\end{multline*}
In this case, we have the same divisor as for $Z_{\left\{  2,3\right\}
}^{(5)}(\boldsymbol{s})$; the differences\textrm{ }are the powers of $|x_{2}|$
and $|x_{3}|$ and the function $\widetilde{\chi}(x_{2})\widetilde{\chi}%
(x_{3})$, that does not contain $(1,1)$ in its support. Hence, the only new
condition will arise from the blow-up at the origin, namely
\begin{equation}
-\operatorname{Re}(s_{12})-\operatorname{Re}(s_{42})-\operatorname{Re}%
(s_{23})-\operatorname{Re}(s_{13})-\operatorname{Re}(s_{43})-2>0. \label{ex6}%
\end{equation}
Hence $Z^{(5)}(\boldsymbol{s})$ is analytic in the region defined by
conditions (\ref{ex1})-(\ref{ex6}), that is,
\[
\begin{aligned} & \operatorname{Re}(s_{ij}) > -1 , \text{ for all ij} \\ & \operatorname{Re}(s_{12})+ \operatorname{Re}(s_{13})+\operatorname{Re}(s_{23}) > -2, \quad \operatorname{Re}(s_{42})+\operatorname{Re}(s_{43})+\operatorname{Re}(s_{23}) > -2 , \\ & \operatorname{Re}(s_{12})+\operatorname{Re}(s_{42})+\operatorname{Re}(s_{23})< -1 , \quad \operatorname{Re}(s_{13})+\operatorname{Re}(s_{43})+\operatorname{Re}(s_{23}) < -1 ,\\ & \sum_{ij} \operatorname{Re}(s_{ij}) < -2 . \end{aligned}
\]
This region of convergence contains the open subset defined by
\[
-\frac{2}{3}<\operatorname{Re}(s_{ij})<-\frac{2}{5}\qquad\text{ for all }ij.
\]
Then, in particular, $Z^{(5)}(s)$ is analytic in the interval $-\frac{2}%
{3}<\operatorname{Re}(s)<-\frac{2}{5}$.

\bigskip

\textbf{Funding:} The first author was partially supported by the University
of Guadalajara. The fourth author was partially supported by Conacyt Grant No.
250845 (Mexico) and by the Debnath Endowed Professorship (UTRGV, USA).

\textbf{Conflicts of Interest:} The authors declare no conflict of interest.

\textbf{ Author Contributions:} all the authors contributed to the manuscript
equally. All authors have read and agreed to the published version of the manuscript.

\end{document}